\documentclass[fleqn,usenatbib]{mnras}

\usepackage[T1]{fontenc}

\DeclareRobustCommand{\VAN}[3]{#2}
\let\VANthebibliography\thebibliography
\def\thebibliography{\DeclareRobustCommand{\VAN}[3]{##3}\VANthebibliography}

\usepackage{graphicx}	
\usepackage{amsmath}	
\usepackage{amssymb}	
\usepackage{pdflscape}

\usepackage{newtxtext,newtxmath}

\title[X-ray properties of early-type stars from T-ReX]{X-ray properties of early-type stars in the Tarantula Nebula from T-ReX}

\author[Crowther et al.]{
Paul A. Crowther$^{1}$\thanks{paul.crowther@sheffield.ac.uk}, Patrick S. Broos$^{2}$, Leisa K. Townsley$^{2}$, Andy M.T. Pollock$^{1}$, Katie A. Tehrani$^{1}$, \newauthor Marc Gagn\'{e}$^{3}$\\
1. Department of Physics and Astronomy, University of Sheffield, Sheffield, S3 7RH, UK\\
2. Department of Astronomy and Astrophysics, 525 Davey Laboratory, Pennsylvania State University, University Park, PA 16802, USA\\
3. Department of Earth and Space Sciences, West Chester University, West Chester, PA 19383, USA
}

\date{Accepted 2022 July 5. Received 2022 July 5; in original form 2022 April 8}

\pubyear{2022}

\begin{document}
\label{firstpage}
\pagerange{\pageref{firstpage}--\pageref{lastpage}}
\maketitle

\begin{abstract}
We reassess the historical $L_{\rm X}/L_{\rm Bol}$ relation for early-type stars from a comparison between T-ReX, the  {\it Chandra} ACIS X-ray survey of the Tarantula Nebula in the LMC, and contemporary spectroscopic analysis of massive stars obtained primarily from {\it VLT}/FLAMES, {\it VLT}/MUSE and {\it HST}/STIS surveys. For 107 sources in common (some host to multiple stars), the majority of which are bolometrically luminous (40\% exceed $10^6 L_{\odot}$), we find an average $\log L_{\rm X} /L_{\rm Bol} = -6.90 \pm 0.65$. Excluding extreme systems  Mk\,34 (WN5h+WN5h), R140a (WC4+WN6+) and VFTS 399 (O9\,IIIn+?), plus four WR sources with anomalously hard X-ray components (R130, R134, R135, Mk~53) and 10 multiple sources within the spatially crowded core of R136a, $\log L_{\rm X} /L_{\rm Bol} = -7.00 \pm 0.49$, in good agreement with Galactic OB stars. No difference is found between single and binary systems, nor between  O, Of/WN and WR stars, although there does appear to be a trend towards harder X-ray emission from O dwarfs, through O (super)giants, Of/WN stars and WR stars. The majority of known OB stars in the Tarantula are not detected in the T-ReX point source catalogue, so we have derived upper limits for all undetected OB stars for which log $L_{\rm Bol}/L_{\odot} \geq 5.0$. A survival analysis using detected and upper-limit log $L_{\rm X}/L_{\rm Bol}$ values indicates no significant difference between luminous O stars in the LMC and the Carina Nebula. This analysis suggests that metallicity does not strongly influence $L_{\rm X}/L_{\rm Bol}$. Plasma temperatures for single, luminous O stars in the Tarantula ($\overline{kT_{m}}=1.0$ keV) are higher than counterparts in Carina ($\overline{kT_{m}}=0.5$ keV).
\end{abstract}




\section{Introduction}

It is well known that Galactic early-type stars are prominent soft ($\sim$keV) X-ray emitters
 \citep{1979ApJ...234L..51H, 1979ApJ...234L..55S}, 
usually attributed to shocks within their intrinsically structured outflows \citep{1970ApJ...159..879L, 1988ApJ...335..914O, 1999ApJ...520..833O}. From early surveys with {\it Einstein}, a linear proportionality of X-ray emission to  bolometric luminosity, $L_{\rm X} \sim 10^{-7} L_{\rm Bol}$, was established for O-type stars \citep{1980ApJ...239L..65L, 1981ApJ...248..279P, 1989ApJ...341..427C} which observations from successor X-ray satellites {\it ROSAT, XMM, Chandra} have supported \citep{1997A&A...322..167B, 2006MNRAS.372..661S, 2009A&A...506.1055N, 2011ApJS..194....7N, 2015ApJS..221....1R}.

It is now established that a high fraction of massive stars lie in binary systems with  short periods \citep{2012Sci...337..444S}. A subset of binaries are known to exhibit enhanced, often harder (few keV), X-ray emission \citep{1987ApJ...320..283P, 1991ApJ...368..241C, 2022arXiv220316842R} attributed to colliding stellar winds, although the majority of binaries do not exhibit excess X-ray emission \citep{2005MNRAS.361..679O}. Further, stellar atmosphere improvements to our understanding of massive stars have resulted in revisions to the properties of early-type stars including their temperatures and luminosities \citep{2002ApJ...579..774C, 2004A&A...415..349R} and a subset of early-type stars are known to possess kG-scale magnetic fields \citep{2017MNRAS.465.2432G}, so historical calibrations should  be used with caution. In addition, $\gamma$ Cas-like Oe and Be stars possess unusually bright, hard X-ray emission \citep{2016AdSpR..58..782S}.

Very few X-ray studies of early-type stars in the Magellanic Clouds have been carried out, including  {\it Chandra} ACIS imaging of the N66 region in the SMC, host to NGC~346 \citep{2003ApJ...586..983N}, ACIS imaging of the N11 region in the LMC \citep{2014ApJS..213...23N} and the 30 Doradus or Tarantula Nebula in the LMC, host to the rich star cluster R136  \citep{2002ApJ...574..762P, Leisa06}. However, early datasets were too shallow (20\,ks for 30 Dor) to draw firm conclusions on the $L_{\rm X}-L_{\rm Bol}$ relation since only a handful of OB stars were detected, although \citet{2014ApJS..213....1T} have provided an updated catalogue of X-ray sources in 30 Dor from 92\,ks {\it Chandra} ACIS observations. \citet{2014ApJS..213...23N} suggest LMC O stars possess similar X-ray properties to Milky Way counterparts, albeit based on stacked observations of undetected O stars in N11. Consequently, X-ray properties of Magellanic Cloud OB stars have historically been assumed in spectroscopic analyses \citep[e.g.][]{2002ApJ...579..774C}, since empirical calibrations at low metallicity have not been available. Since X-rays are believed to be emitted from  shocks embedded in the stellar wind, one would expect $L_{\rm X}$ to depend on mass-loss, which itself dependents on luminosity as $L_{\rm Bol}^{1/\alpha^{\prime}}$ with $\alpha^{\prime}$=0.6 \citep{2008A&ARv..16..209P}.

\begin{figure}
	\centering
	\includegraphics[width=0.8\linewidth,angle=0]{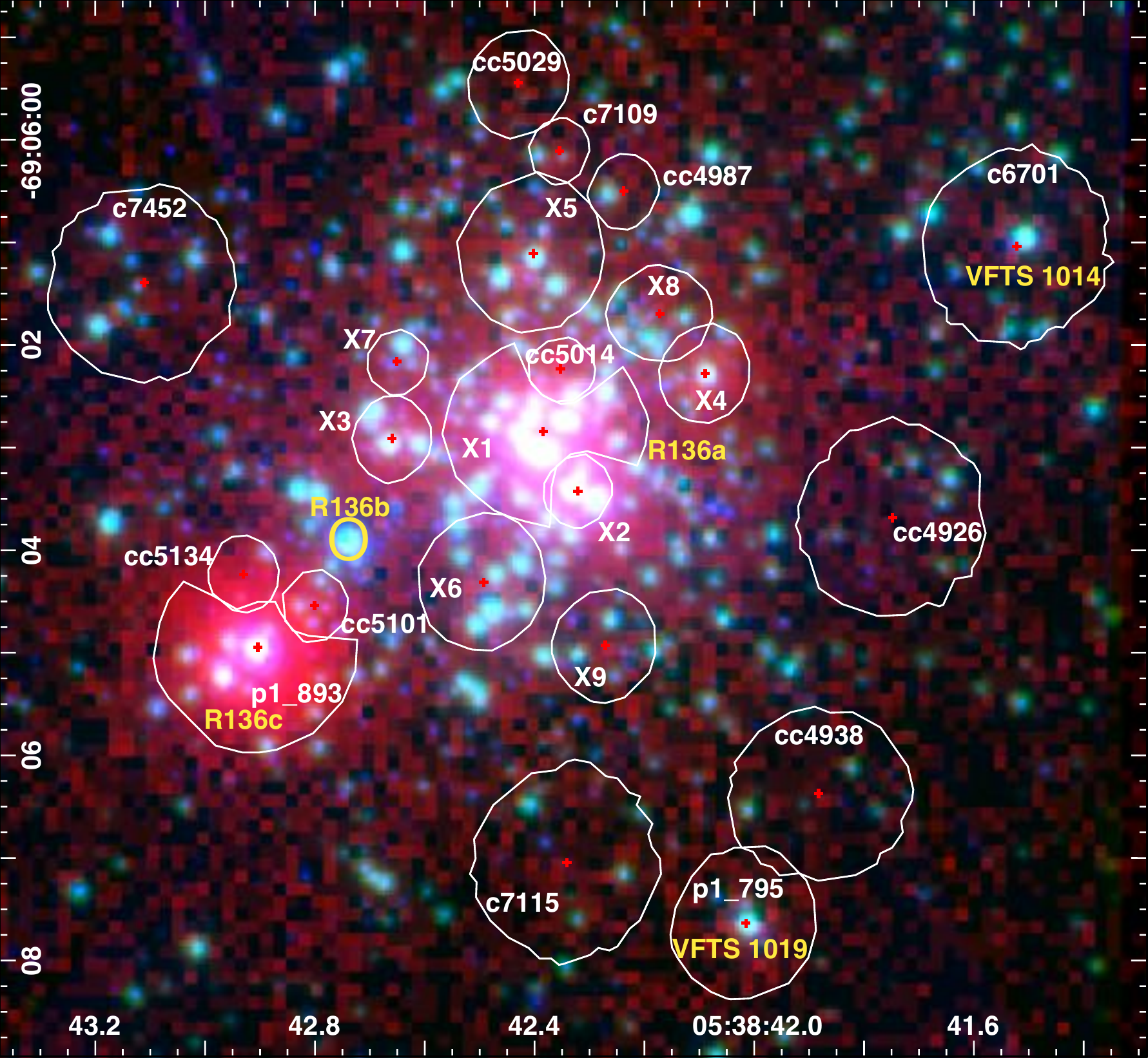}
  \caption{Three colour  10$''\times$10$''$ view of R136 comprising \textit{Chandra} ACIS 0.5--7\,keV (X-ray: red, L. Townsley et al. in prep), {\it HST} WFC3/F555W  \citep[optical: green,][]{2011ApJ...739...27D} and {\it VLT} SPHERE/K$_{s}$  \citep[near-IR: blue,][]{2021MNRAS.503..292K}, with X-ray sources and selected optical counterparts labelled in white and yellow respectively, including 9 X-ray sources associated with R136a.  The white polygons define X-ray event extraction apertures, and include c7452 which is host to  multiple spectroscopically confirmed early-type stars. Optical counterparts to these X-ray sources are listed in Table~\ref{multiple}. The O4\,If/WN8 supergiant R136b, located at an X-ray minimum, is also indicated. North is up and East to the left.} 
	\label{r136a}
\end{figure}

The recent, much deeper (2\,Ms) {\it Chandra} ACIS imaging survey of the Tarantula Nebula in the LMC, 'Tarantula -- Revealed by X-rays' (T-ReX, PI Townsley), finally represents an opportunity to assess whether the canonical Galactic $L_{\rm X} - L_{\rm Bol}$ relation applies to LMC O-type stars, since this region is host to an exceptionally rich massive star content \citep{2019Galax...7...88C}. The Tarantula has been extensively, spectroscopically surveyed with {\it Very Large Telescope (VLT)}/FLAMES  \citep{2011A&A...530A.108E}, {\it VLT}/MUSE \citep{2018A&A...614A.147C} and {\it Hubble Space Telescope (HST)}/STIS \citep{2016MNRAS.458..624C}. To date, \citet{2018MNRAS.474.3228P} have utilised T-ReX to establish Melnick 34 as a colliding wind binary system, while  \citet{2015A&A...579A.131C} have investigated the candidate high mass X-ray binary VFTS 399.

In this paper we briefly introduce the T-ReX point source catalogue in Sect.~\ref{TREX}, present the
$L_{\rm X}-L_{\rm Bol}$ relation for early-type stars in 30 Doradus in Sect.~\ref{LxLbol}, and consider their X-ray hardness in Sect.~\ref{hardness}. 
We investigate non-detections of some 30 Doradus luminous stars  in Sect.~\ref{MLS}, compare their X-ray properties with luminous early-type stars in the Carina Nebula in Sect.~\ref{metallicity_dependence}, and 
present a brief summary in Sect.~\ref{summary}.

\section{T-ReX point source catalogue}\label{TREX}

The 2 Ms {\it Chandra} X-ray Visionary Program T-ReX was executed over 630 days between 2014 May 3 and 2016 Jan 22 using the ACIS instrument \citep{2003SPIE.4851...28G}, centred on R136a the central cluster in the heart of the Tarantula Nebula, with a 16.9$' \times 16.9'$ field of view for each pointing. This dataset also incorporates 92\,ks ACIS observations from 2006 Jan 21--30 \citep{2014ApJS..213....1T}, so represents a {\it hundred-fold} increase on exposure times with respect to early {\it Chandra} observations \citep{2002ApJ...574..762P, Leisa06}.
L.~Townsley et al. (in prep) introduce T-ReX,   while P.~Broos \& L.~Townsley (in prep) have produced a T-ReX point source catalogue (PSC). In brief, data reduction, point source detection and extraction follow \citet{2018ApJS..235...43T} and utilise ACIS Extract \citep{2010ApJ...714.1582B} tools\footnote{doi:10.5281/zenodo.781433}.

\begin{figure}
\centering
	\includegraphics[width=0.95\linewidth]{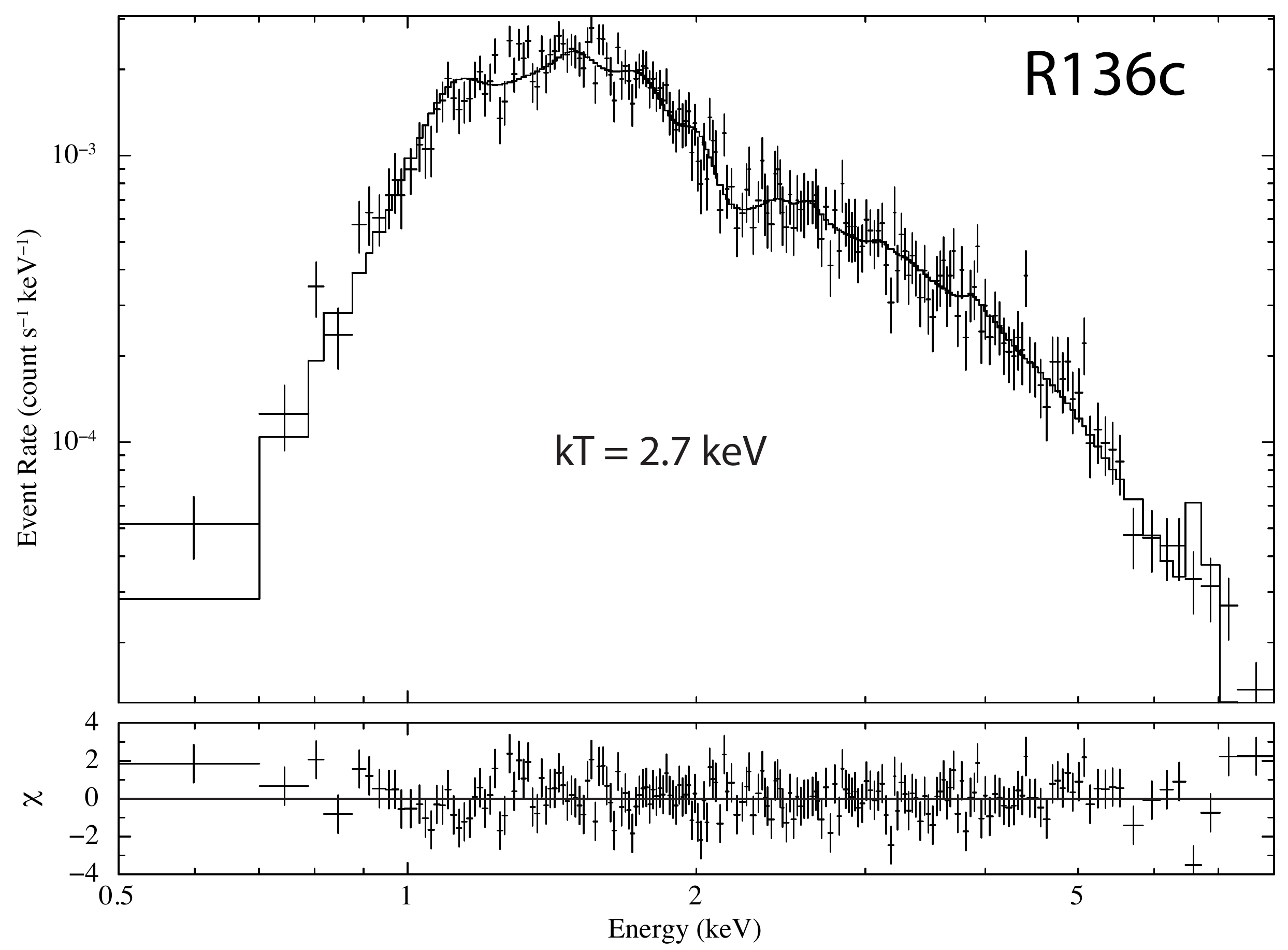}
	\includegraphics[width=0.95\linewidth]{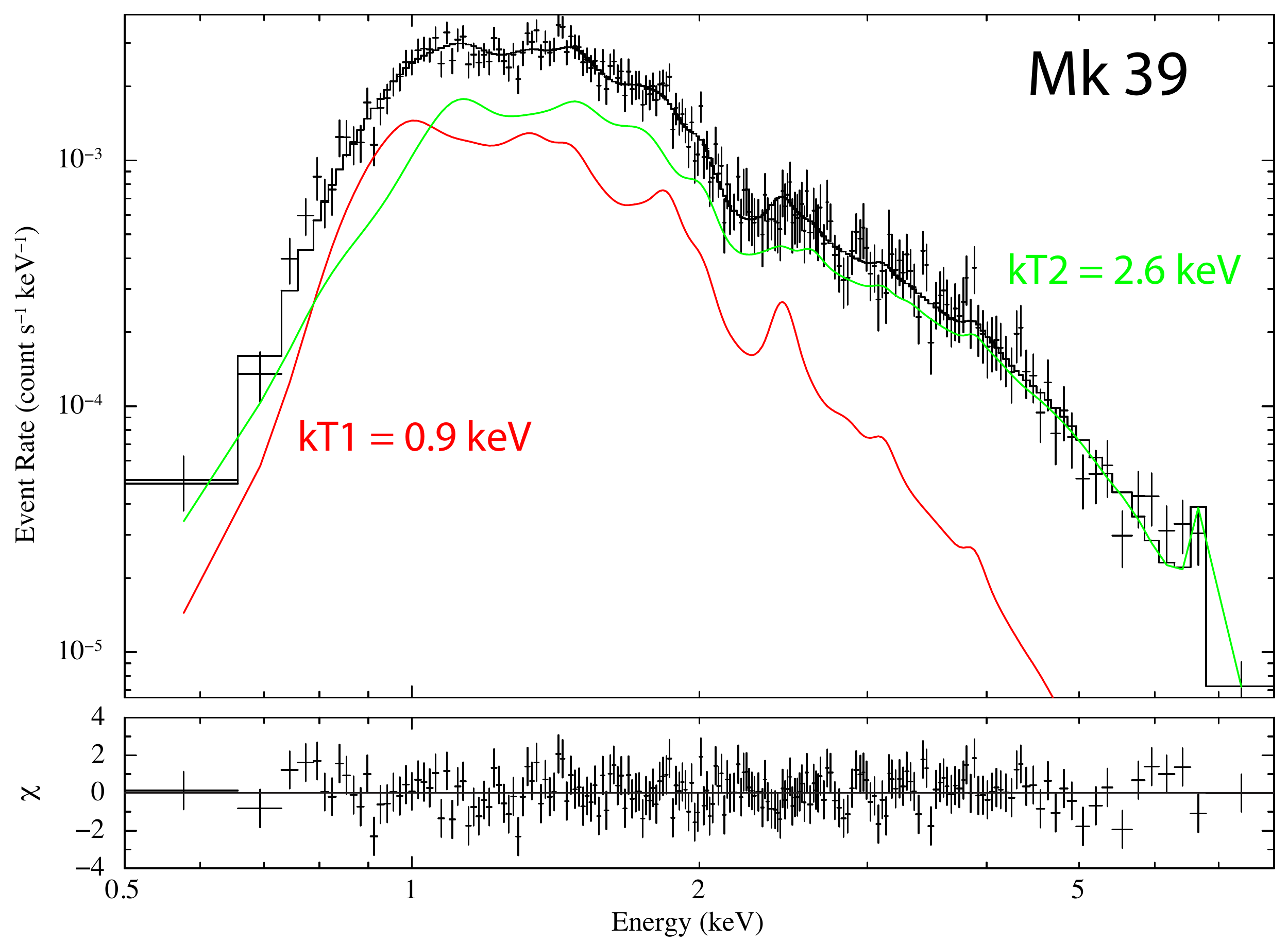}
	\includegraphics[width=0.95\linewidth]{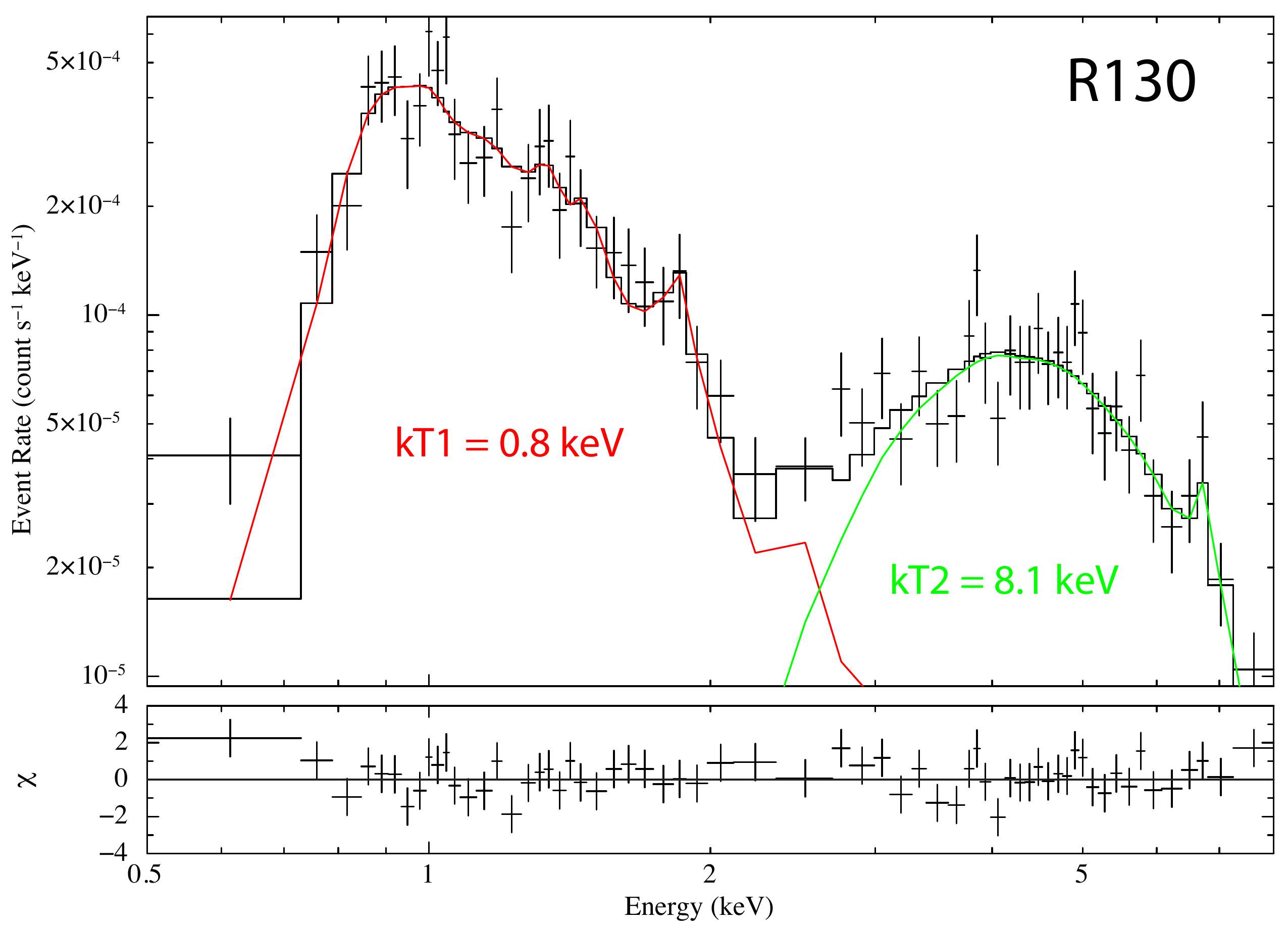}
	\centering
  \caption{Representative XSPEC fits to three X-ray bright sources, involving from top to bottom: R136c (single 2.7\,keV fit in black), Mk\,39 (dual 0.9\,keV red and 2.6\,keV green fit, sum in black), and R130 (dual 0.8\,keV red and 8.1\,keV green fit, sum in black). R130 is one of only four early-type stars in the T-ReX sample which exhibits a very hard X-ray component. XSPEC parameters for these sources are included in Table~\ref{A1}.}
	\label{XSPEC}
\end{figure}

In order to identify massive star counterparts to T-ReX  sources, these were cross-matched against the \citet{2013A&A...558A.134D} catalogue, which involved a compilation of confirmed and candidate early-type
stars in the Tarantula Nebula. This was supplemented by more recent spectroscopic surveys
exploiting {\it VLT}/MUSE \citep{2018A&A...614A.147C} for the NGC~2070 region and {\it HST}/STIS \citep{2016MNRAS.458..624C} for the R136 cluster. In addition, 4 sources inadvertently omitted from \citet{2013A&A...558A.134D} were common to the VLT-FLAMES Tarantula Survey \citep{2011A&A...530A.108E}, namely VFTS 186, 411, 564 and 640.

\begin{table}
\begin{center}
\caption{T-ReX sources host to multiple stellar populations within the core of R136 (X1--9, $r_{\rm d} \leq $2.4$''$ or 0.6 parcsec from R136a1) plus c7452 (5$''$ to the east). 
Catalogue numbers are from WB \citep{1985A&A...150L..18W}, MH \citep{1994AJ....107.1054M}, HSH \citep{1995ApJ...448..179H} and BAT \citep{1999A&AS..137..117B}. Cumulative spectroscopic luminosities from alternative studies \citep{2022arXiv220211080B} agree to within $\pm$0.1 dex.}
\label{multiple}
\begin{tabular}{l@{\hspace{2mm}}r@{\hspace{2mm}}rclc@{\hspace{2mm}}c}
\hline
WB & MH & HSH & BAT & Sp.Type & $\log L/L_{\odot}$ & Ref \\
\hline
  \multicolumn{7}{c}{R136a-X1 (p1\_832); PhotonFlux = $1.12 \times 10^{-5}$ cm$^{-2}$\,s$^{-1}$} \\
 a1 &  498 & 3 & 108 & WN5h & 6.79$\pm$0.10 & 1 \\ 
 a2 &  511 & 5 &  109 & WN5h & 6.75$\pm$0.10 & 1 \\ 
 a4 &  474 & 21 &        & O3\,V((f*)) & 6.24$\pm$0.18 & 1 \\  
 a5 & 519 & 20 & 110  & O2\,I(n)f* & 6.29$\pm$0.10 & 1 \\ 
 a7 & 509 & 24 &        & O3\,III(f*) & 6.25$\pm$0.18 & 1 \\ 
 a8 & 480 & 27 &        & O2--3\,V    & 6.17$\pm$0.10   &  2   \\ 
       & 468      & 30  &        & O6.5\,Vz    & 5.68$\pm$0.14 & 1 \\ 
        & 443     & 35   &       & O3\,V         & 5.74$\pm$0.18 & 1 \\ 
        & 553     & 50   &       & O3--4\,V((f*)) & 5.71$\pm$0.11 & 1 \\ 
        & 542     & 58    &      & O2--3\,V:      & 5.94$\pm$0.16 & 1 \\ 
        & 518     & 66    &      & O2\,V--III(f*) & 5.64$\pm$0.21 & 1 \\ 
         & 533     & 70     &      & O5\,Vz & 5.78$\pm$0.18 & 1 \\ 
 \hline
     \multicolumn{7}{c}{R136a-X2 (cc4970); PhotonFlux = $2.37 \times 10^{-6}$ cm$^{-2}$\,s$^{-1}$} \\
  a3  & 467 &  6 &  106     & WN5h & 6.63$\pm$0.10 & 1 \\ 
  a6 & 454 &19,26 &        & O2\,I(n)f*p+? & 6.27$\pm$0.09 & 1 \\ 
   \hline
  \multicolumn{7}{c}{R136a-X3 (p1\_867); PhotonFlux = $1.59 \times 10^{-6}$ cm$^{-2}$\,s$^{-1}$} \\
      & 608       &  36 &       & O2\,If* & 6.33$\pm$0.11 & 1 \\ 
        & 618     &  46 &       & O2--3\,III(f*) & 6.16$\pm$0.18 & 1 \\ 
        & 592     &  48 &      & O2--3\,III(f*) & 6.05$\pm$0.20 & 1 \\ 
             \hline
   \multicolumn{7}{c}{R136a-X4 (c6981); PhotonFlux = $1.11 \times 10^{-6}$ cm$^{-2}$\,s$^{-1}$ } \\
          & 354  & 42 &        & O3\,V+O3\,V &   5.82$\pm$0.3: & 3, 4 \\ 
            \hline
  \multicolumn{7}{c}{R136a-X5 (c7157); PhotonFlux = $7.67 \times 10^{-7}$ cm$^{-2}$\,s$^{-1}$ } \\
         & 508    &  39 &     & O3\,V+O5.5\,V      & 5.76$\pm$0.3\phantom{0} & 3, 4 \\ 
         & 535  & 40  &        & O3\,V    & 5.88$\pm$0.18 & 1 \\ 
         & 486   & 49   &       & O3\,V     & 5.89$\pm$0.37 & 1 \\ 
            \hline
      \multicolumn{7}{c}{R136a-X6 (c7182); PhotonFlux = $5.86 \times 10^{-7}$ cm$^{-2}$\,s$^{-1}$ } \\
          & 537 & 31    &       & O2\,V((f*)) & 6.01$\pm$0.16 & 1 \\ 
          & 517 & 52 &         & O3--4\,Vz & 5.67$\pm$0.16 & 1 \\ 
          & 551 &  55    &       & O2\,V((f*))z & 5.76$\pm$0.15 & 1 \\ 
           & 536 & 77     &       & O5.5\,V+O5.5\,V & 5.36$\pm$0.3\phantom{0} & 3, 4 \\ 
          & 559 & 94      &      & O4--5\,Vz  & 5.52$\pm$0.23 & 1 \\  
           & 514 & 114    &      & O5--6\,V    & 5.25$\pm$0.21 & 1 \\ 
           \hline
           \multicolumn{7}{c}{R136a-X7 (c7257); PhotonFlux = $4.00 \times 10^{-7}$ cm$^{-2}$\,s$^{-1}$ } \\
           & 602 & 47    &       & O2\.V((f*)) & 6.09$\pm$0.21 & 1 \\ 
            \hline
      \multicolumn{7}{c}{R136a-X8 (c7018); PhotonFlux = $2.07 \times 10^{-7}$ cm$^{-2}$\,s$^{-1}$ } \\
          & 409 & 75    &        & O6\,V & 5.29$\pm$0.22 & 1 \\ 
          & 397 & 108 &         & O\,Vn & 5.04$\pm$0.24 & 1 \\ 
          \hline
            \multicolumn{7}{c}{R136a-X9 (cc4968); PhotonFlux = $1.82 \times 10^{-7}$ cm$^{-2}$\,s$^{-1}$ } \\
          & 434 & 116    &        & O7\,V & 4.84$\pm$0.16 & 1 \\ 
          & 465 & 121 &         & O9.5\,V & 4.86$\pm$0.16 & 1 \\ 
                    \hline
           \multicolumn{7}{c}{c7452; Photon Flux = $2.08 \times 10^{-7}$ cm$^{-2}$\,s$^{-1}$ } \\
           & 744 & 68    &        & O4--5\,Vz & 5.73$\pm$0.22 & 1 \\ 
           & 741 & 102 &         & O2--3\,III & 5.40$\pm$0.3\phantom{0} & 5 \\ 
           &  711 & 129 &          & O             & 4.37$\pm$0.26 & 1 \\ 
\hline
\end{tabular}
\end{center}
\begin{footnotesize}
1: \citet{2020MNRAS.499.1918B}; 2: \citet{2022arXiv220211080B}; 3: \citet{2002ApJ...565..982M}; 4: \citet{2013A&A...558A.134D}; 5 \citet{2016MNRAS.458..624C} 
\end{footnotesize}
\end{table}

In total, P.~Broos \& L.~Townsley (in prep, their table 6) identify 115 sources in common between T-ReX and optical photometric catalogues of hot, luminous stars, with a median separation of 0.15$''$ between X-ray and optical positions. Of these, 108 have optical spectral classifications,  most of which involve a single optical source, with the exception of R140a  \citep{1987ApJ...312..612M} and the rich, crowded R136 cluster, whose brightest X-ray sources are associated with multiple stars. Fig~\ref{r136a} shows a three colour image of R136, including {\it Chandra} ACIS (red, L. Townsley et al. in prep), {\it HST} WFC3/F555W \citep[green,][]{2011ApJ...739...27D} and {\it VLT} SPHERE/K$_{s}$ \citep[blue,][]{2021MNRAS.503..292K}. The white polygons define X-ray event extraction apertures; for uncrowded sources, these represent 90\% encircled energy contours of the Chandra/ACIS point spread function (extraction apertures are reduced for crowded sources). Fig~\ref{r136a} also highlights three X-ray counterparts to optically-bright sources elsewhere in the vicinity of R136a, namely R136c (= VFTS 1025), VFTS 1014 \citep[=HSH 29,][]{1995ApJ...448..179H}  and VFTS 1019 (= HSH 38), plus c7452, which is host to multiple early-type stars. Table~\ref{multiple} provides spectral types and luminosities  of the primary optical counterparts to these 9 X-ray sources within R136a plus c7452.



Optical spectroscopy reveals that the majority of the sources, or primaries in binary systems,  are O-type stars (71 sources), followed by Wolf-Rayet stars (14), B-type stars (7) or Of/WN stars (6). We omit cc4769 a.k.a. VFTS 1003 \citep[= SMB 283,][]{1999A&A...341...98S} from our subsequent discussion since it is a peculiar Be star, hindering spectroscopic analysis \citep{2015A&A...574A..13E}, such that our final sample involves 107 sources, of which 11 are X-ray sources host to multiple populations within R136 or R140a. X-ray properties of all sources are provided in Table~\ref{A1}, sorted by photon flux. For each source, net counts (NetCts) and spectral fits were obtained from all observations of that source. However, in some cases a single observation was used to assess detection significance. Thus, detection significance cannot be inferred from NetCts values. For example, the detection of the last source (cc7873) was determined by a single observation from which 4.9 net counts were extracted, NetCts=4.9, and the detection p-value is 0.0018. The 2\,Ms spectrum of this source has 28 total counts, but the expected background is $\sim$27 counts. 


\begin{table*}
\begin{center}
\caption{Comparison between observed atomic hydrogen column densities (from Lyman $\alpha$) towards early-type stars at a range of radial distances  from R136a ($r_{\rm d}$ in arcmin) and values predicted from the Galactic and LMC calibrations, $\log N(H)/E_{\rm B-V}$ of \citet{1995A&A...293..889P} and \citet{1982A&A...107..247K}, respectively. For R136a, \citet{2022arXiv220211080B} obtained a total Ly$\alpha$ (log) column density of 21.88$\pm$0.07 cm$^{-2}$ from the average of the brightest 29 stars observed with HST/STIS spectroscopy, in good agreement with 21.85$^{+0.10}_{-0.15}$ cm$^{-2}$ from IUE/SWP spectroscopy \citep{1980ApJ...236..769D}.
A Galactic foreground of $E^{\rm MW}_{\rm B-V}$=0.07 or $A^{\rm MW}_{\rm V}$=0.22 mag is adopted in all cases, corresponding to $N(H)^{\rm MW} = 3.8 \times 10^{20}$ cm$^{-2}$ and we adopt  $A^{\rm LMC}_{\rm V} = 3.5 E_{\rm B-V}^{\rm LMC}$ \citep{2013A&A...558A.134D} such that $N(H)^{\rm Tot} = N(H)^{\rm MW} + N(H)^{\rm LMC}$. Overall good agreement is found between the empirical total hydrogen column density and that obtained from calibrations, $\Delta N(H) = \log N(H)_{\rm Ly\alpha}^{\rm Tot} - \log N(H)^{\rm Tot}$, especially for the bulk of T-ReX sources within $r_{\rm d} \leq 7$ arcmin ($\sim$ 100 pc).}
\label{lyman_alpha}
\begin{tabular}{
l@{\hspace{1mm}}l@{\hspace{1mm}}c@{\hspace{2mm}} 
c@{\hspace{1mm}}l@{\hspace{2mm}} 
c@{\hspace{1mm}}c@{\hspace{1mm}}c@{\hspace{2mm}} 
l@{\hspace{2mm}}l}
\hline
Star &  Aliases & $r_{\rm d}$      & $A^{\rm LMC}_{\rm V}$ & Ref & $N(H)^{\rm LMC}$ & $\log N(H)^{\rm Tot}$ & $\log N(H)_{\rm Ly\alpha}^{\rm Tot}$ & Ref & $\Delta N(H)$  \\
        &         & arcmin & mag                  &      & $10^{21}$ cm$^{-2}$ & cm$^{-2}$               & cm$^{-2}$                   & & dex \\
\hline 
R136a & HD~38268, Sk --69$^{\circ}$ 243 & 0.00 & 1.58   & 1 & 9.0 & 21.97 & 21.88 & 2 & --0.09  \\ 
Mk\,42  & P93~922, BAT99-105 & 0.13 & 1.42 & 3 & 8.1 & 21.93 & 21.79 & 4 &  --0.14 \\ 
R144  & HD~38282, Sk --69$^{\circ}$ 246, P93~9037, BAT99-118 & 4.1\phantom{0} & 0.40 & 5 & 2.3 & 21.43 & 21.47 & 6 & +0.05  \\ 
BI\,253 & VFTS\,72 & 7.78 & 0.65 & 3 & 3.7 & 21.61 & 21.68 & 6 & +0.06  \\ 
VFTS\,696 & Sk --68$^{\circ}$ 140 & 9.26 & 0.68 & 4 & 3.9 & 21.63 & 21.78 & 6 & +0.15  \\ 
HDE\,269888 & Sk --69$^{\circ}$ 234, VFTS\,136, BAT99-90 & 9.84 & 0.83 & 7 & 4.8 & 21.71 & 21.52 & 4 & --0.19  \\ 
\hline 
\multicolumn{2}{l}{Average ($\pm$ St. Dev.)}           &        &      &            &           & &  & &--0.03$\pm$0.13 \\
\hline
\end{tabular}
\end{center}
\begin{footnotesize}
1: \citet{2016MNRAS.458..624C} 
2: \citet{2022arXiv220211080B} 
3: \citet{2014A&A...570A..38B} 
4:  \citet{2012ApJ...745..173W} 
5: \citet{2021A&A...650A.147S} 
6: \citet{2019ApJ...871..151R} 
7: \citet{2002A&A...392..653C} 
\end{footnotesize}
\end{table*}


In order to correct observed X-ray fluxes (0.5--8 keV) for absorption by interstellar gas we undertake XSPEC \citep{1996ASPC..101...17A} $\chi^2$ ($\geq$50 net counts) or Cash statistic \citep{1979ApJ...228..939C} ($<$50 net counts) fitting. For 
Mk~34 we provide an update to X-ray luminosities from \citet{2018MNRAS.474.3228P}. \citet{2015A&A...579A.131C} have also analysed initial T-ReX observations of VFTS~399, for which a power law fit is adopted here since high mass X-ray binaries have different X-ray emission mechanisms than OB or WR systems. VFTS~399 aside, all fits used a one-temperature or two-temperature thermal plasma model ({\em vapec}) with frozen metal abundances appropriate for the LMC presented in the Appendix in Table~\ref{abundances}. Spectral fitting restricted plasma temperatures to 0.27 keV $\leq kT \leq$ 9.5 keV.

T\"{u}bingen-Boulder absorption components model the Milky Way ISM ({\em TBabs}$^{\rm MW}$) and LMC ISM ({\em TBvarabs}$^{\rm LMC}$), plus we additionally consider a circumstellar absorption component ({\em TBvarabs}$^{\rm Circ}$), although the latter is excluded from extinction-corrected luminosities for consistency with other studies. For the Milky Way, we adopt a uniform atomic hydrogen column density based on 
\[ N_H^{\rm MW} / E^{\rm\bf MW}_{\rm B-V} = 5.3 \times 10^{21}  \hspace*{0.2cm} {\rm cm}^{-2} {\rm mag}^{-1} \]
\citep{1995A&A...293..889P} 
with $E^{\rm MW}_{\rm B-V}$ = 0.07 mag adopted from \citet{1984ApJ...279..578F}. For the LMC we consider a variable component, generally selected from the literature dust extinctions listed in the Appendix (Table~\ref{A2}) using 
\[ N_H^{\rm LMC}/E^{\rm LMC}_{\rm B-V} = 2  \times 10^{22} \hspace*{0.2cm} {\rm cm}^{-2} {\rm mag}^{-1} \]
\citep{1982A&A...107..247K}, 
with $A^{\rm LMC}_{\rm V}  \sim 3.5 E^{\rm LMC}_{\rm B-V}$ \citep{2013A&A...558A.134D}\footnote{\citet{2013A&A...558A.134D} favour a higher ratio of $A^{\rm LMC}_{\rm V}  \sim 4.2 E^{\rm LMC}_{\rm B-V}$ in the vicinity of R136}. For the total foreground column density $N(H)^{\rm Tot}$ we sum the Milky Way and LMC components, $N(H)^{\rm MW}$ and $N(H)^{\rm LMC}$.

Since there is significant variation in $N_H^{\rm LMC}/E^{\rm LMC}_{\rm B-V}$ \citep[][their fig 5]{2002ApJ...566..857T} we have reassessed the \citet{1982A&A...107..247K} calibration for early-type stars in 30 Doradus. Table~\ref{lyman_alpha} compares 
expected hydrogen column densities to measured   Lyman~$\alpha$  column densities for six sight-lines. Overall we find excellent agreement between the observed hydrogen column densities and the total predicted from calibrations, with  $\log N(H)_{\rm Ly \alpha}^{\rm Tot} - \log N(H)^{\rm Tot} = -0.03\pm$0.13 dex.
Frozen abundances in the {\em TBvarabs} components are appropriate for the LMC (Table~\ref{abundances}). The model is implemented in XSPEC as {\em{(TBvarabs$^{\rm LMC}$*TBabs$^{\rm MW}$)(TBvarabs$^{\rm Circ}$*vapec)} or  {\em{(TBvarabs$^{\rm LMC}$*TBabs$^{\rm LMC}$)(TBvarabs$^{\rm Circ1}$*vapec + TBvarabs$^{\rm Circ2}$*vapec)}. Intrinsic X-ray luminosities assume a distance of 50 kpc to the LMC and set both $N_H^{\rm MW}$ and $N_H^{\rm LMC}$ to zero in XSPEC(which does not support the quantification of uncertainties on corrected luminosities). The average attenuation correction to observed X-ray luminosities is 0.31$\pm$0.17 dex. 

Single temperature models provided satisfactory fits for the majority of X-ray sources (e.g. R136c in top panel of Fig.~\ref{XSPEC}), whilst improved fits were achieved for 27 sources using two plasma temperatures. These are flagged in the Appendix (Table~\ref{A1}) and include Mk\,39 in the central panel of Fig.~\ref{XSPEC}. Four WR systems (R130, R134, R135 and Mk~53) exhibit extremely hard X-ray components, as indicated for R130 in the lower panel of Fig.~\ref{XSPEC} such that corrected X-ray luminosities represent lower limits. These sources possibly exhibit different emission mechanisms which are responsible for very hard plasmas reminiscent of $\gamma$ Cas systems \citep{2016AdSpR..58..782S}  and are therefore excluded. More detailed comparisons are deferred to subsequent studies, including K.~Tehrani et al. (in prep) for WR and Of/WN stars. The X-ray faintest sources with fewer than 10 net counts were fit with a fixed $kT$ = 0.8 keV (close to the median for other sources) so should also be treated with caution. 

\begin{figure*}
\centering
\begin{minipage}[c]{0.75\linewidth}
	\includegraphics[width=0.74\linewidth,angle=-90,bb = 0 60 540 700]{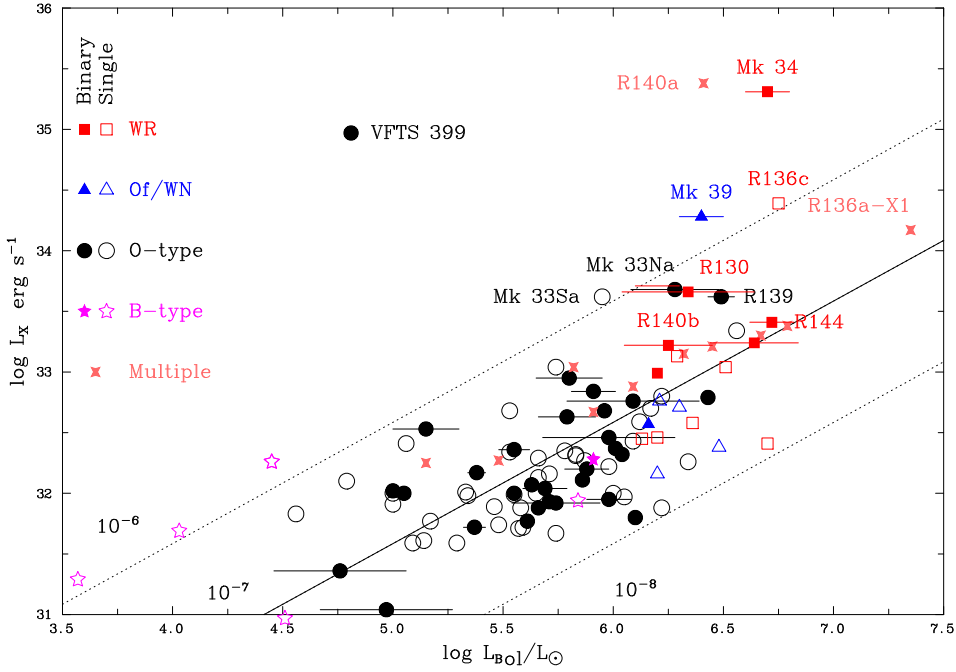}
\end{minipage}\hfill
\begin{minipage}[c]{0.25\linewidth}
\centering
  \caption{Comparison between X-ray and bolometric luminosities for 107 early-type sources from the T-ReX point source catalogue (P.~Broos \& L.~Townsley, in prep). X-ray luminosities are corrected for interstellar extinction ($L_{X}^{tc}$ in Table~\ref{A1}), while references to bolometric luminosities are provided in Tables~\ref{multiple} (R136) and \ref{A2} (non-R136). The solid line indicates the canonical $L_{\rm X} = 10^{-7} L_{\rm Bol}$ relation, with dotted lines offset by $\pm$1 dex. Symbols correspond to the nature of the primary/brightest component, namely WR star (square), Of/WN star (triangle), O-type star (circle), B-type star (star). Spectroscopically confirmed binaries are shown as solid symbols and other sources (single and multiple systems lacking a known binary companion) are shown as open symbols. For clarity, uncertainties are included solely for SB2 systems.} 
	\label{fig1}
	\end{minipage}
\end{figure*}


Table~\ref{literature} compares  X-ray luminosities from T-ReX for a subset of the brightest sources to literature results based on fits to historical {\it Chandra} shallow surveys. Luminosity estimates from early studies suffered from large uncertainties owing to short exposures (20\,ks), such that T-ReX provides more reliable X-ray properties, especially for fainter sources in common. A number of bright PSC sources are known to be X-ray variable \citep[e.g. Mk\,34,][]{2018MNRAS.474.3228P}, detailed discussion of which is deferred to K.~Tehrani et al. (in prep), such that X-ray luminosities considered here are time averaged quantities (XSPEC fits have also been undertaken for Mk~33Na by \citet{2022MNRAS.510.6133B}).

\citet{2019A&A...626A..59C} presented a hardness colour for early-type stars in Westerlund~1, which involved $(h-s)/(h+s)$ where $h$ is is the observed counts in the 2--8 keV range, and $s$ is the observed counts in the 0.5--2 keV range. Here we adapt their hardness colour to calculate a hardness index, $\eta_{2}$, based on {\it intrinsic} luminosities,  in the hard (2--8 keV, $L^{hc}_{\rm X}$) versus soft (0.5--2 keV, $L^{sc}_{\rm X}$) bands , i.e.
\[ 
\eta_{2} = \frac{(L^{hc}_{\rm X}- L^{sc}_{\rm X})}{(L^{sc}_{\rm X} + L^{hc}_{\rm X})},
\] 
such that exclusively soft emitters will have $\eta_{2}$ = --1 and pure hard emitters will have $\eta_{2}$ = +1. Our approach, involving interstellar extinction-corrected quantities, has two main advantages over observed hardness colours . First, observed event rates will evolve over time owing to the degradation of the ACIS soft energy response, whereas our XSPEC fits (on which our luminosities are based) account for these effects. Second, sources will exhibit a range of line-of sight absorptions, particularly when comparing hardness indices across different regions.

Alternatively, armed with a mix of single and dual temperature XSPEC fits we can evaluate a more physically motivated quantity, namely a weighted mean temperature, $T_{\rm m}$ via 
\[ kT_{\rm m} = \frac{\Sigma_i (kT_i \times {\rm EM}_i)}{\Sigma_i {\rm EM_i}}
\]
where EM$_i$ is the volume emission measure.  $\eta_{2}$ and especially $kT_{\rm m}$ are sensitive to well-known degeneracies between column densities and plasma temperatures \citep{2014ApJ...788...90G}, although these are mitigated by optically-derived column densities throughout the present study (circumstellar column densities are not considered when determining X-ray luminosities). Sources for which power law slopes are adopted, such as VFTS 399, permit the determination of a hardness index, though not a plasma temperature.

Table~\ref{A2} in the Appendix provides spectral type information, together with X-ray luminosities, mean plasma temperatures and hardness indices, plus identifications in 30 Doradus-specific catalogues \citep{1985A&A...153..235M, 1993AJ....106..560P, 1994AJ....107.1054M, 1995ApJ...448..179H, 1999A&A...341...98S, 2011A&A...530A.108E, 2018A&A...614A.147C} plus general LMC catalogues \citep{1960MNRAS.121..337F, 1970CoTol..89.....S, 1975A&AS...21..109B, 1992A&AS...92..729S, 1999A&AS..137..117B}. 



\begin{table}
\centering
\caption{Comparison between X-ray luminosities for bright early-type stars in the Tarantula Nebula from T-ReX in common with literature results (LT06 \citet{Leisa06}; SPZ02 \citet{2002ApJ...574..762P}) based on shallower {\it Chandra} surveys (20\,ks), sorted by T-ReX photon flux. X-ray luminosities are obtained from XSPEC fits to T-ReX observations from this work ('X') or adapted from \citet[P18]{2018MNRAS.474.3228P}. Variable bright X-ray sources are flagged as "var".}
\begin{tabular}{r@{\hspace{2mm}}c@{\hspace{1mm}}c@{\hspace{1mm}}r@{\hspace{1mm}}r@{\hspace{2mm}}c@{\hspace{2mm}}r@{\hspace{2mm}}c@{\hspace{2mm}}l}
\hline
Photon flux &  $\log L_{\rm X}$ &  & Ref &            LT06 & $\log L_{\rm X}$ & SPZ02 & $\log L_{\rm X}$ & Source \\
cm$^{-2}$\,s$^{-1}$  & erg\,s$^{-1}$           &         &   &                        & erg\,s$^{-1}$                       &           & erg\,s$^{-1}$                     \\
\hline 
$1.48 \times 10^{-4}$ & 35.31 & var & P18 & 132 & 35.38 & CX5 & 35.26 &  Mk 34\\ 
$1.46 \times 10^{-4}$ & 35.38 & var & X & 51 & 35.25 & CX10 & 35.22 &  R140a \\ 
$4.92 \times 10^{-5}$ & 34.79  & var & X & 27 &  34.06 & ---      & ---        & VFTS 399\\
$1.70 \times 10^{-5}$ & 34.39 & var & X & 102 & 35.04 & CX2 & 34.93 &  R136c \\ 
$1.69 \times 10^{-5}$ & 34.28 & var & X & 36 & 34.22  & CX8 & 34.00 &  Mk 39\\
$1.12 \times 10^{-5}$ & 34.17 & var & X & 83 & 34.31 & CX1 & 34.10 &  R136a-X1 \\
$4.72 \times 10^{-6}$ & 33.62 & var & X & 131 & 34.63 & CX7 & 33.93 &  Mk 33Sa\\
$4.20 \times 10^{-6}$ & 33.62 & var & X & 82 & 33.39 & CX11 & 33.34 &  R139 \\
$4.10 \times 10^{-6}$ & 33.68 & var & X & 133 & 33.99 & CX9 & 33.44 &  Mk 33Na\\
$2.97 \times 10^{-6}$ & 33.41 & var & X & 154 & 33.51 & CX17 & 33.17 &  R144\\
$2.81 \times 10^{-6}$ & 33.66 & --- & X & 10 & 33.48  & ---    & ---       &  R130 \\
$2.37 \times 10^{-6}$ & 33.38 & var & X & 79 & 33.93 & ---    & ---         &  R136a-X2\\
$2.05 \times 10^{-6}$ & 33.34 &---  & X & ---  &   ---     & CX4    & 33.31 &  Mk 42\\
$1.87 \times 10^{-6}$ & 33.24 & var & X & 156 & 33.99 & CX12 & 33.25 &  R145 \\
$1.59 \times 10^{-6}$ & 33.30 & --- & X & 90 & 33.56 &  ---    & --- &  R136a-X3 \\
$1.50 \times 10^{-6}$ & 33.22 & var & X & 49 & 33.57 &    --- & ---         &  R140b \\
$1.08 \times 10^{-6}$ & 33.04 & --- & X & 45 & 33.81 & CX3 & 33.17  &  HSH 28 \\
$4.19 \times 10^{-7}$ & 32.68 & ---  & X & 23 & ---     & CX14 & 32.55 & VFTS 267\\
$1.38 \times 10^{-7}$   & 32.46 &--- & X &   --- &   ---    & CX6     & 32.86 &  R134\\
\hline
\end{tabular}
\label{literature}
\end{table}

\begin{table*}
\centering
\caption{Average $\log L_{\rm X}/L_{\rm Bol}$, hardness index, $\eta_{2}$, and mean plasma temperature, $kT_{m}$, for spectroscopically classified single/binary/multiple T-ReX point sources, including 1$\sigma$ dispersions. Single and binary systems are also separated into the spectral class of the primary. We also provide average properties excluding the following outliers: 3 extreme systems Mk\,34 (WN5h+WN5h, SB2), R140a (WC4+WN6+, multiple), VFTS 399 (O9\,IIIn+?, SB1), plus 4
systems for which XSPEC fitting requires a discrete very hard component: R130 (WN/C+B1\,I, SB2), R134 (WN6(h), single), R135 (WN5:+WN7, SB2) and Mk~53 (WN8(h), single). }\label{table1}
\begin{tabular}{l@{\hspace{4mm}}
r@{\hspace{1.5mm}}c@{\hspace{1.5mm}}c@{\hspace{1.5mm}}c@{\hspace{8mm}}
r@{\hspace{1.5mm}}c@{\hspace{1.5mm}}c@{\hspace{1.5mm}}c@{\hspace{8mm}}
r@{\hspace{1.5mm}}c@{\hspace{1.5mm}}c@{\hspace{1.5mm}}c}
\hline
Sp.Type& N & $\log \overline{L_{\rm X}/L_{\rm Bol}}$ & $\overline{\eta_{2}}$ & $\overline{kT_{m}}$ 
& N & $\log \overline{L_{\rm X}/L_{\rm Bol}}$ & $\overline{\eta_{2}}$ & $\overline{kT_{m}}$
& N & $\log \overline{L_{\rm X}/L_{\rm Bol}}$ & $\overline{\eta_{2}}$ & $\overline{kT_{m}}$\\ 
              &   &                                                               &                                   & keV 
              &   &                                                               &                                   & keV        
             &    &                                                              &                                     & keV\\
\hline
                  & \multicolumn{4}{c}{--- Single --- }  & \multicolumn{4}{c}{--- Binary --- } & \multicolumn{4}{c}{--- Single \& Binary --- } \\
O  all         & 40 & --7.04$\pm$0.44 & --0.62$\pm$0.32       & 1.03$\pm$0.78           & 31 & --6.92$\pm$0.77 & --0.68$\pm$0.37 &  0.84$\pm$0.36& 71 & --6.98$\pm$0.60 & --0.65$\pm$0.34 & 0.95$\pm$0.64\\
O all excl. & 40 &  --7.04$\pm$0.44 & --0.62$\pm$0.32      & 1.03$\pm$0.78            & 30 & --7.03$\pm$0.42 & --0.73$\pm$0.32 & 0.84$\pm$0.36 & 70 & --7.04$\pm$0.43 & --0.67$\pm$0.33 & 0.95$\pm$0.64\\
\\
O\,V-IV & 21 & --7.03$\pm$0.42 & --0.66$\pm$0.32 & 1.08$\pm$0.91 & 17 & --7.00$\pm$0.37 & --0.74$\pm$0.19 & 0.86$\pm$0.30 &  38 & --7.03$\pm$0.37 & --0.71$\pm$0.26 & 0.96$\pm$0.71 \\
O\,V-IV excl. & 21 & --7.03$\pm$0.42  & --0.66$\pm$0.32 & 1.08$\pm$0.91 & 17 & --7.00$\pm$0.37 & --0.74$\pm$0.19 & 0.86$\pm$0.30 & 38 & --7.03$\pm$0.37& --0.71$\pm$0.26 & 0.96$\pm$0.71 \\ 
\\
O\,III--I & 12 & --7.02$\pm$0.51 & --0.68$\pm$0.29 & 0.80$\pm$0.57 & 12 & --6.75$\pm$1.18 & --0.54$\pm$0.53 & 0.88$\pm$0.49 &  24 & --6.88$\pm$0.88 & --0.61$\pm$0.42 & 0.84$\pm$0.52 \\
O\,III--I excl.              & 12 & --7.02$\pm$0.51 & --0.68$\pm$0.29 & 0.80$\pm$0.57 & 11 &  --7.06$\pm$0.56 & --0.65$\pm$0.39  & 0.88$\pm$0.49 & 23 & --7.06$\pm$0.56 & --0.65$\pm$0.39 & 0.84$\pm$0.52\\
\\
WR              &  7& --7.08$\pm$0.61 & +0.26$\pm$0.46 & 4.12$\pm$2.80 & 6  & --6.42$\pm$0.75 & +0.33$\pm$0.32  & 4.36$\pm$3.04 & 13 & --6.78$\pm$0.73 & +0.29$\pm$0.38 & 4.23$\pm$2.79\\
WR excl.      & 5 & --7.00$\pm$0.72 & +0.20$\pm$0.47 &  2.60$\pm$1.22 &  3  & --6.83$\pm$0.19 & +0.23$\pm$0.15  & 2.42$\pm$1.57 & 8 & --6.94$\pm$0.56 & +0.24$\pm$0.38  & 2.96$\pm$1.73\\ 
 \\
Of/WN          &  4 & --7.38$\pm$0.32 & +0.50$\pm$0.18 & 0.86$\pm$0.24 &  2  & --6.44$\pm$1.04 &  --0.33$\pm$0.07   & 1.25$\pm$0.55 & 6 & --7.07$\pm$0.72 & --0.44$\pm$0.17 & 0.99$\pm$0.37 \\
Of/WN excl.  & 4  &--7.38$\pm$0.32  & +0.50$\pm$0.18 & 0.86$\pm$0.24 &  2  & --6.44$\pm$1.04 & --0.33$\pm$0.07  & 1.25$\pm$0.55 & 6 & --7.07$\pm$0.72 & --0.44$\pm$0.17 & 0.99$\pm$0.37 \\
\\
B           &   5$^{a}$   & --6.44$\pm$0.81 & --0.91$\pm$0.05 & 0.63\phantom{$\pm$0.00}  & 1     &  --7.22\phantom{$\pm$0.00}       &  --0.86\phantom{$\pm$0.00}  & 0.76\phantom{$\pm$0.00} & 6 & --6.57$\pm$0.79 & --0.90$\pm$0.05 & 0.70$\pm$0.09 \\
\\
All                 &     56  & --6.97$\pm$0.61 & --0.50$\pm$0.45 & 1.44$\pm$1.62 & 40 & --6.83$\pm$0.77 & --0.51$\pm$0.51  &  1.43$\pm$1.76 & 96 & --6.94$\pm$0.64 & --0.50$\pm$0.47 & 1.44$\pm$1.76 \\
All excl.       &      54    & --7.00$\pm$0.53 & --0.54$\pm$0.43 & 1.17$\pm$0.92 & 36 & --6.99$\pm$0.45 & --0.62$\pm$0.40  & 1.00$\pm$0.69 & 90 &--7.00$\pm$0.49 & --0.57$\pm$0.42 & 1.10$\pm$0.84 \\
\hline
                  & \multicolumn{4}{c}{--- Single+SB1 --- }  & \multicolumn{4}{c}{--- SB2 --- }  &  \multicolumn{4}{c}{--- Multiple --- } \\
All         & 72 & --6.97$\pm$0.70 & --0.53$\pm$0.45 & 1.30$\pm$1.45 & 24 & --6.74$\pm$0.59 & --0.41$\pm$0.53 & 1.83$\pm$0.38 & 11 & --6.56$\pm$0.67 & --0.50$\pm$0.23  & 1.50$\pm$1.11 \\
All excl. & 69 & --7.04$\pm$0.49 & --0.58$\pm$0.42 & 1.09$\pm$0.84 & 21 & --6.84$\pm$0.48 & --0.54$\pm$0.43 & 1.12$\pm$0.85 & 10 &--6.76$\pm$0.19 & --0.50$\pm$0.24 & 1.51$\pm$1.17 \\
\hline
\end{tabular}
(a) VFTS 186 has a measured plasma temperature while other single B stars are adopted to have a median temperature of 0.8 keV.
\end{table*}

\section{Nature of X-ray sources and $L_{\rm X}-L_{\rm Bol}$ relation}\label{LxLbol}

94 of the 107 T-ReX sources have been subject to spectroscopic analysis via a variety of studies, generally outlined in \citet{2019Galax...7...88C}. 50 WR, Of/WN stars and OB-type stars from VFTS have been analysed by \citet{2014A&A...570A..38B}, \citet{2015A&A...575A..70M}, \citet{2017A&A...600A..81R} and \citet{2017A&A...601A..79S}. 5 OB binary systems from the TMBM survey \citep{2017A&A...598A..84A} have been spectroscopically analysed by \citet{2020A&A...634A.118M}. A further 18 OB stars within NGC~2070 observed with {\it VLT}/MUSE have been analysed by \citet{2021A&A...648A..65C}, while \citet{2020MNRAS.499.1918B} have analysed {\it HST}/STIS spectroscopy of individual stars in R136a, involving 8 T-ReX sources. \citet{2012A&A...543A..95R} have analysed {\it HST}/STIS+FOS spectroscopy of Mk 33Sa from \citet{2005ApJ...627..477M} and one further candidate colliding wind binary system Mk 33Na (identified via T-ReX) has been followed up with {\it VLT}/UVES spectroscopic monitoring to confirm its SB2 status and has been analysed by \citet{2022MNRAS.510.6133B}.
The remainder include 8 single or binary Wolf-Rayet stars which have been analysed by \citet{2010MNRAS.408..731C}, \citet{2014A&A...565A..27H}, \citet{2017A&A...598A..85S}, \citet{2019MNRAS.484.2692T}, \citet{2019A&A...627A.151S} and \citet{2021A&A...650A.147S}. Spectral type calibrations are applied for the remaining 14 sources, following \citet{2013A&A...558A.134D}, inevitably leading to higher uncertainties in bolometric luminosities. Table~\ref{A2} in the Appendix provides X-ray and bolometric luminosities for the complete sample, including binary information. X-ray variables are highlighted (with "var") on the basis of Gaussian or Poissonian statistics, although inevitably these are limited to the brightest sources with unattenuated X-ray luminosities of $L_{\rm X} \geq 10^{33}$ erg\,s$^{-1}$. 

\citet{2005MNRAS.361..679O} concluded that, in general, binaries follow the canonical X-ray relationship for single early-type stars, although occasionally colliding wind systems exhibit enhanced, harder X-ray emission, with respect to single stars \citep{2011BSRSL..80..555P, 2012ASPC..465..301G, 2022arXiv220316842R}.  In general, establishing the single or binary nature of early-type stars is notoriously difficult. Fortunately, 76 T-ReX sources were included in VFTS, with spectroscopy across a range of cadences and
whose spectroscopic binary status has been summarised in \citet{2014A&A...564A..40W} and \citet{2015A&A...574A..13E}.
A subset of VFTS sources identified as spectroscopic binaries have been followed up with the Tarantula Massive Binary Monitoring \citep[TMBM,][Shenar et al. in prep]{2017A&A...598A..84A} and B-stars binaries characterization \citep[BBC,][]{2021MNRAS.507.5348V}
surveys in order to establish orbital properties. Excluding complex systems in R136a, 41 sources are confirmed spectroscopic binaries, of which 24 are double-lined spectroscopic binaries (SB2). One source (SMB 151) has been identified as an eclipsing binary OGLE-LMC-ECL-21413 \citep{2011AcA....61..103G} which we include in the statistics of (SB1) spectroscopic binaries.


\begin{figure}
\centering
	\includegraphics[width=0.8\linewidth,angle=-90,bb = 0 90 540 730]{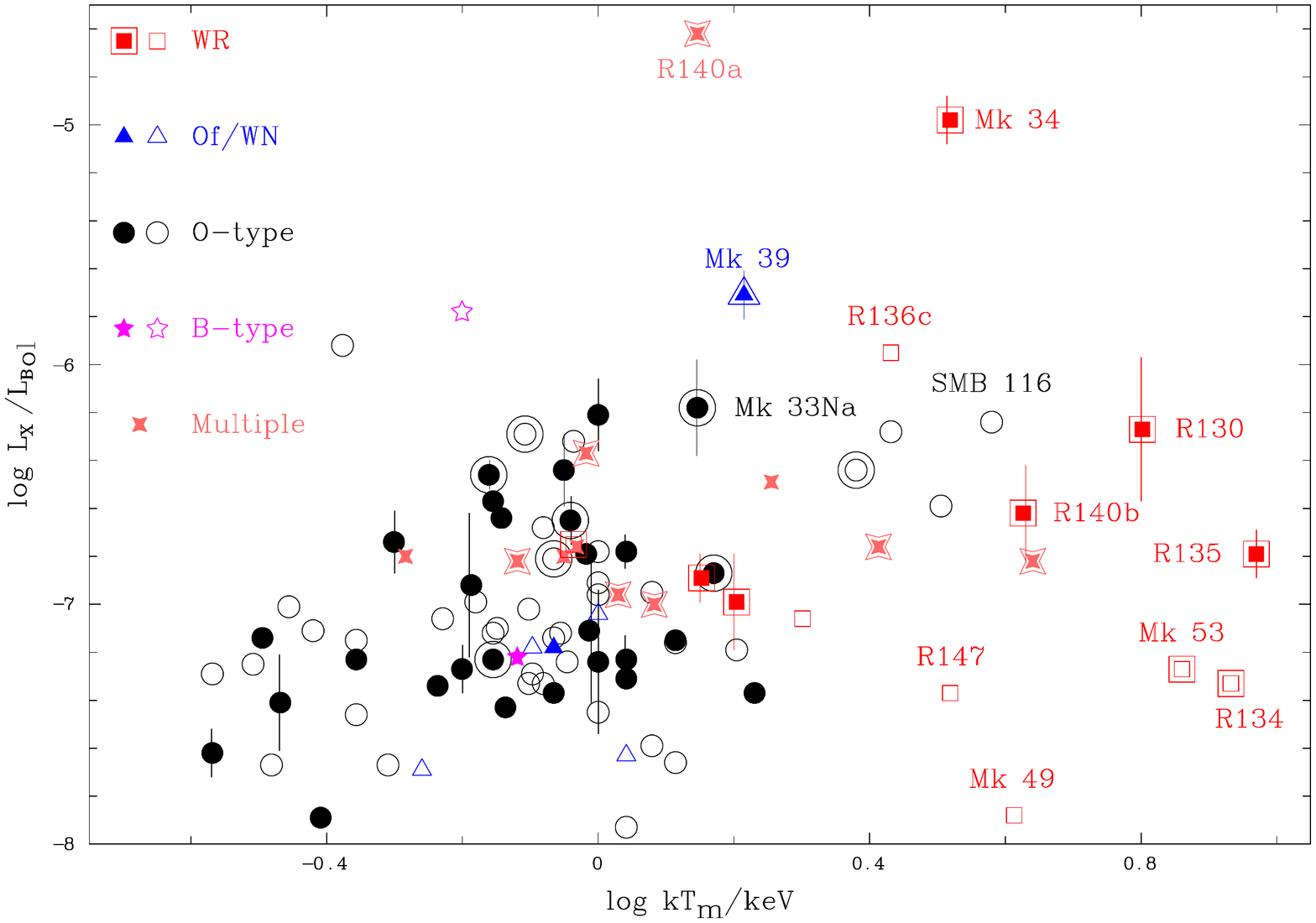}
	\includegraphics[width=0.8\linewidth,angle=-90,bb = 0 90 540 730]{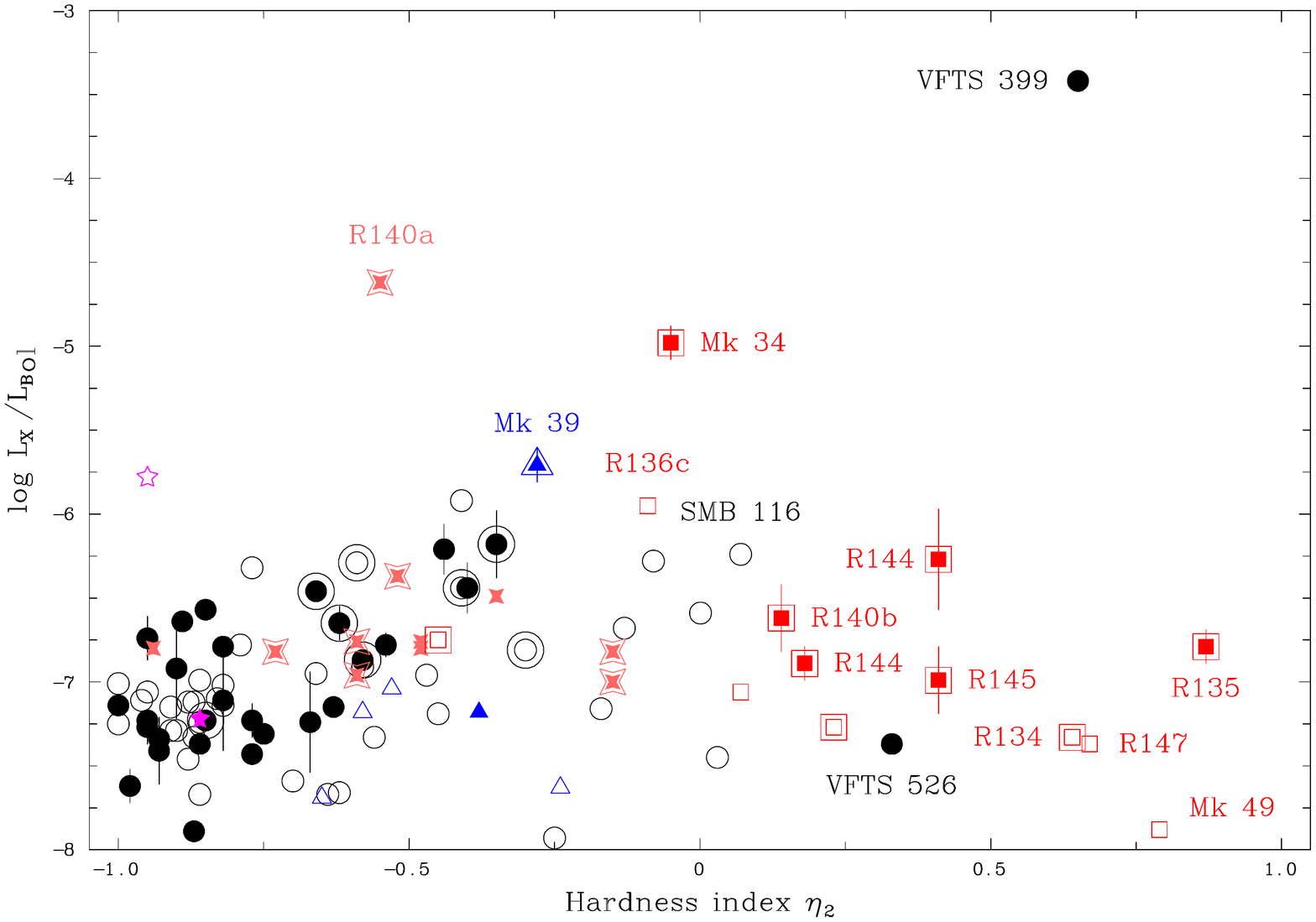}
  \caption{({\it top panel}): Comparison between $\log kT_{\rm m}$ and $\log L_{\rm X}/L_{\rm Bol}$ for early-type sources from the T-ReX point source catalogue (P.~Broos \& L.~Townsley, in prep). Symbols are included in the key (filled symbols are binaries, uncertainties are included solely for SB2 systems, double concentric shapes denote sources with dual temperature plasma fits). Note that spectral fitting to T-ReX sources required $kT \gtrapprox$ 0.3 keV. Several faint sources are excluded from this figure since their X-ray temperatures were adopted; ({\it bottom panel}): As above, except for a comparison between  the hardness index $\eta_2$ and  $\log L_{\rm X}/L_{\rm Bol}$ for T-ReX early-type sources. $\eta_{2}$ compares extinction corrected luminosities in the 2--8 keV to 0.5--2 keV energy range, such that soft emitters would possess an index of --1 and hard emitters would possess +1. } 
	\label{fig2}
\end{figure}

 Figure~\ref{fig1} compares X-ray to bolometric luminosities for all 107 T-ReX sources, with an average ratio of $\log L_{\rm X}/L_{\rm Bol} = -6.90 \pm 0.65$. The rich massive star population of the Tarantula Nebula provides a sample whose O-type sources alone span $4.5 \leq \log L_{\rm Bol}/L_{\odot} \leq 6.5$, while all Of/WN and WR sources exceed $10^{6} L_{\odot}$, with such high luminosity sources providing 40\% of the observed T-ReX sample of early-type stars. Nine X-ray sources lie within the central 2.4$''$ (0.6 parsec) of the R136 star cluster \citep{1998ApJ...493..180M, 2016MNRAS.458..624C} so due to severe crowding the association of X-ray photons with optical counterparts is less secure than elsewhere (recall Fig~\ref{r136a}). 
 
The overwhelming majority of spectroscopically confirmed binaries (filled symbols) do not have higher $L_{\rm X}/L_{\rm Bol}$ ratios than single stars. However, there are some exceptional sources with high X-ray luminosities, namely the candidate X-ray binary VFTS 399 \citep{2015A&A...579A.131C}, the WN5h+WN5h colliding wind binary Mk 34 \citep{2008ApJS..177..216G, 2018MNRAS.474.3228P, 2019MNRAS.484.2692T}, and R140a which is host to WN, WC and OB star(s). 
 \citet{1987ApJ...312..612M} have claimed that the WN star within R140a is an SB1 with a period of 2.7 days, although R140b is now favoured as host to the short period binary system \citep{2019A&A...627A.151S}, while \citet{2001MNRAS.324...18B} have not found any evidence that the WC star is a member of a binary system. R140a aside, other X-ray extreme systems currently lacking a confirmed spectroscopic binary include R136c (WN5h+?, VFTS 1025) for which a tentative 8.2\,day period was proposed by \citet{2009MNRAS.397.2049S}  and Mk 33Sa (O3\,III(f*), HSH 18) for which no spectroscopic evidence of binarity has been identified to date \citep{2005ApJ...627..477M}.


 

Table~\ref{table1} provides a breakdown of average ratios of $\log L_{\rm X}/L_{\rm Bol}$ for single stars, spectroscopic binaries and  multiple  sources within R136a and R140a, with single and binaries split by spectral type of the primary. For the entire sample, multiple systems aside, $\log L_{\rm X}/L_{\rm Bol} = -6.94 \pm 0.66$, in reasonable agreement to previous Galactic results \citep{1989ApJ...341..427C, 2006MNRAS.372..661S, 2015ApJS..221....1R}.  
If one excludes extreme outliers (Mk~34, VFTS 399) plus four systems with extremely hard (8--10 keV) components (R130, R134, R135, Mk~53), a revised average of $\log L_{\rm X}/L_{\rm Bol} = -7.00 \pm 0.49$ is obtained.

Separating the sample into confirmed spectroscopic binaries confirms the results from previous studies \citep{2005MNRAS.361..679O} that binary systems generally do not exhibit excess X-ray emission with respect to single
stars. SB2 systems possess a modest X-ray excess which is much lower than the dispersion, i.e. $\log L_{\rm X}/L_{\rm Bol} = -6.97 \pm 0.61$ (single), {$\log L_{\rm X}/L_{\rm Bol} = -6.96 \pm 0.99$} (SB1) and { $\log L_{\rm X}/L_{\rm Bol} = -6.74 \pm 0.59$} (SB2). If X-ray outliers are omitted, averages are $\log L_{\rm X}/L_{\rm Bol}$ = --7.00$\pm$0.53 (single), --7.19$\pm$0.32 (SB1) and --6.84$\pm$0.48 (SB2). For completeness, the average properties of multiple sources are $\log L_{\rm X}/L_{\rm Bol}$ = --6.76$\pm$0.19, excluding R140a, although these should be treated with caution owing to the challenges of associating X-ray emission to individual sources in such a crowded environment (recall Fig.~\ref{r136a}).

Magnetic OB stars may also be X-ray luminous. Upper limits on magnetic fields have been obtained for some luminous early-type stars in the Tarantula Nebula \citep{2020A&A...635A.163B}. Spectroscopically, O-type stars with kG magnetic fields usually possess a peculiar O?fp spectral type, so \citet{2014A&A...564A..40W} have highlighted seven Onfp stars within the Tarantula Nebula (their fig.~11). Of these, only VFTS 526 is included in the T-ReX PSC, it is also a SB1 system, in spite of which it is X-ray {\it faint}. Consequently we have no evidence from T-ReX that favours excess X-ray emission from potentially magnetic OB stars in the Tarantula Nebula. 

We find no significant dependence of $\log L_{\rm X}/L_{\rm Bol}$ on spectral class for O, Of/WN and WR stars, especially once X-ray outliers have been omitted, although some classes suffer from low number statistics. In particular, Table~\ref{table1} reveals no differences in  $\log L_{\rm X}/L_{\rm Bol}$ between O dwarfs/subgiants and other luminosity classes, in contrast with the findings of \citet{2018A&A...620A..89N} who have investigated the dependence of X-ray properties of Galactic O stars on luminosity class. From Galactic studies, single WC stars are known to be X-ray faint, presumably because of their optically thick winds \citep{2003A&A...402..755O}. There are no single WC stars in the T-ReX point source list, although R140a is known to host WN and WC stars, and R130 is a WN/C+B1\,I binary\footnote{R130 (BAT99-92) has been assigned a variety of subtypes, ranging from WN3b + B1\,Ia \citep{2014A&A...565A..27H} to WC4 + B1\,Ia \citep{2019A&A...627A.151S} to WCE + WN + B1\,I \citep{1989ApJ...337..251C} but WN/C  is favoured \citep{2013A&A...558A.134D} since the blue feature is dominated by He\,{\sc ii} $\lambda$4686 (preferring a WN or WN/C subtype) while C\,{\sc iv} $\lambda\lambda$5801-12 is exceptionally strong (preferring a WC or WN/C subtype). }. 

Only six B stars are detected in T-ReX, with a significant scatter in $\log L_{\rm X}/L_{\rm Bol}$, such that we do not have sufficient statistics to comment on their X-ray properties with respect to previous studies \citep{1997A&A...322..167B,1997ApJ...487..867C, 2011ApJS..194....7N}.

\section{X-ray hardness of T-ReX sources}\label{hardness}

Table~\ref{table1} provides average hardness indices, $\eta_{2}$ and mean plasma temperatures, $kT_{m}$ for T-ReX sources, while Fig.~\ref{fig2} (top panel) compares $kT_{\rm m}$ to $\log L_{\rm X}/L_{\rm Bol}$. Sources with two temperature plasma fits are indicated by two concentric shapes. The average plasma temperature for the entire sample is  $kT_{\rm m}$ = 1.44$\pm$1.61 keV. Fig~\ref{fig2} (bottom panel) compares the hardness index, $\eta_{2}$, of T-ReX sources to $\log L_{\rm X}/L_{\rm Bol}$ with an average value of
$\eta_{2}$ = --0.50$\pm$0.45 for the entire sample. Fig.~\ref{fig2} illustrates that WR systems (shown in red) are predominantly hard emitters, regardless of whether they are known binaries, with an average hardness index of $\eta_{2}$ = +0.29$\pm$0.38. 
Aside from the suspected high-mass X-ray binary VFTS 399 \citep{2015A&A...579A.131C}, half of the six X-ray sources with the hardest indices are apparently single: Mk\,49 (WN6(h)), R134 (WN6(h)) and R147 (WN5h), while the remainder are known SB2 binaries: R135 (WN5:+WN7),  R144 (WN5--6h+WN6--7h) and R145 (WN6h+O3/5If/WN7).

This suggests that high wind densities play a role in harder X-ray emission, since the corresponding hardness indices of O dwarfs/subgiants, O (super)giants, Of/WN stars are $\eta_{2}$ = --0.71$\pm$0.26, --0.61$\pm$0.42 and --0.44$\pm$0.17, respectively. \citet[][their fig.~2]{2019A&A...626A..59C} have previously established that (single and binary) WR stars in the Galactic open cluster Westerlund 1 generally possess harder X-ray emission than OB supergiants, based on a  hardness index constructed from {\it attenuated} fluxes, while  \citet{2018A&A...620A..89N} found Galactic O supergiants possess harder X-ray spectra than dwarfs and giants.

A subset of OB stars are hard X-ray emitters, including SMB 116 \citep[O3--6\,V,][]{1997ApJS..112..457W} suggesting a colliding wind binary origin. The majority of spectroscopic binaries are relatively soft, in common with previous Galactic studies. Indeed, the average hardness index of spectroscopic binaries is statistically identical to that of single stars. The average hardness index of the multiple X-ray sources associated with early-type stars in R136 is --0.50$\pm$0.24, identical to the wider sample, with 136a-X2 (host to R136a3 and an early O supergiant) and c7452 (host to two early O stars) the hardest, and R136a-X7 (host to an O2 dwarf) the softest. None possess X-ray properties which unambiguously flag the presence of a colliding wind binary.

\begin{table}
\centering
\caption{Detection statistics of 1001 spectrally classified early-type stars (or multiple systems whose primary is an early-type star) within the T-ReX field-of-view, drawn from  \citet{2019Galax...7...88C} and references therein,
excluding the spatially crowded central region of R136a.}
\begin{tabular}{l@{\hspace{2mm}}r@{\hspace{2mm}}r@{\hspace{1.5mm}}l@{\hspace{5mm}}r@{\hspace{2mm}}r@{\hspace{1.5mm}}l}
\hline
Subtype/ & \multicolumn{2}{l}{X-ray} & Example & \multicolumn{2}{l}{Non X-ray} & Example \\
Luminosity           & N                 & \%            &                & N      & \%                               & \\
\hline
O-type             & 71  & 14.2\% &                     & 428 & 85.8\% &  BI~253       \\ 
B-type             &   6  &   1.3\% &                      & 469  & 98.7\% &  R142     \\ 
Wolf-Rayet      & 14  & 70.0\% &                      & 6     & 30.0\% & Mk~33Sb       \\ 
Of/WN             & 6     & 85.7\% &                    &  1     & 14.3\% & R136b \\ 
\hline
$\log(L/L_{\odot})<5$        & 10 & 1.5\% &  VFTS399             & 671 & 98.5\%  &       \\ 
$5\leq \log(L/L_{\odot})<6$&50& 17.9\% &             & 229 & 82.1\% &      \\ 
$\log(L/L_{\odot})\geq6$   &37& 90.2\% &             &     4  & 9.8\%  &  VFTS~72     \\ 
\hline
Total              & 97    & 9.7\%    &   & 904 & 90.3\% & \\  
\hline
\end{tabular}
\label{non_detections}
\end{table}

\begin{table*}
\centering
\caption{Count rates (CR) and inferred observed X-ray luminosities $L^{t}_{\rm X}$ for a selection of luminous early-type stars
absent from the T-ReX point source catalogue, based on the maximum-likelihood solution (MLS) approach.  Intrinsic X-ray luminosities, $L^{tc}_{\rm X}$ are estimated from typical attenuation corrections of 0.31$\pm$0.17 dex. Catalogues include R \citep{1960MNRAS.121..337F}, BI \citep{1975A&AS...21..109B}, Mk  \citep{1985A&A...153..235M}, VFTS \citep{2011A&A...530A.108E}, MH \citep{1994AJ....107.1054M} HSH \citep{1995ApJ...448..179H}, SMB \citep{1999A&A...341...98S}, P \citep{1993AJ....106..560P},  CCE \citep{2018A&A...614A.147C}, BAT \citep{1999A&AS..137..117B}}\label{Andy}
\begin{tabular}{c@{\hspace{1mm}}c@{\hspace{1mm}}c@{\hspace{1mm}}c@{\hspace{1mm}}c@{\hspace{1mm}}c@{\hspace{1mm}}c@{\hspace{1mm}}r@{\hspace{1mm}}c@{\hspace{1mm}}c@{\hspace{1mm}}c@{\hspace{1mm}}c@{\hspace{3mm}}l@{\hspace{1mm}}l@{\hspace{1mm}}c@{\hspace{1mm}}c@{\hspace{1mm}}c@{\hspace{1mm}}l@{\hspace{1mm}}l}
\hline
R         & BI & Mk & VFTS & MH & HSH & SMB & P & CCE & BAT  & Spect. & Ref & $\log L_{\rm Bol}$          & Ref & CR        & $\log L^{t}_{\rm X}$  & $\log L^{tc}_{\rm X}/L_{\rm Bol}$ & Nature & Ref\\
            &      &    &           &           &        &      &       &           &         & Type  &       &  $L_{\odot}$ &       & ksec$^{-1}$   & erg\,s$^{-1}$                &  &\\
\hline
 $\cdots$   &253&    $\cdots$    & \phantom{5}72      & $\cdots$ &    $\cdots$        &     $\cdots$      &       $\cdots$     &     $\cdots$     &      $\cdots$          & O2\,V--III & 1 & 6.07$\pm$0.2\phantom{0} & 2 & 0.043$\pm$0.012 & \phantom{$<$}32.08$^{+0.11}_{-0.14}$  &\phantom{$<$}--7.26$^{+0.10}_{-0.14}$  & Single & 1\\[1pt]
142      &   $\cdots$   &    $\cdots$      & 533    & $\cdots$ &     $\cdots$       & 3      & 987  & 2912 &        $\cdots$        & B1.5\,Ia$^+$ & 3 & 5.88$\pm$0.10 & 4 & 0.000$\pm$0.004 & $<$31.05\phantom{+0.00} & $<$--8.1\phantom{+0.00} &  Single & 3\\[1pt]
     $\cdots$         &   $\cdots$   &33Sb&       $\cdots$      & 859 & 34     &      $\cdots$     & 1111 &  $\cdots$ & 115      & WC5   &  5   & 5.9\phantom{0}$\pm$0.3\phantom{0} & 6 & 0.026$\pm$0.009 & \phantom{$<$}31.87$^{+0.13}_{-0.19}$ & \phantom{$<$}--7.31$^{+0.13}_{-0.18}$ & V.Comp & 7\\[1pt]
\hline
\end{tabular}
\newline 
1: \citet{2014A&A...564A..40W}; 
2: \citet{2017A&A...601A..79S};  
3: \citet{2015A&A...574A..13E}; 
4: \citet{2015A&A...575A..70M}; 
5: \citet{1998ApJ...493..180M}; 
6: \citet{2013A&A...558A.134D};  
7: \citet{2001MNRAS.324...18B} 
 \end{table*}

\begin{figure}
\centering
	\includegraphics[width=0.8\linewidth,angle=-90,bb = 0 90 540 730]{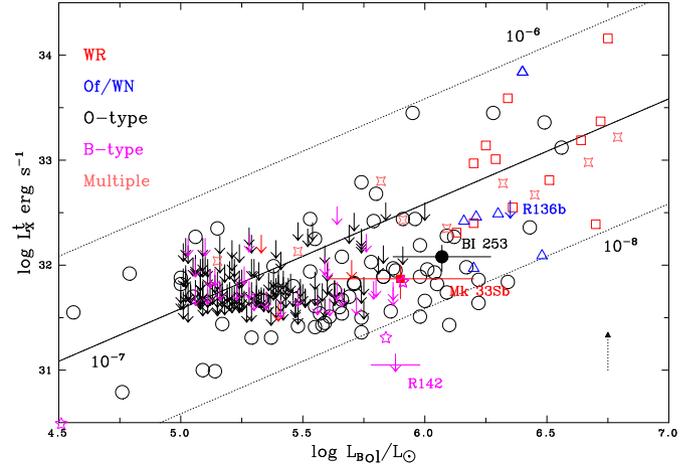}
  \caption{Comparison between {\it observed} X-ray luminosities and bolometric luminosities for 107 early-type sources from the T-ReX point source catalogue, additionally including limits for non-detected luminous ($\log L/L_{\odot} \geq 5$) early-type stars plus inferred X-ray luminosities from count rate analysis (Sect.~\ref{MLS}) of BI~253 (O2\,V--III, solid black circle), R142 (B1.5\,Ia$^{+}$, purple arrow) and Mk\,33Sb (WC5, solid red square). R136b (O4\,If/WN8, blue arrow) is also indicated. Symbols correspond to the nature of the primary/brightest component, namely WR star (square), Of/WN star (triangle), O-type star (circle), B-type star (star). The solid line indicates the canonical $L_{\rm X} = 10^{-7} L_{\rm Bol}$ relation, with dotted lines offset by $\pm$1 dex, and the vertical dotted arrow indicates average X-ray attenuation corrections (star-by-star corrections cannot be determined since intrinsic X-ray spectral energy distributions are unknown).} 
	\label{limits}
\end{figure}

\begin{figure*}
\centering
\begin{minipage}[c]{0.7\linewidth}
	\includegraphics[width=0.69\linewidth,angle=-90,bb = 40 80 540 700]{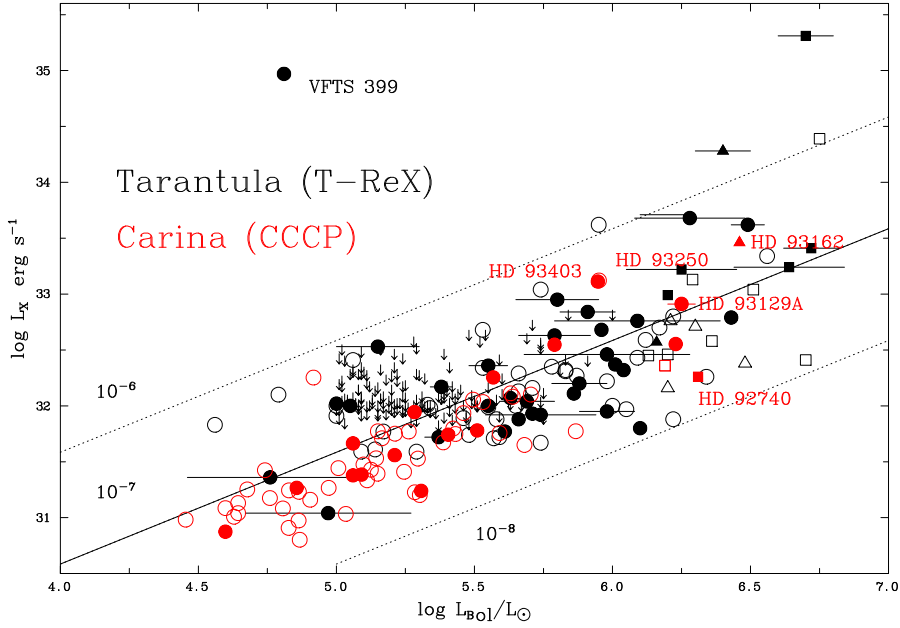}
\end{minipage}\hfill
\begin{minipage}[c]{0.3\linewidth}
\centering
   \caption{Comparison between X-ray and bolometric luminosities for single (open) and binary (filled) O stars (circles), Of/WN stars (triangles) and WN stars (squares) in the Tarantula Nebula from XSPEC fits to T-ReX observations (black) with those in the Carina Nebula from CCCP (red). Overall there are no systematic differences between LMC and Galactic early-type stars, although T-ReX statistics are poor below $\log L/L_{\odot} \sim $ 5.0.  According to  \citet{2013MNRAS.429.3379O} one would anticipate lower $L_{\rm X}/L_{\rm Bol}$ ratios for stars with dense stellar winds, which is not apparent from observations. The solid line indicates the canonical $L_{\rm X} = 10^{-7} L_{\rm Bol}$ relation, with dotted lines offset by $\pm$1 dex. Upper limits to attenuation corrected X-ray luminosities of O stars in the Tarantula have been included (arrows). CCCP results for O stars are drawn from \citet{2011ApJS..194....7N}  and \citet{2011ApJS..194....5G}, with the bolometric luminosity of HD\,93129A updated according to \citet{2019A&A...621A..63G} while CCCP results for Of/WN (HD~93162) and WN stars (HD~92740, HD~93131) are newly presented here.}\label{Carina}
	\end{minipage}
\end{figure*}

\section{T-ReX PSC non-detections}\label{MLS}

The T-ReX point source catalogue represents an exceptionally rich dataset with which to consider X-ray emission from early-type stars in the Tarantula Nebula, yet there are over a thousand hot luminous stars
in this region \citep{2019Galax...7...88C}. In Table~\ref{non_detections} we summarise the fraction of spectroscopically confirmed early-type stars within the T-ReX field-of-view that are detected in the PSC, according to spectral type (of the primary in a binary system) or bolometric luminosity (sum of individual components in a binary), excluding the crowded central region of R136a (recall Fig.~\ref{r136a}). The majority of the highest luminosity stars with $\log L/L_{\odot}\geq 5.6$ (with overwhelmingly O, Of/WN or WR subtypes) are detected in the T-ReX PSC, though the reverse is true for moderate luminosity OB-type stars.


Consequently, we have obtained upper limits (95\% probability) to photon fluxes for all  luminous ($\log L/L_{\odot} \geq 5$) early-type stars which are undetected by T-ReX, in order for us to assess whether these are genuinely X-ray faint\footnote{It is important to emphasise that optical counterparts are excluded if they are offset from X-ray PSC centroids by $\gtrapprox 1''$. P.~Broos \& L.~Townsley (in prep, their table 5) identify PSC sources in close proximity to known early-type stars. Those within 1--2$''$ of luminous stars are noted in Table~S1 in the Supplementary Data}. These are converted to attenuated luminosities from a calibration obtained from PSC detections,
\[\log L_{\rm X}^{t} = \log ({\rm PhotonFlux/cm}^{-2}\,{\rm s}^{-1}) + (38.81\pm0.16). \]
and presented in Table S1 (Supplementary Data). It was not possible to obtain upper limits for sources close to the edge of the T-ReX field of view since the background is highly variable, nor those within the vicinity of the N157B supernova remnant \citep{2006AJ....131.2140T} since diffuse emission is highly variable, although all early-type sources are retained for completeness. Typical upper limits to attenuated luminosities lie in the range $31.6 \leq \log L^{t}_{\rm X}/({\rm erg\,s}^{-1})  \leq$ 32.5. X-ray to bolometric luminosity ratios follow, based on average attenuation corrections of 0.31$\pm$0.17 dex. Star-by-star corrections based on measured optical extinctions cannot be determined since intrinsic X-ray spectral energy distributions are unknown.

By way of example, VFTS 419 (O9:\,V(n))  has a luminosity of  log $L/L_{\odot}$ = 5.07 \citep{2017A&A...601A..79S}, close to the average  of all 499 visually classified O-type stars in the Tarantula. We find an upper limit of $\log L^{t}_{\rm X}/({\rm erg\,s}^{-1}) \leq$  31.71 for VFTS 419, such that $\log L_{\rm X}^{tc}/L_{\rm Bol} \leq -6.64$. Consequently we are not able to assess whether low luminosity O stars are X-ray `normal' or faint. This is further illustrated in Fig.~\ref{limits} where we compare {\it attenuated} X-ray luminosities, or limits, to bolometric luminosities for the full sample. 
 
The absence of 4 very high luminosity stars with $\log L/L_{\odot} \geq 6$ from the T-ReX PSC (Table~\ref{non_detections}) merits further discussion. Of these, one lies at the edge of the survey field (VFTS 3), two lie in regions of very high background (R136b, HSH 59) though one, BI\,253, does permit a useful upper limit of $\log L^{t}_{\rm X}/({\rm erg\,s}^{-1}) \leq$ 32.18, indicating $\log L_{\rm X}/L_{\rm Bol} \leq -7.16$.  R136b represents the only non-detected Of/WN supergiant in T-ReX, though it lies in a complex region (Fig.~\ref{r136a}). 

We have employed one other technique to estimate X-ray luminosities for a subset of very luminous stars excluded from the T-ReX PSC.  We have constructed spatial models incorporating weak target stars and any survey neighbours into multiple overlapping PSFs and a background term and seeking the maximum-likelihood solution (MLS) with respect to all the count rates (CRs) simultaneously. Count rates map into observed luminosities via $\log (L^{t}_{\rm X}/ {\rm erg\,s}^{-1}) = \log ({\rm CR} \times T_{\rm Exp}) + (30.15 \pm 0.18)$, where $T_{\rm Exp}$ is the total exposure time, obtained from a comparison with 16 {\it Chandra} Source Catalog Release 2.0\footnote{https://cxc.cfa.harvard.edu/csc} sources within the T-ReX field.  By way of a sanity check, for 6 sources in common with the T-ReX point source catalogue (Table~\ref{A1}) we find $\log L^{t}_{\rm X}$(MLS) -- $\log L^{t}_{\rm X}$(PSC) = +0.09 $\pm$ 0.06 dex. 

For the MLS technique we have considered the luminous O star BI\,253 (O2\,V--III), the early B supergiant R142 (B1.5\,Ia$^+$) and the WR star Mk\,33Sb (WC5), none of which are included in the T-ReX PSC (Table~\ref{non_detections}). Results of our analysis are presented in Table~\ref{Andy}, and reveal {\it unattenuated} X-ray to bolometric luminosity ratios, $L^{tc}_{X}/L_{\rm Bol}$, for the early O star and WC star which are consistent with early-type stars detected in the T-ReX PSC after folding in typical attenuation corrections. Consequently,  it is currently not possible to conclude whether typical OB stars in 30 Dor are subluminous in X-rays.

We have highlighted these systems in Fig.~\ref{limits}, which further illustrates that the B hypergiant R142 is also X-ray faint. Galactic B stars are known to exhibit a larger dispersion in $L_{\rm X}/L_{\rm Bol}$ than O stars, with a tendency towards a lower value for high luminosity sources \citep{2011ApJS..194....7N}. Still, R142 does appear to be anomalously X-ray faint, in common with Mk\,54 (B0.5\,Ia), the latest luminous B supergiant detected in the T-ReX PSC with $\log L_{\rm X}/L_{\rm Bol} = -7.49$. Of  these three sources one would expect that the WC star Mk~33Sb would be X-ray faint rather than the B hypergiant, since single Galactic WC stars have not been detected in X-rays \citep{2003A&A...402..755O}. Perhaps Mk\,33Sb is an unidentified binary system  \citep[see][]{2001MNRAS.324...18B}. 

\begin{figure}
\centering
	\includegraphics[width=0.75\linewidth,angle=-90,bb = 0 80 540 720]{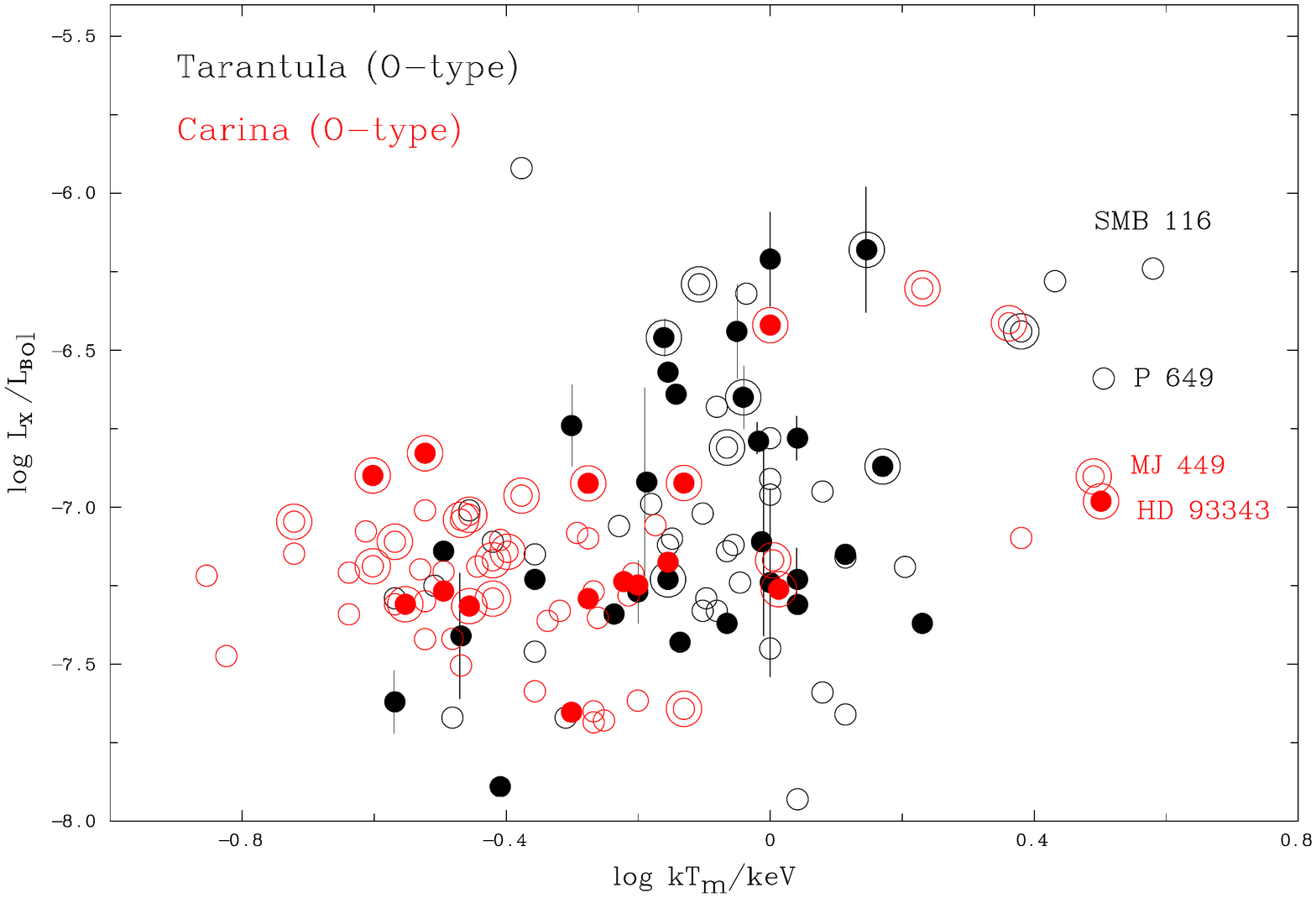}
		\includegraphics[width=0.75\linewidth,angle=-90,bb = 0 80 540 720]{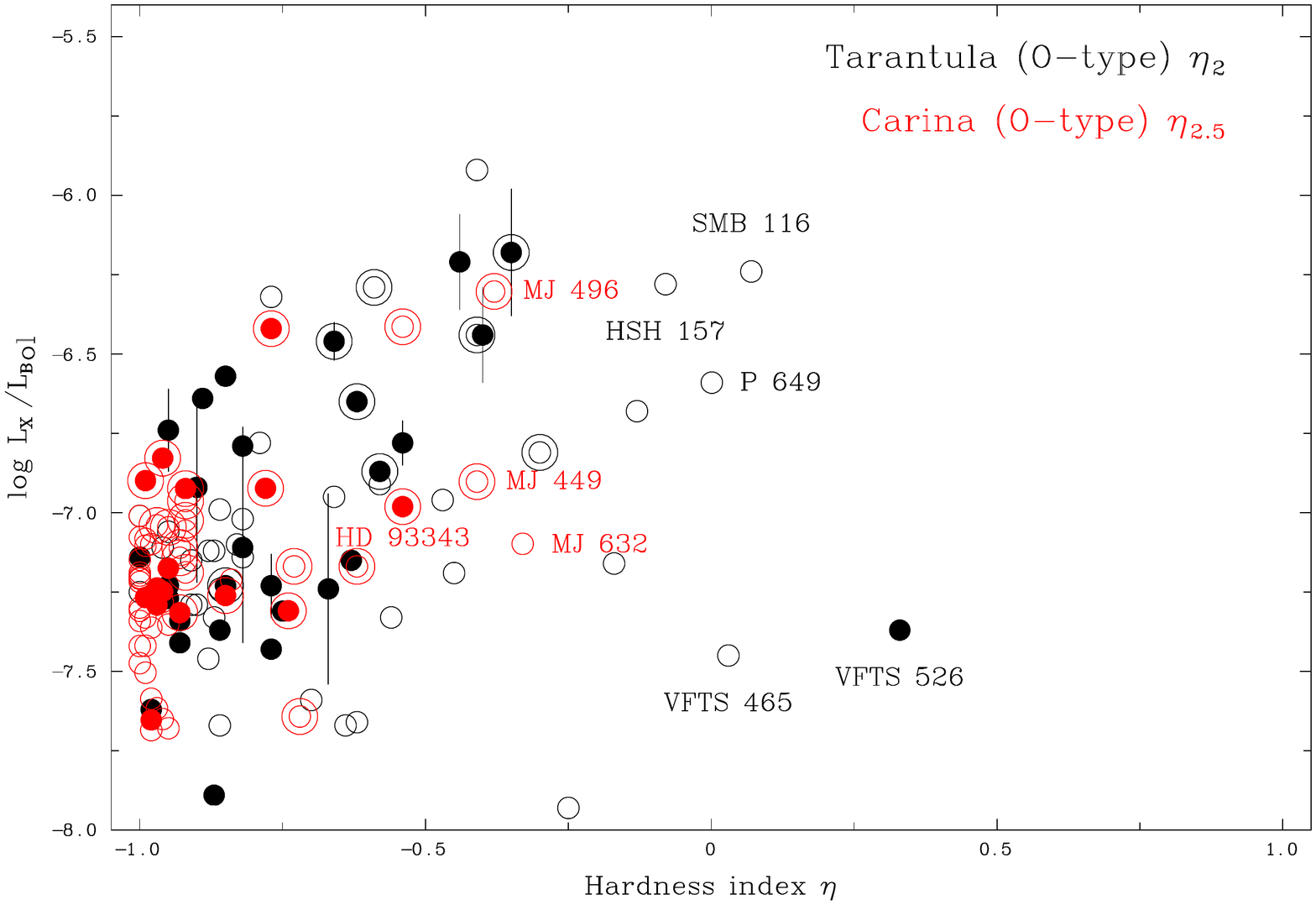}
\centering
  \caption{({\it Top panel}): Comparison between  $\log kT_{\rm m}$ and $\log L_{\rm X}/L_{\rm Bol}$ for single (open) and binary (filled) O stars in the Tarantula Nebula (black) with those in the Carina Nebula (red) from \citet{2011ApJS..194....7N}  and \citet{2011ApJS..194....5G},  in which two plasma temperature components are indicated with two concentric circles. Single O stars in the LMC possess higher plasma temperatures than their counterparts in Carina. Note that 
 spectral fitting to T-ReX sources {\it required} $kT \gtrapprox$ 0.3 keV.  Sources with $kT_{m} \geq$3 keV are labelled. ({\it Bottom panel}): As above, except for a comparison between hardness index, $\eta$, and $\log L_{\rm X}/L_{\rm Bol}$. A modified
index $\eta_{2.5}$ is adopted for Carina O stars since \citet{2011ApJS..194....7N} provided soft, medium and hard luminosities, such that the Galactic $\eta_{2.5}$ index will be systematically softer than the LMC index $\eta_{2}$. }
	\label{Carina-kT}
\end{figure}
\begin{table*}
\begin{center}
\caption{
Comparison between average $\log L_{\rm X}/L_{\rm Bol}$, hardness indices, $\eta$, and mean plasma temperature $kT_{m}$ for {\it luminous} ($\log L_{\rm Bol}/L_{\odot} \geq 5$) OB, Of/WN and WN stars in the Carina Nebula \citep{2011ApJS..194....7N, 2011ApJS..194....5G} versus the Tarantula Nebula from this study. The hardness index for Galactic OB stars, $\eta_{2.5}$ is defined relative to a 2.5 keV threshold instead of 2 keV for LMC stars, so will produce systematically softer indices. Average properties for Carina OB stars are also presented excluding O-type outliers  HD\,93250 (single), HD\,93403 (binary). We include previously unpublished Carina CCCP X-ray results for HD~92740 (WN7ha+O), HD~93131 (WN6ha) and HD~93162 (O2.5\,If*/WN6+O) for completeness (including $\eta_{2}$), adjusting spectroscopic results \citep{2019A&A...625A..57H} to Gaia eDR3 distances following the approach of \citet{2020MNRAS.493.1512R}.
}
\label{MW_vs_LMC}
\begin{tabular}{l@{\hspace{3mm}}
r@{\hspace{2mm}}c@{\hspace{2mm}}c@{\hspace{2mm}}c@{\hspace{2.5mm}}c@{\hspace{10mm}}
r@{\hspace{2mm}}c@{\hspace{2mm}}c@{\hspace{2mm}}c@{\hspace{2.5mm}}c}
\hline
Subtype & N & $\log \overline{L_{\rm Bol}/L_{\odot}}$ & $\log \overline{L_{\rm X}/L_{\rm Bol}}$ & $\overline{\eta_{2.5} }$ & $\overline{kT_{m}}$ & 
                 N & $\log \overline{L_{\rm Bol}/L_{\odot}}$ & $\log \overline{L_{\rm X}/L_{\rm Bol}}$ & $\overline{\eta_{2} }$ & $\overline{kT_{m}}$  \\
\hline
                     & \multicolumn{5}{c}{--- Carina (Milky Way) --- }  & \multicolumn{5}{c}{--- Tarantula (LMC) --- } \\
O single    & 28 & 5.37$\pm$0.25 & --7.25$\pm$0.27 & --0.94$\pm$0.10 &  0.47$\pm$0.39 &                  38 & 5.68$\pm$0.39 & --7.08$\pm$0.41 &   --0.63$\pm$0.31 & 0.99$\pm$0.75 \\
(excl.) & 27 & 5.38$\pm$0.28 & --7.28$\pm$0.21 & --0.96$\pm$0.06 &    0.40$\pm$0.15 &                     38 & 5.68$\pm$0.39            & --7.08$\pm$0.41  &  --0.63$\pm$0.31 & 0.99$\pm$0.75 \\
O binary$^{a}$ & 13 & 5.52$\pm$0.42 & --7.10$\pm$0.31 &  --0.89$\pm$0.13 &  0.77$\pm$0.76 &          28 & 5.77$\pm$0.37 & --7.02$\pm$0.42 & --0.73$\pm$0.28 & 0.84$\pm$0.36 \\
(excl.) & 12 & 5.48$\pm$0.42 & --7.15$\pm$0.24 & --0.90$\pm$0.13 &   0.75$\pm$0.79 &                      28 & 5.77$\pm$0.37 & --7.02$\pm$0.42 & --0.73$\pm$0.28 & 0.84$\pm$0.36 \\
\\
B single &  2 & 5.28$\pm$0.16 & --7.74$\pm$0.15 & --0.66$\pm$0.45 & 1.46$\pm$1.27  &                   1 & 5.84\phantom{$\pm$0.00}  & --7.49\phantom{$\pm$0.00} & --0.87\phantom{$\pm$0.00} & $\cdots$  \\ 
B binary &  0   &       $\cdots$                    &      $\cdots$    & $\cdots$   & $\cdots$               &        1 & 5.91\phantom{$\pm$0.00}                  & --7.22\phantom{$\pm$0.00}   & --0.86\phantom{$\pm$0.00} & 0.76\phantom{$\pm$0.00} \\ 
\hline
Subtype & N & $\log L_{\rm Bol}/L_{\odot}$ & $\log L_{\rm X}/L_{\rm Bol}$ & $\eta_{2}$ & $kT_{m}$ & N & $\log \overline{L_{\rm Bol}/L_{\odot}}$ & $\log \overline{L_{\rm X}/L_{\rm Bol}}$ & $\overline{\eta_{2} }$ & $\overline{kT_{m}}$ \\
               &    &                                                             &                                                               &                                         & keV &
                    &                                                             &                                                                 &                                         & keV\\

\hline
                     & \multicolumn{5}{c}{--- Carina (Milky Way) --- }  & \multicolumn{5}{c}{--- Tarantula (LMC) --- } \\
Of/WN single & 0 & $\cdots$                             & $\cdots$                       & $\cdots$ &  $\cdots$ & 4   &   6.30$\pm$0.13      &    --7.38$\pm$0.32           &      --0.50$\pm$0.18    & 0.86$\pm$0.24 \\ 
Of/WN binary$^{b}$ & 1 & 6.46\phantom{$\pm$0.00} & --6.58\phantom{$\pm$0.00} & --0.44\phantom{$\pm$0.00} &  1.86\phantom{$\pm$0.00}  &    2    &   6.28$\pm$0.17     &   --6.44$\pm$1.04            &    --0.33$\pm$0.07  & 1.25$\pm$0.55    \\
\\
WN single$^{b}$      & 1 & 6.19\phantom{$\pm$0.00} & --7.42\phantom{$\pm$0.00} & --0.38\phantom{$\pm$0.00} & 1.94\phantom{$\pm$0.00} &        8  &     6.42$\pm$0.22  &    --6.77$\pm$1.04           &    +0.16$\pm$0.51   & 4.12$\pm$2.80 \\
WN binary$^{b}$     & 1 & 6.31\phantom{$\pm$0.00} & --7.63\phantom{$\pm$0.00} & --0.40\phantom{$\pm$0.00} & 1.70\phantom{$\pm$0.00} &      5    &    6.50$\pm$0.25 & --6.45$\pm$0.84 & +0.31$\pm$0.35 & 3.97$\pm$3.22 \\
\hline
\end{tabular}
\end{center}
\footnotesize{ (a): Updated luminosity for HD\,93129A from \citet{2019A&A...621A..63G}; (b): X-ray properties of Carina Of/WN and WN stars are based on CCCP observations, with luminosities resulting from Gaia eDR3 distances of 2.48, 2.62 and 2.19 kpc for HD~92740, HD~93131 and HD~93162, respectively.}
\end{table*}

\section{Metallicity dependence of $L_{\rm X}/L_{\rm Bol}$} \label{metallicity_dependence}

X-rays originate in shocks embedded within stellar winds due to instabilities arising from the line driven phenomenon, so one would expect a mass-loss dependence, either $L_{\rm X} \sim \dot{M}$ if the shocks are radiative, or $L_{\rm X} \sim \dot{M}^2$ if they are adiabatic \citep{2013MNRAS.429.3379O}.
To date,  \citet{2014ApJS..213...23N} have suggested LMC O stars possess comparable X-ray emission  to Milky Way O stars from stacked (300\,ks) {\it Chandra} observations of undetected O stars in N11.  At face value our results favour reduced X-ray output from single O stars in the LMC ($\log L_{\rm X}/L_{\rm Bol} = -7.04 \pm 0.44$) versus the Milky Way based on $\log L_{\rm X}/L_{\rm Bol} = -6.72 \pm 0.49$ obtained by \citet{2009A&A...506.1055N} for a large sample of single O stars. However, the latter are a heterogenous sample while the former are strongly biased to high luminosity O stars which possess strong winds.




In order to quantitatively compare X-ray properties of LMC O stars with realistic Galactic O-type counterparts, we focus on the Carina Nebula, which is the closest Galactic analogue to 30 Doradus\footnote{ \citet{2002ApJ...573..191M}, \citet{2014ApJS..213....1T} and \citet{2019AJ....157...29H} have investigated the X-ray properties of luminous stars in NGC~3603, which is the closest Galactic analogue to the R136 cluster \citep{1994ApJ...436..183M} }, comprising a rich population of massive stars  spanning a range of ages in a large star-forming complex. The Carina Nebula has been surveyed in X-rays with the {\it Chandra} Carina Complex Project \citep[CCCP,][]{2011ApJS..194....1T}. \citet{2011ApJS..194....7N} and \citet{2011ApJS..194....5G} have studied the X-ray properties of OB stars in the Carina Nebula. We have supplemented these results with CCCP X-ray results for HD~HD92740 (WN7ha+O), HD~93131 (WN6ha) and HD~93162 (O2.5\,If*/WN6+O) based on Gaia eDR3 distances of 2.48, 2.62 and 2.19 kpc, respectively, following the approach of \citet{2020MNRAS.493.1512R}. Bolometric luminosities for these stars from \citet{2019A&A...625A..57H} have been adjusted to the eDR3 distances. 

Fig.~\ref{Carina} compares the  X-ray luminosities to bolometric luminosities of O stars, Of/WN stars and WN stars in the Tarantula (T-ReX, black) and Carina (CCCP, red). Very few members of Carina significantly exceed the  $10^{-7}$ X-ray to bolometric ratio relation, two single stars  HD\,93250 (O4\,III(fc)), Tr14 MJ\,496 (O8.5\,V) plus  two known binaries HD\,93403 (O5.5\,I+O7\,V) and HD 93162 (O2.5\,If/WN6+O). From inspection,  Galactic and LMC O stars reveal comparable X-ray luminosities, contrary to expectations that weaker winds at reduced metallicity would lead to lower X-ray luminosities \citep{2013MNRAS.429.3379O}. A couple of comments are necessary regarding Fig.~\ref{Carina}. Firstly, CCCP is sensitive to weak X-ray emission from relatively low luminosity O stars, whereas T-ReX has detected exclusively the most extreme X-ray sources in the Tarantula. Indeed only two O stars in Carina exceed $\log L_{\rm Bol}/L_{\odot} =$  6 (both binaries)\footnote{The most extreme O-type system in Carina is HD~93129A \citep{2019A&A...621A..63G} which exhibits normal X-ray properties despite being a spectroscopic binary (O2\,If* + O3\,III(f*)), presumably due to the large physical separation between the two components, diluting any wind-wind collision contribution.} versus 18 sources in 30 Doradus, such that the bulk of the Carina O star sample are lower luminosity, late subtypes ($\log \overline{L_{\rm Bol}/L_{\odot}} = 5.20\pm0.40$)  while there are large numbers of high luminosity, early O subtypes in the T-ReX sample ($\log \overline{L_{\rm Bol}/L_{\odot}} = 5.65\pm0.45$). 

In view of the differences between Carina and Tarantula O star samples, we compare the X-ray properties of {\it luminous} OB stars, Of/WN stars and WN stars in these star-forming regions with $\log L/L_{\odot} \geq 5$ in Table~\ref{MW_vs_LMC}.
 For luminous O stars in Carina $\log L_{\rm X}/L_{\rm Bol} = -7.25\pm 0.27$ for 28 single stars (vs $-7.08\pm 0.42$ for 37 single O stars in the Tarantula) and $\log L_{\rm X}/L_{\rm Bol} = -7.10 \pm 0.31$ for 13 binaries (vs $-7.02 \pm 0.42$ for 28 O-type binaries in the Tarantula), confirming results for the full datasets. Detailed comparisons between the Of/WN and WN stars are severely constrained by their scarcity, although their bolometric luminosities are comparable, and Carina WN stars are relatively X-ray faint. The well known colliding wind binary HD~93162 \citep{2006A&A...445.1093P, 2006A&A...460..777G}  is conspicuous in Carina for its exceptional $L_{\rm X}/L_{\rm Bol}$, yet in comparison with the Tarantula its properties are intermediate between those of Of/WN binaries Mk~39 and Mk~30 (Fig.~\ref{Carina}).

Table~\ref{MW_vs_LMC} also provides mean plasma temperatures for luminous O stars in Carina and the Tarantula Nebula, and reveals softer X-ray emission in the Milky Way ($\overline{kT_{m}} =$ 0.5 keV) than the LMC ($\overline{kT_{m}} = 1.0$ keV).  The upper panel of Fig.~\ref{Carina-kT}  compares $kT_{m}$ and $L_{\rm X}/L_{\rm Bol}$ ratios of Carina O stars to those in the Tarantula. Sources with  $kT_{m} \geq$3 keV are labelled and include HD~93343 (O8\,V+O7--8.5\,V, binary) and MJ~449 (O8.5\,V((f)), single) in Carina, plus SMB~116 (O3--6\,V, single) and P~649 (O8--9\,V, single) in the Tarantula.

\citet{2011ApJS..194....7N} provide extinction-corrected luminosities of Carina stars in the soft ($\leq$ 1 keV), medium (1--2.5 keV, $L_{\rm X}^{mc}$) and hard ($\geq$ 2.5 keV, $L_{\rm X}^{hc}$) bands, it is necessary to use a modified hardness index, $\eta_{2.5}$,
\[
\eta_{2.5} = \frac{L^{hc}_{\rm X}- (L^{sc}_{\rm X} + L^{mc}_{\rm X})}{(L^{sc}_{\rm X} + L^{mc}_{\rm X} + L^{hc}_{\rm X})}.
\]
The lower panel of Fig.~\ref{Carina-kT} compares the hardness indices of Carina and Tarantula O stars, although these are more difficult to directly compare than plasma temperatures. Nevertheless, the average $\eta_{2.5}$ = --0.94$\pm$0.10 for single O stars in Carina  is {\it significantly} softer than the average $\eta_{2}$ = --0.64$\pm$0.32 for single O stars in the Tarantula, too large a difference to be explained solely by their different definitions. 

Overall, there does not appear to be a significant difference between the X-ray luminosities of luminous single O-type stars in the LMC and Galaxy, aside from higher temperatures/harder X-ray emission for single stars at lower metallicity. Table~\ref{MW_vs_LMC} also reinforces higher plasma temperatures and harder $\eta_{2}$ indices for WN stars in the Tarantula than their Carina counterparts.
 
 In Fig.~\ref{KM} we present an estimate
 of the cumulative distribution function (CDF) for  $\log L_{\rm X}/L_{\rm Bol}$, for three samples of luminous O stars:  the T-ReX sample presented here (224 stars, comprising 66 detections from Table~\ref{A2}, BI~253~from Table~\ref{Andy}, plus 157 attenuation-corrected upper limits from Table~S1), the Carina sample (41 stars from \citet{2011ApJS..194....5G}, no upper limits, excluding  the X-ray sub-luminous supergiant HDE~305619) and the combined T-ReX-Carina sample (263 stars). Since these samples contain upper limits, the CDF is calculated using the Kaplan-Meier estimator. 
 
We compare the Tarantula and Carina CDFs using the common method--statistical hypothesis testing, which computes the probability (``p-value'') that the Tarantula and Carina samples would produce CDFs less similar than we observed, under the null hypothesis that Tarantula and Carina have the same $L_{\rm X}/L_{\rm Bol}$ relationship. The measure similarity of our CDFs, we used three two-sample statistical tests (Gehan-Breslow, Prentice, Peto) that are designed for Kaplan-Meier distributions \citep[][see p.173]{1996asst.book.....B}. Since these three hypothesis tests produced large p-values (0.682, 0.831, and 0.827, respectively), we find that these data do not provide enough evidence to support the alternative hypothesis -- that Tarantula and Carina have different $L_{\rm X}/L_{\rm Bol}$ relationships.

Theoretically \citet{1999ApJ...520..833O} have argued  $L_{\rm X} \sim (\dot{M}/v_{\infty})^2$ for optically thin winds, so for $\dot{M} \propto Z^{0.83}$ \citep{2007A&A...473..603M} and  $v_{\infty} \propto Z^{0.13}$ \citep{1992ApJ...401..596L}, one would expect $L_{\rm X} \propto (Z^{0.7})^{2}$ i.e.
$\log L_{\rm X} \propto 1.4 \log Z$, so X-ray luminosities of LMC O stars would be expected to be 0.4 dex lower than Galactic counterparts owing to a half-solar metallicity (Table~\ref{abundances}). Therefore, contrary to  \citet{1999ApJ...520..833O}, we find no evidence supporting lower X-ray luminosities in LMC O stars than Galactic counterparts. \citet{2013MNRAS.429.3379O}  provided an alternative parameterisation in which $L_{\rm X} \propto \dot{M}^{1-m} \propto L_{\rm Bol}^{(1-m)/\alpha^{\prime}}$, where a mixing exponent $m = 1 - \alpha^{\prime} \sim 0.4$ is required to reproduce the empirical Galactic $L_{\rm X} \sim 10^{-7} L_{\rm Bol}$ relationship. This scaling suggests $L_{\rm X} \propto (Z^{0.83})^{0.6} $ or
$\log L_{\rm X} \propto 0.5 \log Z$ based on the empirical mass-loss dependence on $Z$ \citep{2007A&A...473..603M}, such that LMC O stars are expected to be offset by only 0.15 dex from Milky Way counterparts, which is much more challenging to observationally verify. \citet{2013MNRAS.429.3379O} also propose that X-ray luminosities should decrease for optically thick winds with very high mass-loss rates. Since we find no systematic difference between $L_{\rm X}/L_{\rm Bol}$ for single O, Of/WN and WR stars, this result does challenge predictions given our sample includes very luminous stars with dense stellar winds.


\begin{figure*}
\begin{center}
\includegraphics[width=0.33\linewidth]{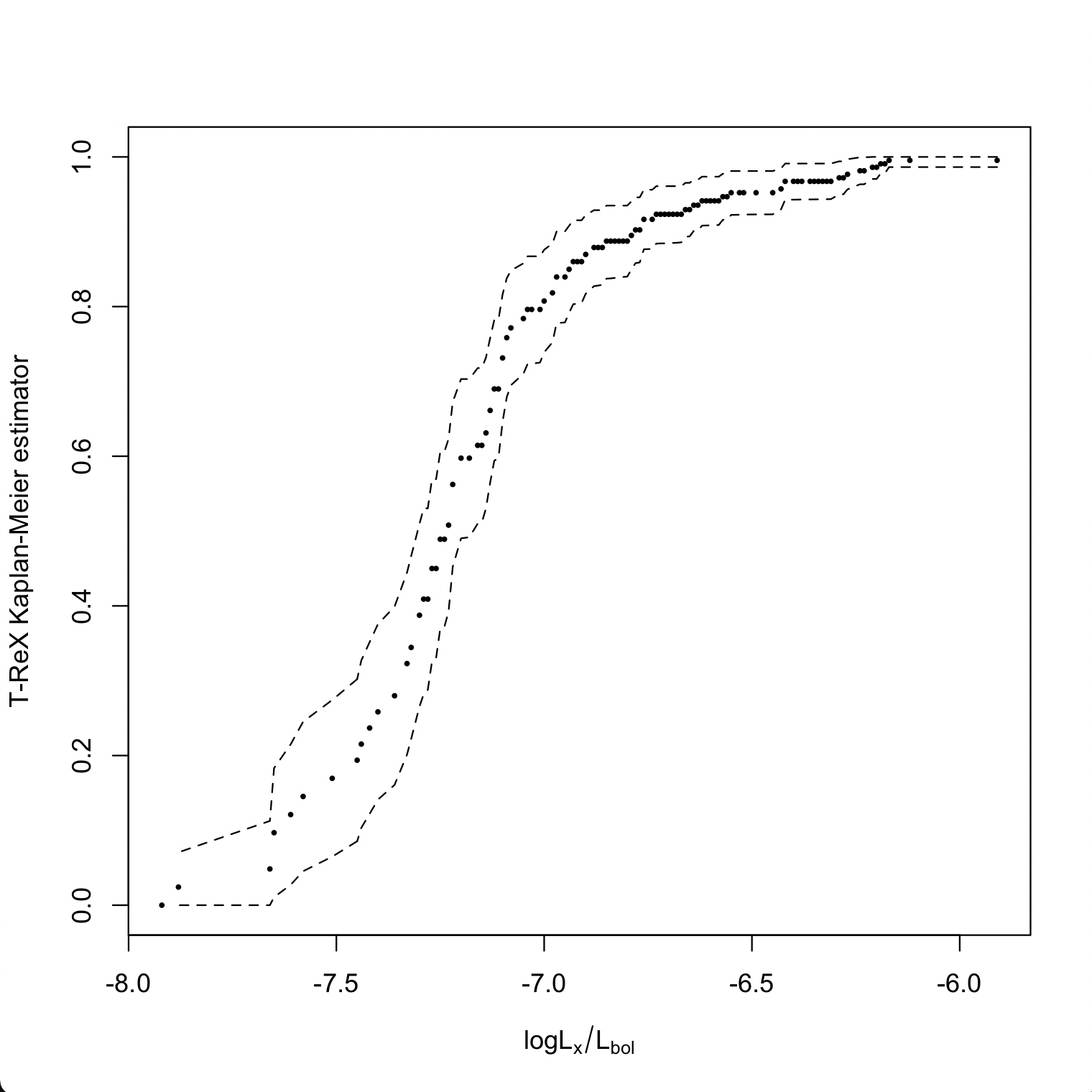}
\includegraphics[width=0.33\linewidth]{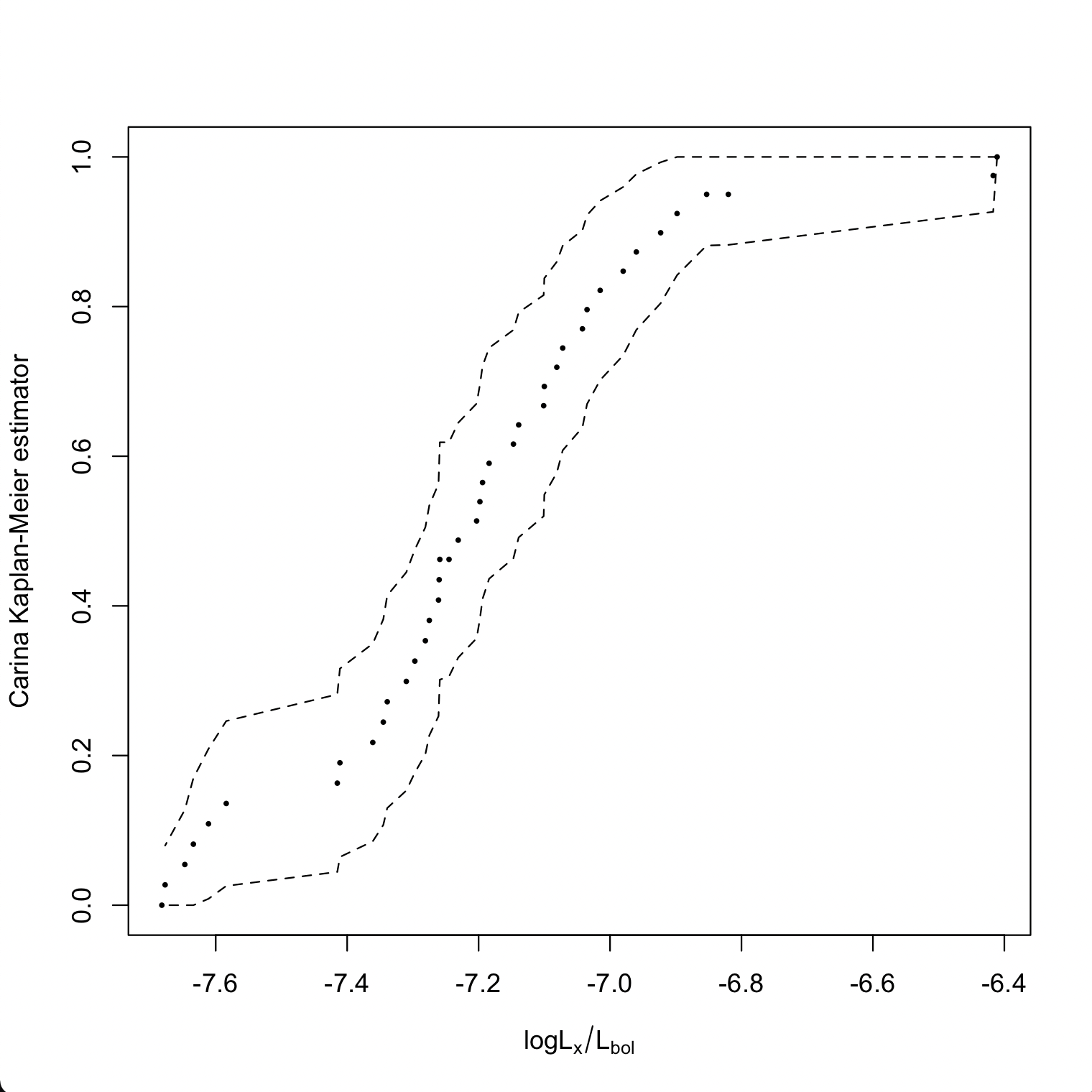}
\includegraphics[width=0.33\linewidth]{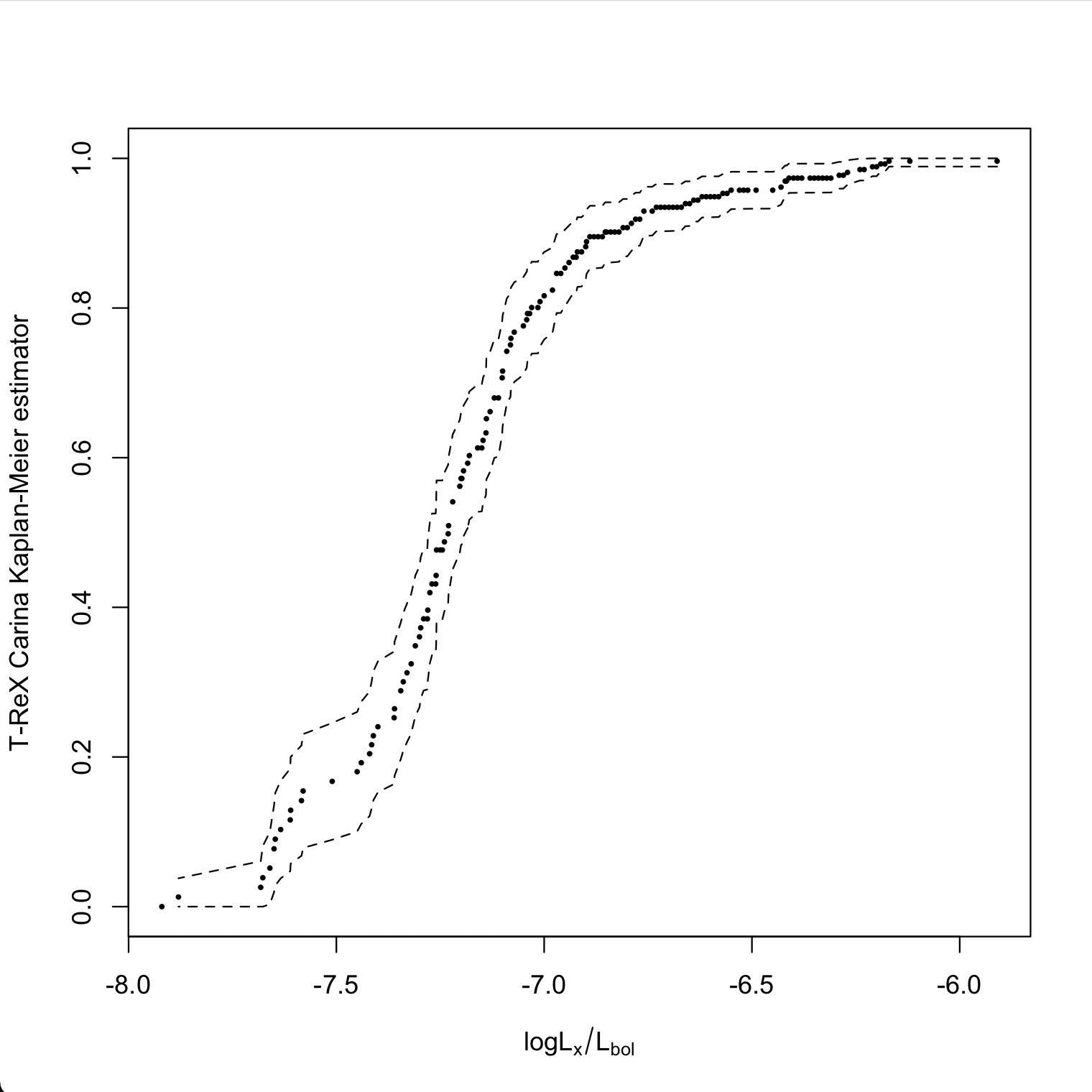}
  \caption{({\it left panel}): Cumulative distribution function (CDF) of $\log L_{\rm X}/L_{\rm Bol}$ for luminous O stars in the T-ReX sample (N=224, 67 detections ncluding BI~253 plus 157 upper limits), with 95\% confidence limits, calculated using the Kaplan-Meier estimator since sample includes upper limits; ({\it centre panel}): CDF for luminous O stars in Carina \citep[N=41,][ excl. HDE~305619]{2011ApJS..194....7N, 2011ApJS..194....5G}; ({\it right panel}): CDF for the joint T-ReX and Carina O star samples (N=265).} 
	\label{KM}
	\end{center}
\end{figure*}

In summary, the null hypothesis is that the T-ReX and Carina $\log L_{\rm X}/L_{\rm Bol}$ values for luminous O stars are drawn from the same parent population. Since these three hypothesis tests produce relatively large p-values, we conclude that there is {\it no} evidence in these data that the Carina and T-ReX samples are drawn from different underlying $\log L_{\rm X}/L_{\rm Bol}$ distributions. This result confirms a similar result for the LMC star forming region N11 \citep{2014ApJS..213...23N}. Further work on low-Z O star wind acceleration may be needed to explain this result.


\section{Summary} \label{summary}

From a comparison between the rich X-ray point source catalogue from T-ReX (P.~Broos \& L.~Townsley, in prep.) to optical photometric and spectroscopic surveys of the Tarantula Nebula we have obtained the X-ray properties
of a rich population of luminous early-type stars at a sub-solar metallicity for the first time, building on previous analyses of shallow X-ray observations \citep{2002ApJ...574..762P, Leisa06, 2014ApJS..213....1T}. Overall, excluding extreme outliers and multiple systems we find  $\log L_{\rm X} /L_{\rm Bol} = -7.00 \pm 0.49$, in good agreement with studies of Galactic early-type stars.

Armed with extensive binary properties of early-type stars in the Tarantula from VFTS \citep{2013A&A...550A.107S, 2014A&A...564A..40W}, TMBM \citep[][Shenar et al, in prep]{2017A&A...598A..84A} and BBC \citep{2021MNRAS.507.5348V}
we have been able to separate T-ReX sources into single, SB1 and SB2. In general, we confirm previous X-ray studies of Galactic OB stars that binaries possess similar X-ray properties to single stars \citep{2005MNRAS.361..679O}.
Exceptional X-ray properties are observed in some systems (Mk\,34, R140a and VFTS 399) and some WR systems are unusually hard X-ray emitters (R130, R134, R135, Mk~53). Indeed, the hardness index increases from O dwarfs, through O (super)giants, Of/WN stars and WN stars, in common with recent Galactic results for OB stars \citep{2018A&A...620A..89N}. 

We find no statistical evidence supporting reduced X-ray luminosities for O stars in the LMC than high luminosity counterparts in the Carina Nebula \citep{2011ApJS..194....7N, 2011ApJS..194....5G}, folding in upper X-ray limits to hundreds of luminous early-type
stars undetected by T-ReX. Consistent results for O, Of/WN and WN stars challenges predictions for X-ray emission from single stars at high luminosity with strong winds \citep{2013MNRAS.429.3379O}. 

We advocate $L_{\rm X} \sim 10^{-7} L_{\rm Bol}$ and (median) $kT_m \sim 0.8$ keV for UV spectroscopic analysis of single LMC O stars from the ULLYSES\footnote{https://ullyses.stsci.edu/} programme since it is well established that X-ray production influences the predicted strength of high ionization metallic resonance lines \citep{1980ApJ...241..300L, 1994A&A...283..525P, 2002ApJ...579..774C}. Further analysis is deferred to future T-ReX studies, including a detailed discussion of X-ray properties of WR and Of/WN stars (K.~Tehrani et al, in prep).

\section*{Acknowledgements}

We would like to thank Lidia Oskinova and the anonymous referee for useful comments which helped to clarify a number of topics in the submitted manuscript, and Tomer Shenar for sharing TMBM results prior to publication.

This work was supported by the {\it Chandra X-ray Observatory} General Observer grants GO5-6080X (PI: L. Townsley) and by GO4-15131X (PI: L. Townsley) and by the Penn State ACIS Instrument Team Contract SV4-74108. All of these were issued by the {\it Chandra} X-ray Centre, which is operated by the Smithsonian Astrophysical Observatory for and on behalf of NASA under contract NAS8-03060. PAC is supported by the Science and Technology Facilities Council research grant ST/V000853/1 (PI. V. Dhillon). We would like to thank Leigh Parrott's contribution to the statistical tests comparing T-ReX and Carina samples. 

This research has made extensive use of NASA's Astrophysics Data System Bibliographic Services, and the SIMBAD database, operated at CDS, Strasbourg, France. For the purpose of open access, the author has applied a Creative Commons Attributation (CC BY) license to any Author Accepted Manuscript version arising.

\section*{Data Availability} 

T-ReX data products are archived in Zenodo collection 10.5281/zenodo.6808367. For every T-ReX point source in this paper we include spectral fitting products and a region file, together with a FITS table and a separate FITS table for upper-limits summarised in the supplementary data. The full T-ReX point source catalog will be added to the Zenodo collection upon publication. Individual T-ReX datasets may also be accessed via the Chandra Data Archive. 

\bibliographystyle{mnras}
\bibliography{LxLbol} 

\begin{thebibliography}{}
\makeatletter
\relax
\def\mn@urlcharsother{\let\do\@makeother \do\$\do\&\do\#\do\^\do\_\do\%\do\~}
\def\mn@doi{\begingroup\mn@urlcharsother \@ifnextchar [ {\mn@doi@}
  {\mn@doi@[]}}
\def\mn@doi@[#1]#2{\def\@tempa{#1}\ifx\@tempa\@empty \href
  {http://dx.doi.org/#2} {doi:#2}\else \href {http://dx.doi.org/#2} {#1}\fi
  \endgroup}
\def\mn@eprint#1#2{\mn@eprint@#1:#2::\@nil}
\def\mn@eprint@arXiv#1{\href {http://arxiv.org/abs/#1} {{\tt arXiv:#1}}}
\def\mn@eprint@dblp#1{\href {http://dblp.uni-trier.de/rec/bibtex/#1.xml}
  {dblp:#1}}
\def\mn@eprint@#1:#2:#3:#4\@nil{\def\@tempa {#1}\def\@tempb {#2}\def\@tempc
  {#3}\ifx \@tempc \@empty \let \@tempc \@tempb \let \@tempb \@tempa \fi \ifx
  \@tempb \@empty \def\@tempb {arXiv}\fi \@ifundefined
  {mn@eprint@\@tempb}{\@tempb:\@tempc}{\expandafter \expandafter \csname
  mn@eprint@\@tempb\endcsname \expandafter{\@tempc}}}

\bibitem[\protect\citeauthoryear{{Almeida} et~al.,}{{Almeida}
  et~al.}{2017}]{2017A&A...598A..84A}
{Almeida} L.~A.,  et~al., 2017, \mn@doi [\aap] {10.1051/0004-6361/201629844},
  \href {https://ui.adsabs.harvard.edu/abs/2017A&A...598A..84A} {598, A84}

\bibitem[\protect\citeauthoryear{{Andrievsky}, {Kovtyukh}, {Korotin}, {Spite}
  \& {Spite}}{{Andrievsky} et~al.}{2001}]{2001A&A...367..605A}
{Andrievsky} S.~M.,  {Kovtyukh} V.~V.,  {Korotin} S.~A.,  {Spite} M.,   {Spite}
  F.,  2001, \mn@doi [\aap] {10.1051/0004-6361:20000407}, \href
  {https://ui.adsabs.harvard.edu/abs/2001A&A...367..605A} {367, 605}

\bibitem[\protect\citeauthoryear{{Arnaud}}{{Arnaud}}{1996}]{1996ASPC..101...17A}
{Arnaud} K.~A.,  1996, in {Jacoby} G.~H.,  {Barnes} J.,  eds,  ASP Conf. Ser. Vol. 101, Astronomical Data Analysis
  Software and Systems V. p.~17

\bibitem[\protect\citeauthoryear{{Asplund}, {Grevesse}, {Sauval}  \&
  {Scott}}{{Asplund} et~al.}{2009}]{2009ARA&A..47..481A}
{Asplund} M.,  {Grevesse} N.,  {Sauval} A.~J.,   {Scott} P.,  2009, \mn@doi
  [\araa] {10.1146/annurev.astro.46.060407.145222}, \href
  {https://ui.adsabs.harvard.edu/abs/2009ARA&A..47..481A} {47, 481}

\bibitem[\protect\citeauthoryear{{Babu} \& {Feigelson}}{{Babu} \&
  {Feigelson}}{1996}]{1996asst.book.....B}
{Babu} G.~J.,  {Feigelson} E.~D.,  1996, {Astrostatistics}

\bibitem[\protect\citeauthoryear{{Bagnulo} et~al.,}{{Bagnulo}
  et~al.}{2020}]{2020A&A...635A.163B}
{Bagnulo} S.,  et~al., 2020, \mn@doi [\aap] {10.1051/0004-6361/201937098},
  \href {https://ui.adsabs.harvard.edu/abs/2020A&A...635A.163B} {635, A163}

\bibitem[\protect\citeauthoryear{{Bartzakos}, {Moffat}  \&
  {Niemela}}{{Bartzakos} et~al.}{2001}]{2001MNRAS.324...18B}
{Bartzakos} P.,  {Moffat} A.~F.~J.,   {Niemela} V.~S.,  2001, \mn@doi [\mnras]
  {10.1046/j.1365-8711.2001.04126.x}, \href
  {https://ui.adsabs.harvard.edu/abs/2001MNRAS.324...18B} {324, 18}

\bibitem[\protect\citeauthoryear{{Bergh\"{o}fer}, {Schmitt}, {Danner}  \&
  {Cassinelli}}{{Bergh\"{o}fer} et~al.}{1997}]{1997A&A...322..167B}
{Bergh\"{o}fer} T.~W.,  {Schmitt} J.~H.~M.~M.,  {Danner} R.,   {Cassinelli}
  J.~P.,  1997, \aap, \href
  {https://ui.adsabs.harvard.edu/abs/1997A&A...322..167B} {322, 167}

\bibitem[\protect\citeauthoryear{{Bestenlehner} et~al.,}{{Bestenlehner}
  et~al.}{2011}]{2011A&A...530L..14B}
{Bestenlehner} J.~M.,  et~al., 2011, \mn@doi [\aap]
  {10.1051/0004-6361/201117043}, \href
  {https://ui.adsabs.harvard.edu/abs/2011A&A...530L..14B} {530, L14}

\bibitem[\protect\citeauthoryear{{Bestenlehner} et~al.,}{{Bestenlehner}
  et~al.}{2014}]{2014A&A...570A..38B}
{Bestenlehner} J.~M.,  et~al., 2014, \mn@doi [\aap]
  {10.1051/0004-6361/201423643}, \href
  {https://ui.adsabs.harvard.edu/abs/2014A&A...570A..38B} {570, A38}

\bibitem[\protect\citeauthoryear{{Bestenlehner} et~al.,}{{Bestenlehner}
  et~al.}{2020}]{2020MNRAS.499.1918B}
{Bestenlehner} J.~M.,  et~al., 2020, \mn@doi [\mnras] {10.1093/mnras/staa2801},
  \href {https://ui.adsabs.harvard.edu/abs/2020MNRAS.499.1918B} {499, 1918}

\bibitem[\protect\citeauthoryear{{Bestenlehner}, {Crowther}, {Broos}, {Pollock}
   \& {Townsley}}{{Bestenlehner} et~al.}{2022}]{2022MNRAS.510.6133B}
{Bestenlehner} J.~M.,  {Crowther} P.~A.,  {Broos} P.~S.,  {Pollock} A. M.~T.,
  {Townsley} L.~K.,  2022, \mn@doi [\mnras] {10.1093/mnras/stab3521}, \href
  {https://ui.adsabs.harvard.edu/abs/2022MNRAS.510.6133B} {510, 6133}

\bibitem[\protect\citeauthoryear{{Bosch}, {Terlevich}, {Melnick}  \&
  {Selman}}{{Bosch} et~al.}{1999}]{1999A&AS..137...21B}
{Bosch} G.,  {Terlevich} R.,  {Melnick} J.,   {Selman} F.,  1999, \mn@doi
  [\aaps] {10.1051/aas:1999480}, \href
  {https://ui.adsabs.harvard.edu/abs/1999A&AS..137...21B} {137, 21}

\bibitem[\protect\citeauthoryear{{Bosch}, {Terlevich}  \& {Terlevich}}{{Bosch}
  et~al.}{2009}]{2009AJ....137.3437B}
{Bosch} G.,  {Terlevich} E.,   {Terlevich} R.,  2009, \mn@doi [\aj]
  {10.1088/0004-6256/137/2/3437}, \href
  {https://ui.adsabs.harvard.edu/abs/2009AJ....137.3437B} {137, 3437}

\bibitem[\protect\citeauthoryear{{Brands} et~al.,}{{Brands}
  et~al.}{2022}]{2022arXiv220211080B}
{Brands} S.~A.,  et~al., 2022, arXiv e-prints, \href
  {https://ui.adsabs.harvard.edu/abs/2022arXiv220211080B} {p. arXiv:2202.11080}

\bibitem[\protect\citeauthoryear{{Breysacher}, {Azzopardi}  \&
  {Testor}}{{Breysacher} et~al.}{1999}]{1999A&AS..137..117B}
{Breysacher} J.,  {Azzopardi} M.,   {Testor} G.,  1999, \mn@doi [\aaps]
  {10.1051/aas:1999240}, \href
  {https://ui.adsabs.harvard.edu/abs/1999A&AS..137..117B} {137, 117}

\bibitem[\protect\citeauthoryear{{Broos}, {Townsley}, {Feigelson}, {Getman},
  {Bauer}  \& {Garmire}}{{Broos} et~al.}{2010}]{2010ApJ...714.1582B}
{Broos} P.~S.,  {Townsley} L.~K.,  {Feigelson} E.~D.,  {Getman} K.~V.,  {Bauer}
  F.~E.,   {Garmire} G.~P.,  2010, \mn@doi [\apj]
  {10.1088/0004-637X/714/2/1582}, \href
  {https://ui.adsabs.harvard.edu/abs/2010ApJ...714.1582B} {714, 1582}

\bibitem[\protect\citeauthoryear{{Broos} et~al.,}{{Broos}
  et~al.}{2011}]{2011ApJS..194....2B}
{Broos} P.~S.,  et~al., 2011, \mn@doi [\apjs] {10.1088/0067-0049/194/1/2},
  \href {https://ui.adsabs.harvard.edu/abs/2011ApJS..194....2B} {194, 2}

\bibitem[\protect\citeauthoryear{{Brunet}, {Imbert}, {Martin}, {Mianes},
  {Pr{\'e}vot}, {Rebeirot}  \& {Rousseau}}{{Brunet}
  et~al.}{1975}]{1975A&AS...21..109B}
{Brunet} J.~P.,  {Imbert} M.,  {Martin} N.,  {Mianes} P.,  {Pr{\'e}vot} L.,
  {Rebeirot} E.,   {Rousseau} J.,  1975, \aaps, \href
  {https://ui.adsabs.harvard.edu/abs/1975A&AS...21..109B} {21, 109}

\bibitem[\protect\citeauthoryear{{Cash}}{{Cash}}{1979}]{1979ApJ...228..939C}
{Cash} W.,  1979, \mn@doi [\apj] {10.1086/156922}, \href
  {https://ui.adsabs.harvard.edu/abs/1979ApJ...228..939C} {228, 939}

\bibitem[\protect\citeauthoryear{{Castro}, {Crowther}, {Evans}, {Mackey},
  {Castro-Rodriguez}, {Vink}, {Melnick}  \& {Selman}}{{Castro}
  et~al.}{2018}]{2018A&A...614A.147C}
{Castro} N.,  {Crowther} P.~A.,  {Evans} C.~J.,  {Mackey} J.,
  {Castro-Rodriguez} N.,  {Vink} J.~S.,  {Melnick} J.,   {Selman} F.,  2018,
  \mn@doi [\aap] {10.1051/0004-6361/201732084}, \href
  {https://ui.adsabs.harvard.edu/abs/2018A&A...614A.147C} {614, A147}

\bibitem[\protect\citeauthoryear{{Castro} et~al.,}{{Castro}
  et~al.}{2021}]{2021A&A...648A..65C}
{Castro} N.,  et~al., 2021, \mn@doi [\aap] {10.1051/0004-6361/202040008}, \href
  {https://ui.adsabs.harvard.edu/abs/2021A&A...648A..65C} {648, A65}

\bibitem[\protect\citeauthoryear{{Chlebowski} \& {Garmany}}{{Chlebowski} \&
  {Garmany}}{1991}]{1991ApJ...368..241C}
{Chlebowski} T.,  {Garmany} C.~D.,  1991, \mn@doi [\apj] {10.1086/169687},
  \href {https://ui.adsabs.harvard.edu/abs/1991ApJ...368..241C} {368, 241}

\bibitem[\protect\citeauthoryear{{Chlebowski}, {Harnden}  \&
  {Sciortino}}{{Chlebowski} et~al.}{1989}]{1989ApJ...341..427C}
{Chlebowski} T.,  {Harnden} F.~R. J.,   {Sciortino} S.,  1989, \mn@doi [\apj]
  {10.1086/167506}, \href
  {https://ui.adsabs.harvard.edu/abs/1989ApJ...341..427C} {341, 427}

\bibitem[\protect\citeauthoryear{{Clark} et~al.,}{{Clark}
  et~al.}{2015}]{2015A&A...579A.131C}
{Clark} J.~S.,  et~al., 2015, \mn@doi [\aap] {10.1051/0004-6361/201424427},
  \href {https://ui.adsabs.harvard.edu/abs/2015A&A...579A.131C} {579, A131}

\bibitem[\protect\citeauthoryear{{Clark}, {Ritchie}  \& {Negueruela}}{{Clark}
  et~al.}{2019}]{2019A&A...626A..59C}
{Clark} J.~S.,  {Ritchie} B.~W.,   {Negueruela} I.,  2019, \mn@doi [\aap]
  {10.1051/0004-6361/201935017}, \href
  {https://ui.adsabs.harvard.edu/abs/2019A&A...626A..59C} {626, A59}

\bibitem[\protect\citeauthoryear{{Cohen}, {Cassinelli}  \&
  {MacFarlane}}{{Cohen} et~al.}{1997}]{1997ApJ...487..867C}
{Cohen} D.~H.,  {Cassinelli} J.~P.,   {MacFarlane} J.~J.,  1997, \mn@doi [\apj]
  {10.1086/304636}, \href
  {https://ui.adsabs.harvard.edu/abs/1997ApJ...487..867C} {487, 867}

\bibitem[\protect\citeauthoryear{{Conti} \& {Massey}}{{Conti} \&
  {Massey}}{1989}]{1989ApJ...337..251C}
{Conti} P.~S.,  {Massey} P.,  1989, \mn@doi [\apj] {10.1086/167101}, \href
  {https://ui.adsabs.harvard.edu/abs/1989ApJ...337..251C} {337, 251}

\bibitem[\protect\citeauthoryear{{Crowther}}{{Crowther}}{2019}]{2019Galax...7...88C}
{Crowther} P.~A.,  2019, \mn@doi [Galaxies] {10.3390/galaxies7040088}, \href
  {https://ui.adsabs.harvard.edu/abs/2019Galax...7...88C} {7, 88}

\bibitem[\protect\citeauthoryear{{Crowther} \& {Smith}}{{Crowther} \&
  {Smith}}{1997}]{1997A&A...320..500C}
{Crowther} P.~A.,  {Smith} L.~J.,  1997, \aap, \href
  {https://ui.adsabs.harvard.edu/abs/1997A&A...320..500C} {320, 500}

\bibitem[\protect\citeauthoryear{{Crowther} \& {Walborn}}{{Crowther} \&
  {Walborn}}{2011}]{2011MNRAS.416.1311C}
{Crowther} P.~A.,  {Walborn} N.~R.,  2011, \mn@doi [\mnras]
  {10.1111/j.1365-2966.2011.19129.x}, \href
  {https://ui.adsabs.harvard.edu/abs/2011MNRAS.416.1311C} {416, 1311}

\bibitem[\protect\citeauthoryear{{Crowther}, {Dessart}, {Hillier}, {Abbott}  \&
  {Fullerton}}{{Crowther} et~al.}{2002a}]{2002A&A...392..653C}
{Crowther} P.~A.,  {Dessart} L.,  {Hillier} D.~J.,  {Abbott} J.~B.,
  {Fullerton} A.~W.,  2002a, \mn@doi [\aap] {10.1051/0004-6361:20020941}, \href
  {https://ui.adsabs.harvard.edu/abs/2002A&A...392..653C} {392, 653}

\bibitem[\protect\citeauthoryear{{Crowther}, {Hillier}, {Evans}, {Fullerton},
  {De Marco}  \& {Willis}}{{Crowther} et~al.}{2002b}]{2002ApJ...579..774C}
{Crowther} P.~A.,  {Hillier} D.~J.,  {Evans} C.~J.,  {Fullerton} A.~W.,  {De
  Marco} O.,   {Willis} A.~J.,  2002b, \mn@doi [\apj] {10.1086/342877}, \href
  {https://ui.adsabs.harvard.edu/abs/2002ApJ...579..774C} {579, 774}

\bibitem[\protect\citeauthoryear{{Crowther}, {Schnurr}, {Hirschi}, {Yusof},
  {Parker}, {Goodwin}  \& {Kassim}}{{Crowther}
  et~al.}{2010}]{2010MNRAS.408..731C}
{Crowther} P.~A.,  {Schnurr} O.,  {Hirschi} R.,  {Yusof} N.,  {Parker} R.~J.,
  {Goodwin} S.~P.,   {Kassim} H.~A.,  2010, \mn@doi [\mnras]
  {10.1111/j.1365-2966.2010.17167.x}, \href
  {https://ui.adsabs.harvard.edu/abs/2010MNRAS.408..731C} {408, 731}

\bibitem[\protect\citeauthoryear{{Crowther} et~al.,}{{Crowther}
  et~al.}{2016}]{2016MNRAS.458..624C}
{Crowther} P.~A.,  et~al., 2016, \mn@doi [\mnras] {10.1093/mnras/stw273}, \href
  {https://ui.adsabs.harvard.edu/abs/2016MNRAS.458..624C} {458, 624}
  
\bibitem[\protect\citeauthoryear{{de Boer}, {Koornneef}  \& {Savage}}{{de Boer}
  et~al.}{1980}]{1980ApJ...236..769D}
{de Boer} K.~S.,  {Koornneef} J.,   {Savage} B.~D.,  1980, \mn@doi [\apj]
  {10.1086/157802}, \href
  {https://ui.adsabs.harvard.edu/abs/1980ApJ...236..769D} {236, 769}

\bibitem[\protect\citeauthoryear{{De Marchi} et~al.,}{{De Marchi}
  et~al.}{2011}]{2011ApJ...739...27D}
{De Marchi} G.,  et~al., 2011, \mn@doi [\apj] {10.1088/0004-637X/739/1/27},
  \href {https://ui.adsabs.harvard.edu/abs/2011ApJ...739...27D} {739, 27}

\bibitem[\protect\citeauthoryear{{Dopita}, {Seitenzahl}, {Sutherland},
  {Nicholls}, {Vogt}, {Ghavamian}  \& {Ruiter}}{{Dopita}
  et~al.}{2019}]{2019AJ....157...50D}
{Dopita} M.~A.,  {Seitenzahl} I.~R.,  {Sutherland} R.~S.,  {Nicholls} D.~C.,
  {Vogt} F. P.~A.,  {Ghavamian} P.,   {Ruiter} A.~J.,  2019, \mn@doi [\aj]
  {10.3847/1538-3881/aaf235}, \href
  {https://ui.adsabs.harvard.edu/abs/2019AJ....157...50D} {157, 50}

\bibitem[\protect\citeauthoryear{{Doran} et~al.,}{{Doran}
  et~al.}{2013}]{2013A&A...558A.134D}
{Doran} E.~I.,  et~al., 2013, \mn@doi [\aap] {10.1051/0004-6361/201321824},
  \href {https://ui.adsabs.harvard.edu/abs/2013A&A...558A.134D} {558, A134}

\bibitem[\protect\citeauthoryear{{Dufton} et~al.,}{{Dufton}
  et~al.}{2018}]{2018A&A...615A.101D}
{Dufton} P.~L.,  et~al., 2018, \mn@doi [\aap] {10.1051/0004-6361/201732440},
  \href {https://ui.adsabs.harvard.edu/abs/2018A&A...615A.101D} {615, A101}

\bibitem[\protect\citeauthoryear{{Dunstall} et~al.,}{{Dunstall}
  et~al.}{2012}]{2012A&A...542A..50D}
{Dunstall} P.~R.,  et~al., 2012, \mn@doi [\aap] {10.1051/0004-6361/201218872},
  \href {https://ui.adsabs.harvard.edu/abs/2012A&A...542A..50D} {542, A50}

\bibitem[\protect\citeauthoryear{{Evans} et~al.,}{{Evans}
  et~al.}{2011}]{2011A&A...530A.108E}
{Evans} C.~J.,  et~al., 2011, \mn@doi [\aap] {10.1051/0004-6361/201116782},
  \href {https://ui.adsabs.harvard.edu/abs/2011A&A...530A.108E} {530, A108}

\bibitem[\protect\citeauthoryear{{Evans} et~al.,}{{Evans}
  et~al.}{2015}]{2015A&A...574A..13E}
{Evans} C.~J.,  et~al., 2015, \mn@doi [\aap] {10.1051/0004-6361/201424414},
  \href {https://ui.adsabs.harvard.edu/abs/2015A&A...574A..13E} {574, A13}

\bibitem[\protect\citeauthoryear{{Feast}, {Thackeray}  \& {Wesselink}}{{Feast}
  et~al.}{1960}]{1960MNRAS.121..337F}
{Feast} M.~W.,  {Thackeray} A.~D.,   {Wesselink} A.~J.,  1960, \mn@doi [\mnras]
  {10.1093/mnras/121.4.337}, \href
  {https://ui.adsabs.harvard.edu/abs/1960MNRAS.121..337F} {121, 337}

\bibitem[\protect\citeauthoryear{{Fitzpatrick} \& {Savage}}{{Fitzpatrick} \&
  {Savage}}{1984}]{1984ApJ...279..578F}
{Fitzpatrick} E.~L.,  {Savage} B.~D.,  1984, \mn@doi [\apj] {10.1086/161923},
  \href {https://ui.adsabs.harvard.edu/abs/1984ApJ...279..578F} {279, 578}

\bibitem[\protect\citeauthoryear{{Foellmi}, {Moffat}  \& {Guerrero}}{{Foellmi}
  et~al.}{2003}]{2003MNRAS.338.1025F}
{Foellmi} C.,  {Moffat} A.~F.~J.,   {Guerrero} M.~A.,  2003, \mn@doi [\mnras]
  {10.1046/j.1365-8711.2003.06161.x}, \href
  {https://ui.adsabs.harvard.edu/abs/2003MNRAS.338.1025F} {338, 1025}

\bibitem[\protect\citeauthoryear{{Gagn{\'e}} et~al.,}{{Gagn{\'e}}
  et~al.}{2011}]{2011ApJS..194....5G}
{Gagn{\'e}} M.,  et~al., 2011, \mn@doi [\apjs] {10.1088/0067-0049/194/1/5},
  \href {https://ui.adsabs.harvard.edu/abs/2011ApJS..194....5G} {194, 5}

\bibitem[\protect\citeauthoryear{{Gagn{\'e}}, {Fehon}, {Savoy}, {Cartagena},
  {Cohen}  \& {Owocki}}{{Gagn{\'e}} et~al.}{2012}]{2012ASPC..465..301G}
{Gagn{\'e}} M.,  {Fehon} G.,  {Savoy} M.~R.,  {Cartagena} C.~A.,  {Cohen}
  D.~H.,   {Owocki} S.~P.,  2012, in {Drissen} L.,  {Robert} C.,  {St-Louis}
  N.,   {Moffat} A.~F.~J.,  eds,  ASP Conf. Ser. Vol. 465, Proceedings of a Scientific Meeting in Honor of
  Anthony F. J. Moffat. p.~301 (\mn@eprint {arXiv} {1205.3510})

\bibitem[\protect\citeauthoryear{{Gamen} et~al.,}{{Gamen}
  et~al.}{2006}]{2006A&A...460..777G}
{Gamen} R.,  et~al., 2006, \mn@doi [\aap] {10.1051/0004-6361:20065618}, \href
  {https://ui.adsabs.harvard.edu/abs/2006A&A...460..777G} {460, 777}

\bibitem[\protect\citeauthoryear{{Garmire}, {Bautz}, {Ford}, {Nousek}  \&
  {Ricker}}{{Garmire} et~al.}{2003}]{2003SPIE.4851...28G}
{Garmire} G.~P.,  {Bautz} M.~W.,  {Ford} P.~G.,  {Nousek} J.~A.,   {Ricker}
  George~R. J.,  2003, in {Truemper} J.~E.,  {Tananbaum} H.~D.,  eds,  Society
  of Photo-Optical Instrumentation Engineers (SPIE) Conference Series Vol.
  4851, X-Ray and Gamma-Ray Telescopes and Instruments for Astronomy.. pp
  28--44, \mn@doi{10.1117/12.461599}

\bibitem[\protect\citeauthoryear{{Garnett}}{{Garnett}}{1999}]{1999IAUS..190..266G}
{Garnett} D.~R.,  1999, in {Chu} Y.~H.,  {Suntzeff} N.,  {Hesser} J.,
  {Bohlender} D.,  eds,  Vol. 190, New Views of the Magellanic Clouds. p.~266

\bibitem[\protect\citeauthoryear{{Gayley}}{{Gayley}}{2014}]{2014ApJ...788...90G}
{Gayley} K.~G.,  2014, \mn@doi [\apj] {10.1088/0004-637X/788/1/90}, \href
  {https://ui.adsabs.harvard.edu/abs/2014ApJ...788...90G} {788, 90}

\bibitem[\protect\citeauthoryear{{Graczyk} et~al.,}{{Graczyk}
  et~al.}{2011}]{2011AcA....61..103G}
{Graczyk} D.,  et~al., 2011, \actaa, \href
  {https://ui.adsabs.harvard.edu/abs/2011AcA....61..103G} {61, 103}

\bibitem[\protect\citeauthoryear{{Gruner} et~al.,}{{Gruner}
  et~al.}{2019}]{2019A&A...621A..63G}
{Gruner} D.,  et~al., 2019, \mn@doi [\aap] {10.1051/0004-6361/201833178}, \href
  {https://ui.adsabs.harvard.edu/abs/2019A&A...621A..63G} {621, A63}

\bibitem[\protect\citeauthoryear{{Grunhut} et~al.,}{{Grunhut}
  et~al.}{2017}]{2017MNRAS.465.2432G}
{Grunhut} J.~H.,  et~al., 2017, \mn@doi [\mnras] {10.1093/mnras/stw2743}, \href
  {https://ui.adsabs.harvard.edu/abs/2017MNRAS.465.2432G} {465, 2432}

\bibitem[\protect\citeauthoryear{{Guerrero} \& {Chu}}{{Guerrero} \&
  {Chu}}{2008}]{2008ApJS..177..216G}
{Guerrero} M.~A.,  {Chu} Y.-H.,  2008, \mn@doi [\apjs] {10.1086/587059}, \href
  {https://ui.adsabs.harvard.edu/abs/2008ApJS..177..216G} {177, 216}

\bibitem[\protect\citeauthoryear{{Hainich} et~al.,}{{Hainich}
  et~al.}{2014}]{2014A&A...565A..27H}
{Hainich} R.,  et~al., 2014, \mn@doi [\aap] {10.1051/0004-6361/201322696},
  \href {https://ui.adsabs.harvard.edu/abs/2014A&A...565A..27H} {565, A27}

\bibitem[\protect\citeauthoryear{{Hamann} et~al.,}{{Hamann}
  et~al.}{2019}]{2019A&A...625A..57H}
{Hamann} W.~R.,  et~al., 2019, \mn@doi [\aap] {10.1051/0004-6361/201834850},
  \href {https://ui.adsabs.harvard.edu/abs/2019A&A...625A..57H} {625, A57}

\bibitem[\protect\citeauthoryear{{Harnden} F.~R. et~al.,}{{Harnden}
  et~al.}{1979}]{1979ApJ...234L..51H}
{Harnden} F.~R. J.,  et~al., 1979, \mn@doi [\apjl] {10.1086/183107}, \href
  {https://ui.adsabs.harvard.edu/abs/1979ApJ...234L..51H} {234, L51}

\bibitem[\protect\citeauthoryear{{H{\'e}nault-Brunet}
  et~al.,}{{H{\'e}nault-Brunet} et~al.}{2012}]{2012A&A...546A..73H}
{H{\'e}nault-Brunet} V.,  et~al., 2012, \mn@doi [\aap]
  {10.1051/0004-6361/201219471}, \href
  {https://ui.adsabs.harvard.edu/abs/2012A&A...546A..73H} {546, A73}

\bibitem[\protect\citeauthoryear{{Hill}, {Andrievsky}  \& {Spite}}{{Hill}
  et~al.}{1995}]{1995A&A...293..347H}
{Hill} V.,  {Andrievsky} S.,   {Spite} M.,  1995, \aap, \href
  {https://ui.adsabs.harvard.edu/abs/1995A&A...293..347H} {293, 347}

\bibitem[\protect\citeauthoryear{{Huenemoerder}, {Schulz}  \&
  {Nichols}}{{Huenemoerder} et~al.}{2019}]{2019AJ....157...29H}
{Huenemoerder} D.~P.,  {Schulz} N.~S.,   {Nichols} J.~S.,  2019, \mn@doi [\aj]
  {10.3847/1538-3881/aaf380}, \href
  {https://ui.adsabs.harvard.edu/abs/2019AJ....157...29H} {157, 29}

\bibitem[\protect\citeauthoryear{{Hunter}, {Shaya}, {Holtzman}, {Light},
  {O'Neil}  \& {Lynds}}{{Hunter} et~al.}{1995}]{1995ApJ...448..179H}
{Hunter} D.~A.,  {Shaya} E.~J.,  {Holtzman} J.~A.,  {Light} R.~M.,  {O'Neil}
  Earl~J. J.,   {Lynds} R.,  1995, \mn@doi [\apj] {10.1086/175950}, \href
  {https://ui.adsabs.harvard.edu/abs/1995ApJ...448..179H} {448, 179}

\bibitem[\protect\citeauthoryear{{Hunter} et~al.,}{{Hunter}
  et~al.}{2007}]{2007A&A...466..277H}
{Hunter} I.,  et~al., 2007, \mn@doi [\aap] {10.1051/0004-6361:20066148}, \href
  {https://ui.adsabs.harvard.edu/abs/2007A&A...466..277H} {466, 277}

\bibitem[\protect\citeauthoryear{{Khorrami} et~al.,}{{Khorrami}
  et~al.}{2021}]{2021MNRAS.503..292K}
{Khorrami} Z.,  et~al., 2021, \mn@doi [\mnras] {10.1093/mnras/stab388}, \href
  {https://ui.adsabs.harvard.edu/abs/2021MNRAS.503..292K} {503, 292}

\bibitem[\protect\citeauthoryear{{Koornneef}}{{Koornneef}}{1982}]{1982A&A...107..247K}
{Koornneef} J.,  1982, \aap, \href
  {https://ui.adsabs.harvard.edu/abs/1982A&A...107..247K} {107, 247}

\bibitem[\protect\citeauthoryear{{Korn}, {Becker}, {Gummersbach}  \&
  {Wolf}}{{Korn} et~al.}{2000}]{2000A&A...353..655K}
{Korn} A.~J.,  {Becker} S.~R.,  {Gummersbach} C.~A.,   {Wolf} B.,  2000, \aap,
  \href {https://ui.adsabs.harvard.edu/abs/2000A&A...353..655K} {353, 655}

\bibitem[\protect\citeauthoryear{{Lebouteiller}, {Bernard-Salas}, {Brandl},
  {Whelan}, {Wu}, {Charmandaris}, {Devost}  \& {Houck}}{{Lebouteiller}
  et~al.}{2008}]{2008ApJ...680..398L}
{Lebouteiller} V.,  {Bernard-Salas} J.,  {Brandl} B.,  {Whelan} D.~G.,  {Wu}
  Y.,  {Charmandaris} V.,  {Devost} D.,   {Houck} J.~R.,  2008, \mn@doi [\apj]
  {10.1086/587503}, \href
  {https://ui.adsabs.harvard.edu/abs/2008ApJ...680..398L} {680, 398}

\bibitem[\protect\citeauthoryear{{Leitherer}, {Robert}  \&
  {Drissen}}{{Leitherer} et~al.}{1992}]{1992ApJ...401..596L}
{Leitherer} C.,  {Robert} C.,   {Drissen} L.,  1992, \mn@doi [\apj]
  {10.1086/172089}, \href
  {https://ui.adsabs.harvard.edu/abs/1992ApJ...401..596L} {401, 596}

\bibitem[\protect\citeauthoryear{{Long} \& {White}}{{Long} \&
  {White}}{1980}]{1980ApJ...239L..65L}
{Long} K.~S.,  {White} R.~L.,  1980, \mn@doi [\apjl] {10.1086/183293}, \href
  {https://ui.adsabs.harvard.edu/abs/1980ApJ...239L..65L} {239, L65}

\bibitem[\protect\citeauthoryear{{Lucy} \& {Solomon}}{{Lucy} \&
  {Solomon}}{1970}]{1970ApJ...159..879L}
{Lucy} L.~B.,  {Solomon} P.~M.,  1970, \mn@doi [\apj] {10.1086/150365}, \href
  {https://ui.adsabs.harvard.edu/abs/1970ApJ...159..879L} {159, 879}

\bibitem[\protect\citeauthoryear{{Lucy} \& {White}}{{Lucy} \&
  {White}}{1980}]{1980ApJ...241..300L}
{Lucy} L.~B.,  {White} R.~L.,  1980, \mn@doi [\apj] {10.1086/158342}, \href
  {https://ui.adsabs.harvard.edu/abs/1980ApJ...241..300L} {241, 300}

\bibitem[\protect\citeauthoryear{{Mahy} et~al.,}{{Mahy}
  et~al.}{2020}]{2020A&A...634A.118M}
{Mahy} L.,  et~al., 2020, \mn@doi [\aap] {10.1051/0004-6361/201936151}, \href
  {https://ui.adsabs.harvard.edu/abs/2020A&A...634A.118M} {634, A118}

\bibitem[\protect\citeauthoryear{{Malumuth} \& {Heap}}{{Malumuth} \&
  {Heap}}{1994}]{1994AJ....107.1054M}
{Malumuth} E.~M.,  {Heap} S.~R.,  1994, \mn@doi [\aj] {10.1086/116917}, \href
  {https://ui.adsabs.harvard.edu/abs/1994AJ....107.1054M} {107, 1054}

\bibitem[\protect\citeauthoryear{{Massey} \& {Hunter}}{{Massey} \&
  {Hunter}}{1998}]{1998ApJ...493..180M}
{Massey} P.,  {Hunter} D.~A.,  1998, \mn@doi [\apj] {10.1086/305126}, \href
  {https://ui.adsabs.harvard.edu/abs/1998ApJ...493..180M} {493, 180}

\bibitem[\protect\citeauthoryear{{Massey}, {Penny}  \& {Vukovich}}{{Massey}
  et~al.}{2002}]{2002ApJ...565..982M}
{Massey} P.,  {Penny} L.~R.,   {Vukovich} J.,  2002, \mn@doi [\apj]
  {10.1086/324783}, \href
  {https://ui.adsabs.harvard.edu/abs/2002ApJ...565..982M} {565, 982}

\bibitem[\protect\citeauthoryear{{Massey}, {Puls}, {Pauldrach}, {Bresolin},
  {Kudritzki}  \& {Simon}}{{Massey} et~al.}{2005}]{2005ApJ...627..477M}
{Massey} P.,  {Puls} J.,  {Pauldrach} A.~W.~A.,  {Bresolin} F.,  {Kudritzki}
  R.~P.,   {Simon} T.,  2005, \mn@doi [\apj] {10.1086/430417}, \href
  {https://ui.adsabs.harvard.edu/abs/2005ApJ...627..477M} {627, 477}

\bibitem[\protect\citeauthoryear{{Massey}, {Morrell}, {Neugent}, {Penny},
  {DeGioia-Eastwood}  \& {Gies}}{{Massey} et~al.}{2012}]{2012ApJ...748...96M}
{Massey} P.,  {Morrell} N.~I.,  {Neugent} K.~F.,  {Penny} L.~R.,
  {DeGioia-Eastwood} K.,   {Gies} D.~R.,  2012, \mn@doi [\apj]
  {10.1088/0004-637X/748/2/96}, \href
  {https://ui.adsabs.harvard.edu/abs/2012ApJ...748...96M} {748, 96}

\bibitem[\protect\citeauthoryear{{McEvoy} et~al.,}{{McEvoy}
  et~al.}{2015}]{2015A&A...575A..70M}
{McEvoy} C.~M.,  et~al., 2015, \mn@doi [\aap] {10.1051/0004-6361/201425202},
  \href {https://ui.adsabs.harvard.edu/abs/2015A&A...575A..70M} {575, A70}

\bibitem[\protect\citeauthoryear{{Melnick}}{{Melnick}}{1985}]{1985A&A...153..235M}
{Melnick} J.,  1985, \aap, \href
  {https://ui.adsabs.harvard.edu/abs/1985A&A...153..235M} {153, 235}

\bibitem[\protect\citeauthoryear{{Moffat}, {Niemela}, {Phillips}, {Chu}  \&
  {Seggewiss}}{{Moffat} et~al.}{1987}]{1987ApJ...312..612M}
{Moffat} A. F.~J.,  {Niemela} V.~S.,  {Phillips} M.~M.,  {Chu} Y.-H.,
  {Seggewiss} W.,  1987, \mn@doi [\apj] {10.1086/164906}, \href
  {https://ui.adsabs.harvard.edu/abs/1987ApJ...312..612M} {312, 612}

\bibitem[\protect\citeauthoryear{{Moffat}, {Drissen}  \& {Shara}}{{Moffat}
  et~al.}{1994}]{1994ApJ...436..183M}
{Moffat} A. F.~J.,  {Drissen} L.,   {Shara} M.~M.,  1994, \mn@doi [\apj]
  {10.1086/174891}, \href
  {https://ui.adsabs.harvard.edu/abs/1994ApJ...436..183M} {436, 183}

\bibitem[\protect\citeauthoryear{{Moffat} et~al.,}{{Moffat}
  et~al.}{2002}]{2002ApJ...573..191M}
{Moffat} A.~F.~J.,  et~al., 2002, \mn@doi [\apj] {10.1086/340491}, \href
  {https://ui.adsabs.harvard.edu/abs/2002ApJ...573..191M} {573, 191}

\bibitem[\protect\citeauthoryear{{Mokiem} et~al.,}{{Mokiem}
  et~al.}{2007}]{2007A&A...473..603M}
{Mokiem} M.~R.,  et~al., 2007, \mn@doi [\aap] {10.1051/0004-6361:20077545},
  \href {https://ui.adsabs.harvard.edu/abs/2007A&A...473..603M} {473, 603}

\bibitem[\protect\citeauthoryear{{Naz{\'e}}}{{Naz{\'e}}}{2009}]{2009A&A...506.1055N}
{Naz{\'e}} Y.,  2009, \mn@doi [\aap] {10.1051/0004-6361/200912659}, \href
  {https://ui.adsabs.harvard.edu/abs/2009A&A...506.1055N} {506, 1055}

\bibitem[\protect\citeauthoryear{{Naz{\'e}}, {Hartwell}, {Stevens}, {Manfroid},
  {Marchenko}, {Corcoran}, {Moffat}  \& {Skalkowski}}{{Naz{\'e}}
  et~al.}{2003}]{2003ApJ...586..983N}
{Naz{\'e}} Y.,  {Hartwell} J.~M.,  {Stevens} I.~R.,  {Manfroid} J.,
  {Marchenko} S.,  {Corcoran} M.~F.,  {Moffat} A.~F.~J.,   {Skalkowski} G.,
  2003, \mn@doi [\apj] {10.1086/367831}, \href
  {https://ui.adsabs.harvard.edu/abs/2003ApJ...586..983N} {586, 983}

\bibitem[\protect\citeauthoryear{{Naz{\'e}} et~al.,}{{Naz{\'e}}
  et~al.}{2011}]{2011ApJS..194....7N}
{Naz{\'e}} Y.,  et~al., 2011, \mn@doi [\apjs] {10.1088/0067-0049/194/1/7},
  \href {https://ui.adsabs.harvard.edu/abs/2011ApJS..194....7N} {194, 7}

\bibitem[\protect\citeauthoryear{{Naz{\'e}}, {Wang}, {Chu}, {Gruendl}  \&
  {Oskinova}}{{Naz{\'e}} et~al.}{2014}]{2014ApJS..213...23N}
{Naz{\'e}} Y.,  {Wang} Q.~D.,  {Chu} Y.-H.,  {Gruendl} R.,   {Oskinova} L.,
  2014, \mn@doi [\apjs] {10.1088/0067-0049/213/2/23}, \href
  {https://ui.adsabs.harvard.edu/abs/2014ApJS..213...23N} {213, 23}

\bibitem[\protect\citeauthoryear{{Nebot G{\'o}mez-Mor{\'a}n} \&
  {Oskinova}}{{Nebot G{\'o}mez-Mor{\'a}n} \&
  {Oskinova}}{2018}]{2018A&A...620A..89N}
{Nebot G{\'o}mez-Mor{\'a}n} A.,  {Oskinova} L.~M.,  2018, \mn@doi [\aap]
  {10.1051/0004-6361/201833453}, \href
  {https://ui.adsabs.harvard.edu/abs/2018A&A...620A..89N} {620, A89}

\bibitem[\protect\citeauthoryear{{Oskinova}}{{Oskinova}}{2005}]{2005MNRAS.361..679O}
{Oskinova} L.~M.,  2005, \mn@doi [\mnras] {10.1111/j.1365-2966.2005.09229.x},
  \href {https://ui.adsabs.harvard.edu/abs/2005MNRAS.361..679O} {361, 679}

\bibitem[\protect\citeauthoryear{{Oskinova}, {Ignace}, {Hamann}, {Pollock}  \&
  {Brown}}{{Oskinova} et~al.}{2003}]{2003A&A...402..755O}
{Oskinova} L.~M.,  {Ignace} R.,  {Hamann} W.~R.,  {Pollock} A.~M.~T.,   {Brown}
  J.~C.,  2003, \mn@doi [\aap] {10.1051/0004-6361:20030300}, \href
  {https://ui.adsabs.harvard.edu/abs/2003A&A...402..755O} {402, 755}

\bibitem[\protect\citeauthoryear{{Owocki} \& {Cohen}}{{Owocki} \&
  {Cohen}}{1999}]{1999ApJ...520..833O}
{Owocki} S.~P.,  {Cohen} D.~H.,  1999, \mn@doi [\apj] {10.1086/307500}, \href
  {https://ui.adsabs.harvard.edu/abs/1999ApJ...520..833O} {520, 833}

\bibitem[\protect\citeauthoryear{{Owocki}, {Castor}  \& {Rybicki}}{{Owocki}
  et~al.}{1988}]{1988ApJ...335..914O}
{Owocki} S.~P.,  {Castor} J.~I.,   {Rybicki} G.~B.,  1988, \mn@doi [\apj]
  {10.1086/166977}, \href
  {https://ui.adsabs.harvard.edu/abs/1988ApJ...335..914O} {335, 914}

\bibitem[\protect\citeauthoryear{{Owocki}, {Sundqvist}, {Cohen}  \&
  {Gayley}}{{Owocki} et~al.}{2013}]{2013MNRAS.429.3379O}
{Owocki} S.~P.,  {Sundqvist} J.~O.,  {Cohen} D.~H.,   {Gayley} K.~G.,  2013,
  \mn@doi [\mnras] {10.1093/mnras/sts599}, \href
  {https://ui.adsabs.harvard.edu/abs/2013MNRAS.429.3379O} {429, 3379}

\bibitem[\protect\citeauthoryear{{Pallavicini}, {Golub}, {Rosner}, {Vaiana},
  {Ayres}  \& {Linsky}}{{Pallavicini} et~al.}{1981}]{1981ApJ...248..279P}
{Pallavicini} R.,  {Golub} L.,  {Rosner} R.,  {Vaiana} G.~S.,  {Ayres} T.,
  {Linsky} J.~L.,  1981, \mn@doi [\apj] {10.1086/159152}, \href
  {https://ui.adsabs.harvard.edu/abs/1981ApJ...248..279P} {248, 279}

\bibitem[\protect\citeauthoryear{{Parker}}{{Parker}}{1993}]{1993AJ....106..560P}
{Parker} J.~W.,  1993, \mn@doi [\aj] {10.1086/116661}, \href
  {https://ui.adsabs.harvard.edu/abs/1993AJ....106..560P} {106, 560}

\bibitem[\protect\citeauthoryear{{Pauldrach}, {Kudritzki}, {Puls}, {Butler}  \&
  {Hunsinger}}{{Pauldrach} et~al.}{1994}]{1994A&A...283..525P}
{Pauldrach} A.~W.~A.,  {Kudritzki} R.~P.,  {Puls} J.,  {Butler} K.,
  {Hunsinger} J.,  1994, \aap, \href
  {https://ui.adsabs.harvard.edu/abs/1994A&A...283..525P} {283, 525}

\bibitem[\protect\citeauthoryear{{Pawlak} et~al.,}{{Pawlak}
  et~al.}{2016}]{2016AcA....66..421P}
{Pawlak} M.,  et~al., 2016, \actaa, \href
  {https://ui.adsabs.harvard.edu/abs/2016AcA....66..421P} {66, 421}

\bibitem[\protect\citeauthoryear{{Pittard}}{{Pittard}}{2011}]{2011BSRSL..80..555P}
{Pittard} J.,  2011, Bulletin de la Societe Royale des Sciences de Liege, \href
  {https://ui.adsabs.harvard.edu/abs/2011BSRSL..80..555P} {80, 555}

\bibitem[\protect\citeauthoryear{{Pollock}}{{Pollock}}{1987}]{1987ApJ...320..283P}
{Pollock} A.~M.~T.,  1987, \mn@doi [\apj] {10.1086/165539}, \href
  {https://ui.adsabs.harvard.edu/abs/1987ApJ...320..283P} {320, 283}

\bibitem[\protect\citeauthoryear{{Pollock} \& {Corcoran}}{{Pollock} \&
  {Corcoran}}{2006}]{2006A&A...445.1093P}
{Pollock} A.~M.~T.,  {Corcoran} M.~F.,  2006, \mn@doi [\aap]
  {10.1051/0004-6361:20053496}, \href
  {https://ui.adsabs.harvard.edu/abs/2006A&A...445.1093P} {445, 1093}

\bibitem[\protect\citeauthoryear{{Pollock}, {Crowther}, {Tehrani}, {Broos}  \&
  {Townsley}}{{Pollock} et~al.}{2018}]{2018MNRAS.474.3228P}
{Pollock} A.~M.~T.,  {Crowther} P.~A.,  {Tehrani} K.,  {Broos} P.~S.,
  {Townsley} L.~K.,  2018, \mn@doi [\mnras] {10.1093/mnras/stx2879}, \href
  {https://ui.adsabs.harvard.edu/abs/2018MNRAS.474.3228P} {474, 3228}

\bibitem[\protect\citeauthoryear{{Portegies Zwart}, {Pooley}  \&
  {Lewin}}{{Portegies Zwart} et~al.}{2002}]{2002ApJ...574..762P}
{Portegies Zwart} S.~F.,  {Pooley} D.,   {Lewin} W. H.~G.,  2002, \mn@doi
  [\apj] {10.1086/340996}, \href
  {https://ui.adsabs.harvard.edu/abs/2002ApJ...574..762P} {574, 762}

\bibitem[\protect\citeauthoryear{{Predehl} \& {Schmitt}}{{Predehl} \&
  {Schmitt}}{1995}]{1995A&A...293..889P}
{Predehl} P.,  {Schmitt} J.~H.~M.~M.,  1995, \aap, \href
  {https://ui.adsabs.harvard.edu/abs/1995A&A...293..889P} {500, 459}

\bibitem[\protect\citeauthoryear{{Puls}, {Vink}  \& {Najarro}}{{Puls}
  et~al.}{2008}]{2008A&ARv..16..209P}
{Puls} J.,  {Vink} J.~S.,   {Najarro} F.,  2008, \mn@doi [\aapr]
  {10.1007/s00159-008-0015-8}, \href
  {https://ui.adsabs.harvard.edu/abs/2008A&ARv..16..209P} {16, 209}

\bibitem[\protect\citeauthoryear{{Ram{\'\i}rez-Agudelo}
  et~al.,}{{Ram{\'\i}rez-Agudelo} et~al.}{2017}]{2017A&A...600A..81R}
{Ram{\'\i}rez-Agudelo} O.~H.,  et~al., 2017, \mn@doi [\aap]
  {10.1051/0004-6361/201628914}, \href
  {https://ui.adsabs.harvard.edu/abs/2017A&A...600A..81R} {600, A81}

\bibitem[\protect\citeauthoryear{{Rate} \& {Crowther}}{{Rate} \&
  {Crowther}}{2020}]{2020MNRAS.493.1512R}
{Rate} G.,  {Crowther} P.~A.,  2020, \mn@doi [\mnras] {10.1093/mnras/stz3614},
  \href {https://ui.adsabs.harvard.edu/abs/2020MNRAS.493.1512R} {493, 1512}

\bibitem[\protect\citeauthoryear{{Rauw}}{{Rauw}}{2022}]{2022arXiv220316842R}
{Rauw} G.,  2022, arXiv e-prints, \href
  {https://ui.adsabs.harvard.edu/abs/2022arXiv220316842R} {p. arXiv:2203.16842}

\bibitem[\protect\citeauthoryear{{Rauw} et~al.,}{{Rauw}
  et~al.}{2015}]{2015ApJS..221....1R}
{Rauw} G.,  et~al., 2015, \mn@doi [\apjs] {10.1088/0067-0049/221/1/1}, \href
  {https://ui.adsabs.harvard.edu/abs/2015ApJS..221....1R} {221, 1}

\bibitem[\protect\citeauthoryear{{Repolust}, {Puls}  \& {Herrero}}{{Repolust}
  et~al.}{2004}]{2004A&A...415..349R}
{Repolust} T.,  {Puls} J.,   {Herrero} A.,  2004, \mn@doi [\aap]
  {10.1051/0004-6361:20034594}, \href
  {https://ui.adsabs.harvard.edu/abs/2004A&A...415..349R} {415, 349}

\bibitem[\protect\citeauthoryear{{Rivero Gonz{\'a}lez}, {Puls}, {Massey}  \&
  {Najarro}}{{Rivero Gonz{\'a}lez} et~al.}{2012}]{2012A&A...543A..95R}
{Rivero Gonz{\'a}lez} J.~G.,  {Puls} J.,  {Massey} P.,   {Najarro} F.,  2012,
  \mn@doi [\aap] {10.1051/0004-6361/201218955}, \href
  {https://ui.adsabs.harvard.edu/abs/2012A&A...543A..95R} {543, A95}

\bibitem[\protect\citeauthoryear{{Roman-Duval} et~al.,}{{Roman-Duval}
  et~al.}{2019}]{2019ApJ...871..151R}
{Roman-Duval} J.,  et~al., 2019, \mn@doi [\apj] {10.3847/1538-4357/aaf8bb},
  \href {https://ui.adsabs.harvard.edu/abs/2019ApJ...871..151R} {871, 151}

\bibitem[\protect\citeauthoryear{{Sab{\'\i}n-Sanjuli{\'a}n}
  et~al.,}{{Sab{\'\i}n-Sanjuli{\'a}n} et~al.}{2017}]{2017A&A...601A..79S}
{Sab{\'\i}n-Sanjuli{\'a}n} C.,  et~al., 2017, \mn@doi [\aap]
  {10.1051/0004-6361/201629210}, \href
  {https://ui.adsabs.harvard.edu/abs/2017A&A...601A..79S} {601, A79}

\bibitem[\protect\citeauthoryear{{Sana}, {Rauw}, {Naz{\'e}}, {Gosset}  \&
  {Vreux}}{{Sana} et~al.}{2006}]{2006MNRAS.372..661S}
{Sana} H.,  {Rauw} G.,  {Naz{\'e}} Y.,  {Gosset} E.,   {Vreux} J.~M.,  2006,
  \mn@doi [\mnras] {10.1111/j.1365-2966.2006.10847.x}, \href
  {https://ui.adsabs.harvard.edu/abs/2006MNRAS.372..661S} {372, 661}

\bibitem[\protect\citeauthoryear{{Sana} et~al.,}{{Sana}
  et~al.}{2012}]{2012Sci...337..444S}
{Sana} H.,  et~al., 2012, \mn@doi [Science] {10.1126/science.1223344}, \href
  {https://ui.adsabs.harvard.edu/abs/2012Sci...337..444S} {337, 444}

\bibitem[\protect\citeauthoryear{{Sana} et~al.,}{{Sana}
  et~al.}{2013}]{2013A&A...550A.107S}
{Sana} H.,  et~al., 2013, \mn@doi [\aap] {10.1051/0004-6361/201219621}, \href
  {https://ui.adsabs.harvard.edu/abs/2013A&A...550A.107S} {550, A107}

\bibitem[\protect\citeauthoryear{{Sanduleak}}{{Sanduleak}}{1970}]{1970CoTol..89.....S}
{Sanduleak} N.,  1970, Contributions from the Cerro Tololo Inter-American
  Observatory, \href {https://ui.adsabs.harvard.edu/abs/1970CoTol..89.....S}
  {89}

\bibitem[\protect\citeauthoryear{{Schild} \& {Testor}}{{Schild} \&
  {Testor}}{1992}]{1992A&AS...92..729S}
{Schild} H.,  {Testor} G.,  1992, \aaps, \href
  {https://ui.adsabs.harvard.edu/abs/1992A&AS...92..729S} {92, 729}

\bibitem[\protect\citeauthoryear{{Schneider} et~al.,}{{Schneider}
  et~al.}{2018}]{2018Sci...359...69S}
{Schneider} F.~R.~N.,  et~al., 2018, \mn@doi [Science]
  {10.1126/science.aan0106}, \href
  {https://ui.adsabs.harvard.edu/abs/2018Sci...359...69S} {359, 69}

\bibitem[\protect\citeauthoryear{{Schnurr}, {Moffat}, {St-Louis}, {Morrell}  \&
  {Guerrero}}{{Schnurr} et~al.}{2008}]{2008MNRAS.389..806S}
{Schnurr} O.,  {Moffat} A.~F.~J.,  {St-Louis} N.,  {Morrell} N.~I.,
  {Guerrero} M.~A.,  2008, \mn@doi [\mnras] {10.1111/j.1365-2966.2008.13584.x},
  \href {https://ui.adsabs.harvard.edu/abs/2008MNRAS.389..806S} {389, 806}

\bibitem[\protect\citeauthoryear{{Schnurr}, {Chen{\'e}}, {Casoli}, {Moffat}  \&
  {St-Louis}}{{Schnurr} et~al.}{2009}]{2009MNRAS.397.2049S}
{Schnurr} O.,  {Chen{\'e}} A.~N.,  {Casoli} J.,  {Moffat} A.~F.~J.,
  {St-Louis} N.,  2009, \mn@doi [\mnras] {10.1111/j.1365-2966.2009.15060.x},
  \href {https://ui.adsabs.harvard.edu/abs/2009MNRAS.397.2049S} {397, 2049}

\bibitem[\protect\citeauthoryear{{Selman}, {Melnick}, {Bosch}  \&
  {Terlevich}}{{Selman} et~al.}{1999}]{1999A&A...341...98S}
{Selman} F.,  {Melnick} J.,  {Bosch} G.,   {Terlevich} R.,  1999, \aap, \href
  {https://ui.adsabs.harvard.edu/abs/1999A&A...341...98S} {341, 98}

\bibitem[\protect\citeauthoryear{{Seward}, {Forman}, {Giacconi}, {Griffiths},
  {Harnden}, {Jones}  \& {Pye}}{{Seward} et~al.}{1979}]{1979ApJ...234L..55S}
{Seward} F.~D.,  {Forman} W.~R.,  {Giacconi} R.,  {Griffiths} R.~E.,  {Harnden}
  F.~R. J.,  {Jones} C.,   {Pye} J.~P.,  1979, \mn@doi [\apjl]
  {10.1086/183108}, \href
  {https://ui.adsabs.harvard.edu/abs/1979ApJ...234L..55S} {234, L55}

\bibitem[\protect\citeauthoryear{{Shenar} et~al.,}{{Shenar}
  et~al.}{2017}]{2017A&A...598A..85S}
{Shenar} T.,  et~al., 2017, \mn@doi [\aap] {10.1051/0004-6361/201629621}, \href
  {https://ui.adsabs.harvard.edu/abs/2017A&A...598A..85S} {598, A85}

\bibitem[\protect\citeauthoryear{{Shenar} et~al.,}{{Shenar}
  et~al.}{2019}]{2019A&A...627A.151S}
{Shenar} T.,  et~al., 2019, \mn@doi [\aap] {10.1051/0004-6361/201935684}, \href
  {https://ui.adsabs.harvard.edu/abs/2019A&A...627A.151S} {627, A151}

\bibitem[\protect\citeauthoryear{{Shenar} et~al.,}{{Shenar}
  et~al.}{2021}]{2021A&A...650A.147S}
{Shenar} T.,  et~al., 2021, \mn@doi [\aap] {10.1051/0004-6361/202140693}, \href
  {https://ui.adsabs.harvard.edu/abs/2021A&A...650A.147S} {650, A147}

\bibitem[\protect\citeauthoryear{{Shenar} et~al.,}{{Shenar}
  et~al.}{2022}]{Shenar2022}
{Shenar} T.,  et~al., 2022, Nature Astronomy

\bibitem[\protect\citeauthoryear{{Smith}, {Shara}  \& {Moffat}}{{Smith}
  et~al.}{1990}]{1990ApJ...348..471S}
{Smith} L.~F.,  {Shara} M.~M.,   {Moffat} A. F.~J.,  1990, \mn@doi [\apj]
  {10.1086/168256}, \href
  {https://ui.adsabs.harvard.edu/abs/1990ApJ...348..471S} {348, 471}

\bibitem[\protect\citeauthoryear{{Smith}, {Nota}, {Pasquali}, {Leitherer},
  {Clampin}  \& {Crowther}}{{Smith} et~al.}{1998}]{1998ApJ...503..278S}
{Smith} L.~J.,  {Nota} A.,  {Pasquali} A.,  {Leitherer} C.,  {Clampin} M.,
  {Crowther} P.~A.,  1998, \mn@doi [\apj] {10.1086/305980}, \href
  {https://ui.adsabs.harvard.edu/abs/1998ApJ...503..278S} {503, 278}

\bibitem[\protect\citeauthoryear{{Smith}, {Lopes de Oliveira}  \&
  {Motch}}{{Smith} et~al.}{2016}]{2016AdSpR..58..782S}
{Smith} M.~A.,  {Lopes de Oliveira} R.,   {Motch} C.,  2016, \mn@doi [Advances
  in Space Research] {10.1016/j.asr.2015.12.032}, \href
  {https://ui.adsabs.harvard.edu/abs/2016AdSpR..58..782S} {58, 782}

\bibitem[\protect\citeauthoryear{{Tehrani}, {Crowther}, {Bestenlehner},
  {Littlefair}, {Pollock}, {Parker}  \& {Schnurr}}{{Tehrani}
  et~al.}{2019}]{2019MNRAS.484.2692T}
{Tehrani} K.~A.,  {Crowther} P.~A.,  {Bestenlehner} J.~M.,  {Littlefair} S.~P.,
   {Pollock} A.~M.~T.,  {Parker} R.~J.,   {Schnurr} O.,  2019, \mn@doi [\mnras]
  {10.1093/mnras/stz147}, \href
  {https://ui.adsabs.harvard.edu/abs/2019MNRAS.484.2692T} {484, 2692}

\bibitem[\protect\citeauthoryear{{Toribio San Cipriano},
  {Dom{\'\i}nguez-Guzm{\'a}n}, {Esteban}, {Garc{\'\i}a-Rojas}, {Mesa-Delgado},
  {Bresolin}, {Rodr{\'\i}guez}  \& {Sim{\'o}n-D{\'\i}az}}{{Toribio San
  Cipriano} et~al.}{2017}]{2017MNRAS.467.3759T}
{Toribio San Cipriano} L.,  {Dom{\'\i}nguez-Guzm{\'a}n} G.,  {Esteban} C.,
  {Garc{\'\i}a-Rojas} J.,  {Mesa-Delgado} A.,  {Bresolin} F.,  {Rodr{\'\i}guez}
  M.,   {Sim{\'o}n-D{\'\i}az} S.,  2017, \mn@doi [\mnras]
  {10.1093/mnras/stx328}, \href
  {https://ui.adsabs.harvard.edu/abs/2017MNRAS.467.3759T} {467, 3759}

\bibitem[\protect\citeauthoryear{{Townsley}, {Broos}, {Feigelson}, {Brandl},
  {Chu}, {Garmire}  \& {Pavlov}}{{Townsley}
  et~al.}{2006a}]{2006AJ....131.2140T}
{Townsley} L.~K.,  {Broos} P.~S.,  {Feigelson} E.~D.,  {Brandl} B.~R.,  {Chu}
  Y.-H.,  {Garmire} G.~P.,   {Pavlov} G.~G.,  2006a, \mn@doi [\aj]
  {10.1086/500532}, \href
  {https://ui.adsabs.harvard.edu/abs/2006AJ....131.2140T} {131, 2140}

\bibitem[\protect\citeauthoryear{{Townsley}, {Broos}, {Feigelson}, {Garmire}
  \& {Getman}}{{Townsley} et~al.}{2006b}]{Leisa06}
{Townsley} L.~K.,  {Broos} P.~S.,  {Feigelson} E.~D.,  {Garmire} G.~P.,
  {Getman} K.~V.,  2006b, \mn@doi [\aj] {10.1086/500535}, \href
  {https://ui.adsabs.harvard.edu/abs/2006AJ....131.2164T} {131, 2164}

\bibitem[\protect\citeauthoryear{{Townsley} et~al.,}{{Townsley}
  et~al.}{2011}]{2011ApJS..194....1T}
{Townsley} L.~K.,  et~al., 2011, \mn@doi [\apjs] {10.1088/0067-0049/194/1/1},
  \href {https://ui.adsabs.harvard.edu/abs/2011ApJS..194....1T} {194, 1}

\bibitem[\protect\citeauthoryear{{Townsley}, {Broos}, {Garmire}, {Bouwman},
  {Povich}, {Feigelson}, {Getman}  \& {Kuhn}}{{Townsley}
  et~al.}{2014}]{2014ApJS..213....1T}
{Townsley} L.~K.,  {Broos} P.~S.,  {Garmire} G.~P.,  {Bouwman} J.,  {Povich}
  M.~S.,  {Feigelson} E.~D.,  {Getman} K.~V.,   {Kuhn} M.~A.,  2014, \mn@doi
  [\apjs] {10.1088/0067-0049/213/1/1}, \href
  {https://ui.adsabs.harvard.edu/abs/2014ApJS..213....1T} {213, 1}

\bibitem[\protect\citeauthoryear{{Townsley}, {Broos}, {Garmire}, {Anderson},
  {Feigelson}, {Naylor}  \& {Povich}}{{Townsley}
  et~al.}{2018}]{2018ApJS..235...43T}
{Townsley} L.~K.,  {Broos} P.~S.,  {Garmire} G.~P.,  {Anderson} G.~E.,
  {Feigelson} E.~D.,  {Naylor} T.,   {Povich} M.~S.,  2018, \mn@doi [\apjs]
  {10.3847/1538-4365/aaaf67}, \href
  {https://ui.adsabs.harvard.edu/abs/2018ApJS..235...43T} {235, 43}

\bibitem[\protect\citeauthoryear{{Trundle}, {Dufton}, {Hunter}, {Evans},
  {Lennon}, {Smartt}  \& {Ryans}}{{Trundle} et~al.}{2007}]{2007A&A...471..625T}
{Trundle} C.,  {Dufton} P.~L.,  {Hunter} I.,  {Evans} C.~J.,  {Lennon} D.~J.,
  {Smartt} S.~J.,   {Ryans} R.~S.~I.,  2007, \mn@doi [\aap]
  {10.1051/0004-6361:20077838}, \href
  {https://ui.adsabs.harvard.edu/abs/2007A&A...471..625T} {471, 625}

\bibitem[\protect\citeauthoryear{{Tsamis}, {Barlow}, {Liu}, {Danziger}  \&
  {Storey}}{{Tsamis} et~al.}{2003}]{2003MNRAS.338..687T}
{Tsamis} Y.~G.,  {Barlow} M.~J.,  {Liu} X.~W.,  {Danziger} I.~J.,   {Storey}
  P.~J.,  2003, \mn@doi [\mnras] {10.1046/j.1365-8711.2003.06081.x}, \href
  {https://ui.adsabs.harvard.edu/abs/2003MNRAS.338..687T} {338, 687}

\bibitem[\protect\citeauthoryear{{Tumlinson} et~al.,}{{Tumlinson}
  et~al.}{2002}]{2002ApJ...566..857T}
{Tumlinson} J.,  et~al., 2002, \mn@doi [\apj] {10.1086/338112}, \href
  {https://ui.adsabs.harvard.edu/abs/2002ApJ...566..857T} {566, 857}

\bibitem[\protect\citeauthoryear{{Villase{\~n}or} et~al.,}{{Villase{\~n}or}
  et~al.}{2021}]{2021MNRAS.507.5348V}
{Villase{\~n}or} J.~I.,  et~al., 2021, \mn@doi [\mnras]
  {10.1093/mnras/stab2197}, \href
  {https://ui.adsabs.harvard.edu/abs/2021MNRAS.507.5348V} {507, 5348}

\bibitem[\protect\citeauthoryear{{Walborn} \& {Blades}}{{Walborn} \&
  {Blades}}{1997}]{1997ApJS..112..457W}
{Walborn} N.~R.,  {Blades} J.~C.,  1997, \mn@doi [\apjs] {10.1086/313043},
  \href {https://ui.adsabs.harvard.edu/abs/1997ApJS..112..457W} {112, 457}

\bibitem[\protect\citeauthoryear{{Walborn} et~al.,}{{Walborn}
  et~al.}{2014}]{2014A&A...564A..40W}
{Walborn} N.~R.,  et~al., 2014, \mn@doi [\aap] {10.1051/0004-6361/201323082},
  \href {https://ui.adsabs.harvard.edu/abs/2014A&A...564A..40W} {564, A40}

\bibitem[\protect\citeauthoryear{{Weigelt} \& {Baier}}{{Weigelt} \&
  {Baier}}{1985}]{1985A&A...150L..18W}
{Weigelt} G.,  {Baier} G.,  1985, \aap, \href
  {https://ui.adsabs.harvard.edu/abs/1985A&A...150L..18W} {150, L18}

\bibitem[\protect\citeauthoryear{{Welty}, {Xue}  \& {Wong}}{{Welty}
  et~al.}{2012}]{2012ApJ...745..173W}
{Welty} D.~E.,  {Xue} R.,   {Wong} T.,  2012, \mn@doi [\apj]
  {10.1088/0004-637X/745/2/173}, \href
  {https://ui.adsabs.harvard.edu/abs/2012ApJ...745..173W} {745, 173}


\makeatother
\end{thebibliography}





\section*{Supplementary Material} 

Supplementary Material are available at MNRAS online. \\

\noindent {\bf Table S1}. Upper limits to observed X-ray luminosities $L^{t}_{\rm X}$ for luminous ($\log L/L_{\odot} \geq 5$) early-type stars excluded from the T-ReX point source catalogue, in RA order. 

\appendix

\section{Thermal plasma fits and properties of early-type stars of T-ReX sources}

\begin{landscape}
\begin{table}
\caption{Thermal plasma fits to X-ray spectroscopy for T-ReX sources, sorted by decreasing photon flux. Notes: Quantities marked with an asterisk ($\ast$) were frozen in the fit. Quantities marked with $\ddagger$ are at the upper (9.5 keV) or lower (0.27 keV) limit chosen for the model; uncertainties are not computed when fit parameters are at such limits. Uncertainties represent 90\% confidence intervals. X-ray luminosities in the soft ($s$) band (0.5--2 keV), hard ($h$) band (2--8 keV), total ($t$) band (0.5--8 keV) are derived from the model spectrum. Absorption-corrected luminosities, superscripted with a $c$, are calculated with {\it only} the Galactic and LMC ISM absorption components removed, no correction is made for any circumstellar absorption ($N(H)^{1}$, $N(H)^{2}$). For two-temperature fits, luminosities are shown for each thermal plasma component so that their relative contribution can be assessed. 
}
\label{A1}
\begin{center}
 \begin{tabular}{l@{\hspace{2mm}}l@{\hspace{2mm}}c@{\hspace{2mm}}r@{\hspace{2mm}}r@{\hspace{2mm}}l@{\hspace{2mm}}l@{\hspace{2mm}}l@{\hspace{2mm}}l@{\hspace{2mm}}l@{\hspace{2mm}}l@{\hspace{2mm}}r@{\hspace{2mm}}r
 @{\hspace{2mm}}r@{\hspace{2mm}}r@{\hspace{2mm}}r@{\hspace{2mm}}r@{\hspace{2mm}}l@{\hspace{2mm}}l@{\hspace{2mm}}l@{\hspace{2mm}}l}   
\hline
\multicolumn{5}{c}{T-ReX Source} & 
\multicolumn{7}{c}{Spectral Fit} & 
 \multicolumn{8}{c}{X-ray Luminosities}\\
\multicolumn{5}{c}{\hrulefill} &
\multicolumn{7}{c}{\hrulefill} &    
\multicolumn{8}{c}{\hrulefill} \\
Label & CXOU J & Photon Flux & NetCts & $\chi_{\nu}^2$ & $N(H)^{\rm LMC}$ & $N(H)^{1}$ & $kT^1$ & $EM^1$ & $N(H)^{2}$ &  $kT^2$ & $EM^2$ &  
$L^{1,t}_{\rm X}$ & $L^{1,tc}_{\rm X}$ & $L^{2,t}_{\rm X}$ & $L^{2,tc}_{\rm X}$ & $L^{sc}_{\rm X}$ & $L^{hc}_{\rm X}$ & $L^t_{\rm X}$    & $L^{tc}_{\rm X}$  \\ [1pt]
 & & cm$^{-2}$\,s$^{-1}$ & count & & $10^{22}$cm$^{-2}$ & $10^{22}$cm$^{-2}$ & keV & log cm$^{-3}$ &  
                        $10^{22}$cm$^{-2}$ & keV & log cm$^{-3}$ &  
\multicolumn{8}{c}{log erg s$^{-1}$} \\
\hline
p1\_995$^{a}$ & 053844.25$-$690605.9 & $1.48 \times 10^{-4}$ & 74445 &  1.05        & $1.05_{-0.04}^{+0.04}$  & $0.0*$ & $1.05_{-0.03}^{+0.03}$ & $57.62_{-0.14}^{+0.14}$ & $0.0*$ & $3.92_{-0.12}^{+0.12}$ & $58.15_{-0.03}^{+0.03}$ & 34.17 & 34.55 & 35.07 & 35.23 & 35.03 & 34.99 & 35.12 & 35.31  \\ %
p1\_752$^{a}$& 053841.59$-$690513.4 & $1.46 \times 10^{-4}$ & 54076 &   1.13      & $1.15_{-0.07}^{+0.07}$    & $0.0*$   & $0.63_{-0.02}^{+0.02}$         &  $58.15_{-0.18}^{+0.18}$    & $0.0*$       & $2.19_{-0.06}^{+0.06}$       & $58.14_{-0.04}^{+0.04}$     & 34.46     &35.06      & 34.86    &     35.10  & 35.27 & 34.73 &  35.01    & 35.38   \\ %
p1\_610 & 053833.43$-$691159.0 & $4.92 \times 10^{-5}$ & 22077 & 1.01 & \multicolumn{7}{l}{Power law fit to entire dataset vs  \citet{2015A&A...579A.131C} fit to partial dataset}& $\cdots$ & $\cdots$ & $\cdots$ & $\cdots$ & 34.22&34.89& $\cdots$ &34.98  \\ 
p1\_893 & 053842.90$-$690604.9 & $1.70 \times 10^{-5}$ &  6955 &  1.04 & $1.4*$   &$0.0*$               & $2.7_{-0.1}^{+0.1}$      & $ 57.4_{-0.01}^{+0.01}$ & $\cdots$ & $\cdots$   & $\cdots$   &  $\cdots$ & $\cdots$ & $\cdots$ & $\cdots$ &   34.13 &   34.05 &   34.16 &   34.39  \\ [1pt] 
p1\_698 & 053840.22$-$690559.8 &  $1.69 \times 10^{-5}$ & 7988 &  0.86 &$0.83*$                &$1.0_{-0.2}^{+0.2}$    & $0.88_{-0.10}^{+0.08}$   & $ 57.2_{-0.2}^{+0.1}$   & $0.0*$  & $2.6_{-0.2}^{+0.3}$   & $ 57.1_{-0.07}^{+0.06}$ &    33.55 &   33.80 &   33.93 &   34.11 &   34.09 &   33.84 &   34.08 &   34.28 \\ [1pt] 
p1\_832 & 053842.38$-$690602.8 &  $1.12 \times 10^{-5}$  & 4453 &  1.10 &$0.92*$                &$0.67_{-0.03}^{\cdots}$                 & $0.47_{\cdots}^{+0.1}$   & $ 57.4_{-0.3}^{\cdots}$ & $0.80_{\cdots}^{+1.1}$     & $2.1_{-0.3}^{+0.6}$   & $ 57.0_{-0.1}^{+0.2}$ &    33.44 &   33.93 &   33.69 &   33.80 &   34.04 &   33.58 &   33.88 &   34.17 \\ [1pt] 
p1\_979 & 053844.12$-$690556.6 &  $4.72 \times 10^{-6}$ & 2181 &  1.18 &$0.62*$               &$1.7_{-0.3}^{+0.2}$            & $0.3*$    & $ 57.7_{-0.2}^{+0.2}$ & $1.4_{-0.8}^{+1.2}$                        & $1.6_{-0.2}^{+0.2}$ & $56.7_{-0.10}^{+0.09}$                   
& 33.00   &  33.30 & 33.26 & 33.34  &  33.47 &   33.09 &   33.45 &   33.62 \\ [1pt] 
p1\_830 & 053842.34$-$690458.1 &  $4.20 \times 10^{-6}$ &  1681 &  0.94 &$0.76*$                &$1.0_{-0.3}^{+0.7}$         & $0.49_{-0.3}^{+0.1}$     & $ 57.0_{-0.3}^{+0.4}$ & $1.4_{-0.6}^{+0.9}$      & $1.1_{-0.2}^{+0.2}$   & $ 56.7_{-0.3}^{+0.2}$  &    33.01 &   33.36 &   33.11 &   33.27 &   33.54 &   32.86 &   33.36 &   33.62 \\ [1pt] 
p1\_1000 & 053844.33$-$690554.7 &  $4.10 \times 10^{-6}$ & 1866 &  0.92 &$0.90*$                &$1.1_{-0.3}^{+0.2}$      & $0.73_{-0.08}^{+0.08}$   & $ 56.8_{-0.2}^{+0.2}$   & $0.0*$                  & $3.5_{-0.9}^{+3.7}$ & $ 56.3_{-0.2}^{+0.1}$  &    33.07 &   33.38 &   33.22 &   33.38 &   33.51 &   33.19 &   33.45 &   33.68 \\ [1pt] 
p1\_1194 & 053853.36$-$690200.9 & $2.97 \times 10^{-6}$ &  1233 &  0.86 &$0.23*$               &$1.1_{-0.6}^{+0.4}$         & $0.66_{-0.2}^{+0.1}$     & $ 56.5_{-0.4}^{+0.2}$   & $5.1_{-1.8}^{+2.3}$        & $1.8_{-0.3}^{+0.4}$   & $ 56.8_{-0.1}^{+0.2}$  &    32.83 &   32.95 &   33.22 &   33.23 &   33.02 &   33.18 &   33.37 &   33.41 \\ [1pt] 
p1\_296 & 053749.06$-$690508.1    & $2.81 \times 10^{-6}$ &  1174 &  0.97 &$0.53*$               &$0.55_{-0.2}^{+0.2}$      & $0.78_{-0.11}^{+0.09}$   & $ 56.4_{-0.1}^{+0.1}$ & $34_{-6.2}^{+8.5}$ & $8.1_{-3.6}^{\cdots}$ & $ 56.9_{-0.09}^{+0.20}$ &    32.93 &   33.19 &   33.48 &   33.49 &   33.14 &   33.51 &   33.59 &   33.66 \\ [1pt] 
cc4970 & 053842.32$-$690603.4 &  $2.37 \times 10^{-6}$ &  496 &  1.02 &$0.93*$                &$1.6_{-0.9}^{+0.7}$         & $0.86_{-0.5}^{+0.3}$      & $ 56.8_{-0.5}^{+0.2}$ &$10_{\cdots}^{\cdots}$  & $2.3_{\cdots}^{\cdots}$ & $56.3_{-6.9}^{+0.9}$ &  33.05 & 33.26 & 32.73 & 32.76 &   33.14 &   33.01 &   33.22 &   33.38 \\ [1pt] 
 p1\_812 & 053842.11$-$690555.2 & $2.05 \times 10^{-6}$ &  934 &  0.87 &$0.81*$    &$1.2_{-0.2}^{+0.5}$                     & $0.51_{-0.2}^{+0.2}$      & $ 56.8_{-0.4}^{+0.3}$ & $0.72_{\cdots}^{+3.7}$  & $2.6_{-0.9}^{+2.4}$ & $56.1_{-0.3}^{+0.3}$ &  32.76 & 33.09 & 32.88 & 32.97 &   33.18 &   32.82&   33.12 &   33.34  \\ [1pt] 
p1\_1256 & 053857.06$-$690605.6 &  $1.87 \times 10^{-6}$ &  784 &  1.23 &$0.67*$               &$3.1_{-0.6}^{+0.6}$          & $1.6_{-0.2}^{+0.3}$      & $ 56.8_{-0.10}^{+0.10}$ & $\cdots$    & $\cdots$ & $\cdots$  & $\cdots$  &  $\cdots$ & $\cdots$ & $\cdots$ &   32.71 &   33.09 &   33.19 &   33.24  \\  [1pt] 
p1\_1088 & 053847.48$-$690025.1 &  $1.74 \times 10^{-6}$ &   694 &  1.24 &$0.42*$               &$1.3_{-0.4}^{+0.3}$         & $0.77_{-0.11}^{+0.08}$   & $ 56.6_{-0.9}^{+0.2}$ & $5.6_{\cdots}^{\cdots}$  & $1.5_{\cdots}^{\cdots}$ & $56.0_{-6.5}^{+0.6}$  &  32.92 & 33.06 & 32.29 & 32.31 &   32.99 &   32.57 &   33.01 &   33.13  \\  [1pt] 
p1\_867 & 053842.65$-$690602.9 &  $1.59 \times 10^{-6}$ &  522 &  1.67 &$1.0*$                &$0.97_{-0.2}^{+0.4}$         & $0.65_{-0.1}^{+0.1}$     & $ 56.7_{-0.17}^{+0.10}$   & $0.0*$ & $9.5\ddagger$  & $55.4_{-5.9}^{+0.4}$   &  32.82 & 33.21 & 32.47 & 32.58 &   33.20 &   32.62 &   32.98 &   33.30  \\ [1pt] 
 p1\_754 & 053841.62$-$690515.1 &  $1.50 \times 10^{-6}$ & 397 &  1.12 &$0.41*$    &$0.0*$                     & $0.78_{-0.5}^{+0.1}$      & $ 55.5_{-1.6}^{+0.2}$ & $0.94_{\cdots}^{+2.2}$ & $5.1_{-2.4}^{\cdots}$ & $56.1_{-0.2}^{+0.2}$  & 32.37 & 32.65 & 33.06 & 33.09 &   32.86 &   32.98 &   33.14 &   33.22  \\ [1pt] 
c6981 & 053842.08$-$690602.2 & $1.11 \times 10^{-6}$ &   337 &  0.95 &$0.70*$                &$0.0*$                     & $0.63_{\cdots}^{+0.4}$      & $ 55.5_{-0.5}^{\cdots}$ & $1.9_{-1.1}^{+2.1}$ & $1.0_{-0.3}^{+0.6}$ & $56.4_{-0.3}^{+0.2}$  & 32.16 & 32.64 & 32.70 & 32.83 &   32.92 &   32.42 &   32.81 &   33.04  \\ [1pt] 
p1\_745 & 053841.49$-$690556.9 & $1.08 \times 10^{-6}$ &   474 &  0.36 &$0.79*$                &$1.3_{-0.3}^{+0.6}$         & $0.58_{-0.4}^{+0.2}$     & $ 56.7_{-0.2}^{+0.4}$   & $21_{\cdots}^{\cdots}$  & $1.1^{+6.4}_{\cdots}$ & $ 56.5_{-7.0}^{\cdots}$   & 32.70 &   32.99 &   32.08 &   32.10 &   32.94 &   32.35 &   32.79 &   33.04  \\ [1pt] 
c7157 & 053842.40$-$690601.1 &  $7.67 \times 10^{-7}$ &  345 &  0.65 &$0.86*$                &$0.0*$           & $0.69_{-0.5}^{+0.1}$     & $ 55.9_{-0.1}^{\cdots}$ & $0.0*$      & $8.6_{-6.5}^{\cdots}$ & $ 55.4_{-0.2}^{+0.3}$   &     32.48 &   33.01 &   32.50 &   32.60 &   33.05 &   32.47 &   32.79 &   33.15  \\ [1pt] 
p1\_1234 & 053855.52$-$690426.7 & $7.14 \times 10^{-7}$  &   303 &  0.61 &$2.4*$                &$0.78_{\cdots}^{+1.0}$   & $2.0_{-0.4}^{+0.6}$      & $ 56.2_{-0.1}^{+0.2}$ & $\cdots$ & $\cdots$  & $\cdots$ & $\cdots$ & $\cdots$ & $\cdots$ & $\cdots$ &   32.70 &   32.76 &   32.81 &   33.04  \\  [1pt] 
p1\_663 & 053838.02$-$690543.3 &  $6.90 \times 10^{-7}$ &   216 &  1.10 &$0.96*$                &$0.89_{\cdots}^{+1.2}$     & $0.49_{-0.3}^{+0.4}$   & $ 56.3_{-0.8}^{+0.4}$   & $3.0_{\cdots}^{+30.5}$     &  $1.9_{-0.9}^{\cdots}$ & $ 55.9_{-1.5}^{+0.5}$  &    32.31 &   32.76 &   32.44 &   32.50 &   32.79 &   32.42 &   32.68 &   32.95 &    \\  [1pt] 
p1\_924 & 053843.20$-$690614.4 &   $6.01 \times 10^{-7}$ &  261 &  1.24 &$0.79*$                &$1.4_{-0.4}^{+0.5}$        & $0.80_{-0.1}^{+0.2}$     & $ 56.3_{-0.2}^{+0.2}$   & $\cdots$ & $\cdots$  & $\cdots$ & $\cdots$ & $\cdots$ & $\cdots$ & $\cdots$ &   32.61 &   32.04 &   32.49 &   32.71  \\  [1pt] 
c7182 & 053842.49$-$690604.3 &  $5.86 \times 10^{-7}$ &   259 &  1.65 &$0.96*$   &$0.0*$                     & $0.27\ddagger$      & $ 56.3_{-0.2}^{+0.1}$ &$0.0*$ & $2.7_{-0.9}^{+1.8}$  & $55.7_{-0.1}^{+0.1}$ & 32.15 & 33.06 & 32.51 & 32.70 &   33.15 &   32.36 &   32.67 &   33.21  \\ [1pt] 
p1\_786 & 053841.87$-$690614.4 & $4.97 \times 10^{-7} $ &  194 &  0.65 &$1.2*$                &$0.95_{-0.7}^{+0.7}$       & $1.0_{-0.4}^{+0.3}$     & $ 56.1_{-0.2}^{+0.4}$   & $\cdots$ & $\cdots$ & $\cdots$ & $\cdots$ & $\cdots$ & $\cdots$ & $\cdots$ &   33.64 &   32.12 &   32.46 &   32.76  \\  [1pt] 
p1\_911 & 053843.09$-$690546.9 & $4.66 \times 10^{-7}$ &    204 &  1.20 &$0.74*$               &$2.0_{-0.4}^{+0.6}$                     & $0.86_{-0.2}^{+0.2}$      & $ 56.2_{-0.2}^{+0.2}$ & $\cdots$ & $\cdots$ & $\cdots$ & $\cdots$ & $\cdots$ & $\cdots$ & $\cdots$ &   32.41 &   32.06 &   32.42 &   32.57  \\  [1pt] 
p1\_441 & 053759.53$-$690901.5 & $4.58 \times 10^{-7}$ &  182 &  1.05 &$0.30*$               &$0.51^{\cdots}_{-0.4}$    & $0.50_{-0.2}^{+0.2}$     & $ 56.0_{-0.4}^{+0.7}$      & $\cdots$ & $\cdots$  & $\cdots$ & $\cdots$ & $\cdots$ & $\cdots$ & $\cdots$ &   32.62 &   31.13 &   32.42 &   32.63  \\  [1pt] 
p1\_480 & 053813.97$-$690747.7 & $4.19 \times 10^{-7}$ &    169 &      0.85 & $0.54*$ & $0.0*$                     & $0.7_{\cdots}^{+0.2}$      & $ 55.3_{-0.3}^{+0.2}$ & $0.055_{\cdots}^{+4.3}$ & $2.1_{-0.9}^{+2.1}$ & $55.4_{-0.2}^{+0.4}$ & 32.01 & 32.38 & 32.24 & 32.38 &   32.58 &   32.02 &   32.44 &   32.68  \\  [1pt] 
c7257 & 053842.65$-$690602.1 & $4.00 \times 10^{-7}$ &    93 &       0.44 &$1.1*$                &$0.53_{\cdots}^{+1.4}$         & $0.52_{\cdots}^{+0.3}$   & $ 56.2_{-0.4}^{\cdots}$   & $\cdots$ & $
\cdots$  & $\cdots$ & $\cdots$ & $\cdots$ & $\cdots$ & $\cdots$ &   32.87 &   31.43 &   32.32 &   32.88  \\  [1pt] 
p1\_1186 & 053852.72$-$690643.2 & $4.00 \times 10^{-7}$ &  153 &  1.16 &$0.79*$  &$0.0*$                     & $1.1_{\cdots}^{+0.3}$      & $ 55.4_{-0.5}^{+0.1}$    & $0.0*$ & $5.0_{\cdots}^{\cdots}$   & $55.1_{-0.5}^{+0.5}$ & 32.15 & 32.48 & 32.13 & 32.25 &   32.53 &   32.16 &   32.44 &   32.68  \\ [1pt] 
c5349 & 053833.82$-$690957.1 & $3.98 \times 10^{-7}$ & 147 &  1.75 &$0.77*$   &$0.0*$                    & $0.27\ddagger$      & $ 55.8_{-6.3}^{+0.2}$  & $0.0*$ & $2.2_{-0.5}^{+1.1}$  & $55.5_{-0.2}^{+0.1}$ & 31.78 & 32.55 & 32.33 & 32.52 &   32.75 &   32.12 &   32.44 &   32.84 \\ [1pt] 
p1\_795 & 053842.01$-$690607.6 & $3.94 \times 10^{-7}$ & 158 &  0.85  &$0.95*$                &$0.42_{-0.4}^{+0.5}$                     & $0.65_{-0.2}^{+0.2}$   & $ 55.9_{-0.2}^{+0.4}$   & $\cdots$ & $\cdots$  & $\cdots$ & $\cdots$ & $\cdots$ & $\cdots$ & $\cdots$ &   32.74 &   31.49 &   32.28 &   32.76  \\  [1pt] 
c6974 & 053842.09$-$690545.5 & $3.82 \times 10^{-7}$ &  161 & 0.52  &$0.79*$                &$1.3_{\cdots}^{+0.6}$                     & $1.0_{-0.3}^{+0.9}$      & $ 56.0_{-0.3}^{+0.2}$    & $\cdots$ & $\cdots$ & $\cdots$  & $\cdots$ & $\cdots$ & $\cdots$ & $\cdots$ &   32.39 &   31.99 &   32.35 &   32.53  \\ [1pt] 
p1\_611 & 053833.62$-$690450.4 & $3.72 \times 10^{-7}$ &    140  &  2.05 &$0.44*$               &$0.0*$ & $0.70_{\cdots}^{\cdots}$            & $ 54.7_{-0.4}^{+1.8}$     & $49\ddagger$ & $9.5\ddagger$  & $56.4_{-0.1}^{+0.2}$ & 31.52 & 31.83 & 32.96 & 32.96 &   31.81 &   32.96 &   32.97 &   32.99  \\ [1pt] 
p1\_749 & 053841.55$-$690519.4 & $3.46 \times 10^{-7}$ &   113 &  1.50 &$0.67*$               &$0.0*$      & $0.41_{\cdots}^{+0.2}$     & $ 55.7_{-1.0}^{+0.3}$    & $0.0*$ & $1.6_{-0.6}^{\cdots}$  & $55.2_{-0.4}^{+0.3}$ & 32.10 & 32.66 & 32.00 & 32.22 &   32.76 &   31.70 &   32.36 &   32.79  \\  [1pt] 
\hline
\end{tabular}
\end{center}
\footnotesize{
a: Most observations of p1\_995 and p1\_752 suffer from an instrumental problem known as photon pile-up.
Those piled spectra were corrected using a model of the ACIS detectors \citep[see Appendix of][]{2011ApJS..194....2B}}
\end{table}
\end{landscape}

\addtocounter{table}{-1}

\begin{landscape}
\begin{table}
\caption{(continued)}
\begin{center}
 \begin{tabular}{l@{\hspace{2mm}}l@{\hspace{2mm}}c@{\hspace{2mm}}r@{\hspace{2mm}}r@{\hspace{2mm}}l@{\hspace{2mm}}l@{\hspace{2mm}}l@{\hspace{2mm}}l@{\hspace{2mm}}l@{\hspace{2mm}}l@{\hspace{2mm}}r@{\hspace{2mm}}r
 @{\hspace{2mm}}r@{\hspace{2mm}}r@{\hspace{2mm}}r@{\hspace{2mm}}r@{\hspace{2mm}}l@{\hspace{2mm}}l@{\hspace{2mm}}l@{\hspace{2mm}}l}   
\hline
\multicolumn{5}{c}{T-ReX Source} & 
\multicolumn{7}{c}{Spectral Fit} & 
 \multicolumn{8}{c}{X-ray Luminosities}\\
\multicolumn{5}{c}{\hrulefill} &
\multicolumn{7}{c}{\hrulefill} &    
\multicolumn{8}{c}{\hrulefill} \\
Label & CXOU J & Photon Flux & NetCts & $\chi_{\nu}^2$ & $N(H)^{\rm LMC}$ & $N(H)^{1}$ & $kT^1$ & $EM^1$ & $N(H)^{2}$ &  $kT^2$ & $EM^2$ &  
$L^{1,t}_{\rm X}$ & $L^{1,tc}_{\rm X}$ & $L^{2,t}_{\rm X}$ & $L^{2,tc}_{\rm X}$ & $L^{sc}_{\rm X}$ & $L^{hc}_{\rm X}$ & $L^t_{\rm X}$    & $L^{tc}_{\rm X}$  \\ [1pt]
%
 & & cm$^{-2}$\,s$^{-1}$ & count & & $10^{22}$cm$^{-2}$ & $10^{22}$cm$^{-2}$ & keV & log cm$^{-3}$ &  
                        $10^{22}$cm$^{-2}$ & keV & log cm$^{-3}$ &  
\multicolumn{8}{c}{log erg s$^{-1}$} \\
\hline
p1\_722 & 053841.21$-$690258.3 & $3.44 \times 10^{-7}$ &    93 &  0.64 &$0.35*$               &$0.84_{\cdots}^{+0.9}$                     & $1.1_{-0.5}^{+0.5}$      & $ 55.6_{-0.2}^{+0.4}$  & $\cdots$ & $\cdots$    & $\cdots$  &  $\cdots$ & $\cdots$ & $\cdots$ & $\cdots$ &   32.25 &   31.74 &   32.25 &   32.36  \\  [1pt]
c10289 & 053911.27$-$690201.2 & $3.23 \times 10^{-7}$ &   100 &  0.66 &$0.76*$                &$2.9_{-2.2}^{+3.3}$          & $3.3_{-1.4}^{\cdots}$    & $ 55.8_{-0.3}^{+0.3}$     & $\cdots$ & $\cdots$ &$\cdots$ & $\cdots$ & $\cdots$ & $\cdots$ & $\cdots$ &   31.79 &   32.50 &   32.55 &   32.58  \\ [1pt] 
p1\_955 & 053843.67$-$690547.8 & $2.77 \times 10^{-7}$ &  116 &  0.74 &$0.51*$               &$0.0*$                     & $0.97_{-0.1}^{+0.1}$     & $ 55.4_{-0.09}^{+0.07}$   & $\cdots$ & $\cdots$  & $\cdots$ & $\cdots$ & $\cdots$ & $\cdots$ & $\cdots$ &   32.42 &   31.42 &   32.19 &   32.46  \\  [1pt]
c893 & 053708.89$-$690720.9 & $2.47 \times 10^{-7}$ &   74 &  0.55 &$0.55*$               &$0.31_{\cdots}^{+0.9}$                      & $0.70_{\cdots}^{+0.5}$     & $ 55.7_{-0.4}^{+1.6}$    & $\cdots$ & $\cdots$   & $\cdots$ & $\cdots$ & $\cdots$ & $\cdots$ & $\cdots$ &   32.56 &   31.35 &   32.27 &   32.59  \\  [1pt]
c7131 & 053842.37$-$690615.1 & $2.45 \times 10^{-7}$ &    90 &  0.46 &$1.3*$                &$2.2_{-1.0}^{+1.7}$         & $0.55_{-0.2}^{+0.3}$     & $ 56.4_{-0.5}^{+0.5}$     & $\cdots$ & $\cdots$ & $\cdots$ & $\cdots$ & $\cdots$ & $\cdots$ & $\cdots$ &   32.30 &   31.63 &   32.09 &   32.38  \\  [1pt]
p1\_1033 & 053844.93$-$690554.1 & $2.18 \times 10^{-7}$ &    96 &  1.44 &$0.90*$                &$1.2_{-1.2}^{+0.9}$       & $1.0_{-0.5}^{+0.6}$      & $ 55.7_{-0.3}^{+0.5}$    &$\cdots$ & $\cdots$  & $\cdots$  & $\cdots$ & $\cdots$ & $\cdots$ & $\cdots$ &   32.16 &   31.73 &   32.08 &   32.29  \\[1pt] 
p1\_1035 & 053845.11$-$690508.4 &$2.08 \times 10^{-7}$ & 50 &  1.02 &$0.79*$                &$0.0*$     & $3.8_{-1.6}^{\cdots}$    & $ 55.3_{-0.1}^{+0.1}$    & $\cdots$ & $\cdots$  & $\cdots$  &  $\cdots$ & $\cdots$ & $\cdots$ & $\cdots$ &   32.08 &   32.14 &   32.27 &   32.41  \\[1pt] 
c7452 & 053843.11$-$690601.3 & $2.08 \times 10^{-7}$ & 124 &  1.01 & $1.1*$               & $0.062_{\cdots}^{+4.3}$ & $0.63_{\cdots}^{\cdots}$ & $55.3_{-5.8}^{+1.1}$ & $0.49_{\cdots}^{\cdots}$ & $8.1_{\cdots}^{\cdots}$ & $55.3_{-5.8}^{+0.4}$ & 31.67 & 32.32 & 32.34 & 32.42 & 32.43 & 32.30 & 32.43 & 32.67 \\[1pt]
c7018 & 053842.17$-$690601.6 & $2.07 \times 10^{-7}$ &   45 &  0.95 &$0.82*$                &$1.5_{-0.6}^{+1.2}$       & $0.89_{-0.3}^{+0.8}$      & $ 55.8_{-0.3}^{+0.2}$     & $\cdots$ & $\cdots$   & $\cdots$   &  $\cdots$ & $\cdots$ & $\cdots$ & $\cdots$ &   32.14 &   31.68 &   32.07 &   32.27  \\ [1pt] 
c6701 & 053841.52$-$690601.0 & $2.03 \times 10^{-7}$ &   61  &  0.90 &$1.2*$                &$0.0*$    & $0.35_{\cdots}^{+0.6}$   & $ 55.9_{-0.6}^{+0.7}$       & $\cdots$ & $\cdots$   & $\cdots$  &  $\cdots$ & $\cdots$ & $\cdots$ & $\cdots$ &   32.80 &   30.43 &   31.86 &   32.80  \\  [1pt]
p1\_987 & 053844.23$-$690547.1 & $1.99 \times 10^{-7}$ &    84  &  0.40 &$0.67*$               &$0.41_{\cdots}^{+0.6}$     & $0.79_{-0.5}^{+0.4}$     & $ 55.5_{-0.3}^{+1.0}$    & $\cdots$ & $\cdots$   & $\cdots$  &  $\cdots$ & $\cdots$ & $\cdots$ & $\cdots$ &   32.31 &   31.29 &   32.03 &   32.35  \\  [1pt]
c6062 & 053839.15$-$690621.2 & $1.92 \times 10^{-7}$ &   66 &  1.08 &$0.67*$               &$4.2_{-2.8}^{+6.4}$    & $4.1_{-2.6}^{\cdots}$    & $ 55.6_{-0.3}^{+0.5}$      & $\cdots$ & $\cdots$     & $\cdots$  &  $\cdots$ & $\cdots$ & $\cdots$ & $\cdots$ &   31.41 &   32.36 &   32.39 &   32.41  \\ [1pt]
c4876 & 053828.43$-$691119.4 & $1.90 \times 10^{-7}$ &   69  &  1.02 &$0.31*$               &$0.045_{\cdots}^{+2.9}$   & $0.96_{\cdots}^{+0.3}$     & $ 55.1_{-0.2}^{\cdots}$   & $\cdots$ & $\cdots$    & $\cdots$ &  $\cdots$ & $\cdots$ & $\cdots$ & $\cdots$ &   32.13 &   31.14 &   31.99 &   32.17  \\ [1pt]
p1\_682 & 053839.36$-$690606.5 & $1.88 \times 10^{-7}$ &   67 &  1.39 &$1.2*$                &$0.0*$    & $0.59_{-0.3}^{+0.3}$   & $ 55.6_{-0.2}^{+0.6}$  & $\cdots$ & $\cdots$     & $\cdots$   &  $\cdots$ & $\cdots$ & $\cdots$ & $\cdots$ &   32.69 &   31.06 &   31.98 &   32.70  \\ [1pt]
cc4968 & 053842.27$-$690604.9 & $1.82 \times 10^{-7}$ &   66 &  0.62 &$0.76*$                &$0.0*$     & $1.8_{-0.7}^{+4.5}$      & $ 55.3_{-0.3}^{+0.2}$       & $\cdots$ & $\cdots$    & $\cdots$  &  $\cdots$ & $\cdots$ & $\cdots$ & $\cdots$ &   32.08 &   31.76 &   32.03 &   32.25  \\ [1pt]
p1\_760 & 053841.74$-$690625.0 & $1.77 \times 10^{-7}$ &    72 &  1.55 &$0.94*$                &$0.0*$    & $1.1_{-0.4}^{+0.3}$     & $ 55.3_{-0.13}^{+0.10}$   & $\cdots$ & $\cdots$      & $\cdots$   &  $\cdots$ & $\cdots$ & $\cdots$ & $\cdots$ &   32.26 &   31.39 &   31.93 &   32.32  \\[1pt] 
p1\_866 & 053842.63$-$690556.2 & $1.72 \times 10^{-7}$ &   62 &  1.30 &$1.4*$                &$1.4_{\cdots}^{\cdots}$    & $0.83_{-0.6}^{+0.8}$     & $ 55.7_{-0.5}^{+1.5}$    & $\cdots$ & $\cdots$    & $\cdots$  &  $\cdots$ & $\cdots$ & $\cdots$ & $\cdots$ &   32.11 &   31.55 &   31.89 &   32.22  \\ [1pt] 
c8160 & 053846.17$-$690617.2 & $1.70 \times 10^{-7}$ &    64 &  0.16 &$0.67*$               &$0.63_{\cdots}^{+1.1}$    & $0.44_{\cdots}^{+0.5}$   & $ 55.8_{-0.7}^{\cdots}$   & $\cdots$ & $\cdots$     & $\cdots$  &  $\cdots$ & $\cdots$ & $\cdots$ & $\cdots$ &   32.36 &   30.79 &   31.96 &   32.37  \\  [1pt]
p1\_637 & 053836.40$-$690657.5 & $1.70 \times 10^{-7}$ &   67 &  1.16 &$1.1*$                &$0.0*$     & $1.6_{-0.4}^{+1.0}$    & $ 55.2_{-0.2}^{+0.1}$    & $49\ddagger$ & $9.5\ddagger$    & $55.6_{-6.1}^{+0.4}$  &  31.89 & 32.19 & 32.10 & 32.11&   32.04 &   32.24 &   32.31 &   32.45  \\  [1pt]
c6030 & 053838.82$-$690649.6 &$1.60 \times 10^{-7}$ &    50 & 1.27 &$1.2*$                &$1.6_{\cdots}^{+1.4}$  & $1.1_{-0.5}^{+1.0}$      & $ 55.6_{-0.4}^{+0.4}$     & $\cdots$ & $\cdots$     & $\cdots$  &  $\cdots$ & $\cdots$ & $\cdots$ & $\cdots$ &   31.95 &   31.74 &   31.97 &   32.16  \\ [1pt]
c7410 & 053843.04$-$690611.2 & $1.55 \times 10^{-7}$ &    69  &  0.55 &$0.97*$                &$0.56_{\cdots}^{+1.0}$   & $0.71_{-0.4}^{\cdots}$     & $ 55.5_{-0.4}^{+1.0}$    & $\cdots$ & $\cdots$     & $\cdots$   &  $\cdots$ & $\cdots$ & $\cdots$ & $\cdots$ &   32.28 &   31.22 &   31.89 &   32.32  \\ [1pt]
p1\_695 & 053840.17$-$690551.2 &$1.55 \times 10^{-7}$  &    63  &  0.29 &$1.1*$                &$0.0*$   & $1.6_{-0.4}^{+0.9}$      & $ 55.3_{-0.1}^{+0.1}$     & $\cdots$ & $\cdots$     & $\cdots$   &  $\cdots$ & $\cdots$ & $\cdots$ & $\cdots$ &   32.13 &   31.71 &   31.97 &   32.27  \\ [1pt]
c6973 & 053842.05$-$6906 &$1.51 \times 10^{-7}$ &    39  &  0.74 &$1.1*$                &$1.2_{-0.7}^{+0.7}$   & $0.49_{-0.1}^{+0.2}$   & $ 56.0_{-0.5}^{+0.4}$    & $\cdots$ & $\cdots$    & $\cdots$  &  $\cdots$ & $\cdots$ & $\cdots$ & $\cdots$ &   32.23 &   31.01 &   31.84 &   32.26  \\ [1pt]
c6443 & 053840.79$-$690525.1 & $1.48 \times 10^{-7}$ &   46  &  0.89 &$0.90*$                &$0.0*$   & $1.0_{-0.2}^{+0.2}$     & $ 55.3_{-0.1}^{+0.1}$      & $\cdots$ & $\cdots$  & $\cdots$   &  $\cdots$ & $\cdots$ & $\cdots$ & $\cdots$ &   32.29 &   31.36 &   31.94 &   32.34 \\ [1pt]
c8112 & 053845.69$-$690622.5 & $1.42 \times 10^{-7}$ &    55  &  1.31 &$0.90*$                &$0.31_{\cdots}^{+0.9}$   & $0.76_{-0.4}^{+0.3}$     & $ 55.3_{-0.3}^{+0.6}$      & $\cdots$ & $\cdots$    & $\cdots$   &  $\cdots$ & $\cdots$ & $\cdots$ & $\cdots$ &   32.25 &   31.12 &   31.84 &   32.28  \\ [1pt]
c7859 & 053844.41$-$690536.2 & $1.40 \times 10^{-7}$ &   28  &  0.64 &$0.79*$                &$4.4_{-2.3}^{\cdots}$    & $0.33_{\cdots}^{+0.3}$      & $ 57.0_{-7.6}^{+1.7}$       & $\cdots$ & $\cdots$     & $\cdots$    &  $\cdots$ & $\cdots$ & $\cdots$ & $\cdots$ &   31.88 &   31.21 &   31.82 &   31.97  \\ [1pt]
c6370 & 053840.54$-$690557.1 & $1.38 \times 10^{-7}$ &   35 &  0.96 &$0.75*$               &$0.0*$    & $1.1_{-0.6}^{\cdots}$       & $ 54.7_{-0.4}^{+0.3}$    & $25_{-14.2}^{\cdots}$ & $9.5\ddagger$                   & $55.6_{\cdots}^{+0.7}$     &  31.45 & 31.76 & 32.35 & 32.36 &   31.70 &   32.37 &   32.40 &   32.46  \\ [1pt]
p1\_553 & 053823.69$-$690503.4 & $1.37 \times 10^{-7}$ &   60  &  0.49 &$0.30*$               &$0.0*$    & $0.63_{-0.2}^{+0.2}$   & $ 55.1_{-0.2}^{+0.3}$   & $\cdots$ & $\cdots$    & $\cdots$  &  $\cdots$ & $\cdots$ & $\cdots$ & $\cdots$ &   32.19 &   30.63 &   31.95 &   32.20  \\ [1pt]
c7907 & 053844.55$-$690451.1 & $1.31 \times 10^{-7}$ &     49 &   0.87 &$0.63*$               &$0.59_{\cdots}^{+1.0}$    & $0.38_{-0.2}^{+0.2}$     & $ 55.8_{-0.5}^{+1.2}$     & $\cdots$ & $\cdots$   & $\cdots$    &  $\cdots$ & $\cdots$ & $\cdots$ & $\cdots$ &   32.30 &   30.50 &   31.89 &   32.31  \\ [1pt]
p1\_718 & 053840.97$-$690556.0 & $1.22 \times 10^{-7}$ &   45 &   0.83 &$0.71*$               &$0.27_{\cdots}^{+0.6}$    & $0.86_{-0.3}^{+0.3}$     & $ 55.2_{-0.3}^{+0.3}$     & $\cdots$ & $\cdots$     & $\cdots$ &  $\cdots$ & $\cdots$ & $\cdots$ & $\cdots$ &   32.12 &   31.12 &   31.83 &   32.16  \\ [1pt]
p1\_766 & 053841.80$-$690532.2 & $1.09 \times 10^{-7}$ &   31 &   1.02 &$0.62*$               &$3.2_{-2.0}^{+1.1}$     & $0.83_{-0.2}^{+0.2}$     & $ 55.8_{-0.2}^{+0.1}$    & $\cdots$ & $\cdots$   & $\cdots$ &  $\cdots$ & $\cdots$ & $\cdots$ & $\cdots$ &   31.66 &   31.55 &   31.82 &   31.91  \\ [1pt]
p1\_672 & 053838.76$-$690613.0 & $1.03 \times 10^{-7}$ &    29 &   1.20 &$0.87*$                &$0.0_{\cdots}^{\cdots}$    & $1.3_{\cdots}^{\cdots}$      & $ 55.1_{\cdots}^{\cdots}$    & $\cdots$ & $\cdots$      & $\cdots$ &  $\cdots$ & $\cdots$ & $\cdots$ & $\cdots$ &   31.98 &   31.34 &   31.76 &   32.07  \\ [1pt]
p1\_937 & 053843.29$-$690616.4 & $1.00 \times 10^{-7}$ &    39  &  1.07 &$0.75*$   &$0.0*$    & $0.31_{-0.05}^{+0.06}$     & $ 55.6_{-0.2}^{+0.2}$    & $\cdots$ & $\cdots$     & $\cdots$    &  $\cdots$ & $\cdots$ & $\cdots$ & $\cdots$ &   32.43 &   29.83 &   31.74 &   32.43  \\ [1pt]
cc5383 & 053844.97$-$690507.8 &$9.87 \times 10^{-8}$ &    20 &  0.76 &$0.50*$               &$1.2_{\cdots}^{+1.9}$  & $1.3_{-0.5}^{+1.9}$    & $ 55.2_{-0.4}^{+0.4}$     & $\cdots$ & $\cdots$      & $\cdots$ &  $\cdots$ & $\cdots$ & $\cdots$ & $\cdots$ &   31.66 &   31.51 &   31.80 &   31.89 \\ [1pt]
p1\_1062 & 053846.26$-$690559.3 & $8.68 \times 10^{-8}$ &   32  &  1.45 &$0.79*$                &$1.0_{\cdots}^{+2.3}$  & $0.44_{\cdots}^{+0.5}$   & $ 55.7_{-0.8}^{\cdots}$    & $\cdots$ & $\cdots$    & $\cdots$    &  $\cdots$ & $\cdots$ & $\cdots$ & $\cdots$ &   31.97 &   30.62 &   31.61 &   31.99  \\ [1pt]
\hline
\end{tabular}
\end{center}
\end{table}
\end{landscape}

\addtocounter{table}{-1}

\begin{landscape}
\begin{table}
\caption{(continued)}
\begin{center}
 \begin{tabular}{l@{\hspace{2mm}}l@{\hspace{2mm}}c@{\hspace{2mm}}r@{\hspace{2mm}}r@{\hspace{2mm}}l@{\hspace{2mm}}l@{\hspace{2mm}}l@{\hspace{2mm}}l@{\hspace{2mm}}l@{\hspace{2mm}}l@{\hspace{2mm}}r@{\hspace{2mm}}r
 @{\hspace{2mm}}r@{\hspace{2mm}}r@{\hspace{2mm}}r@{\hspace{2mm}}r@{\hspace{2mm}}l@{\hspace{2mm}}l@{\hspace{2mm}}l@{\hspace{2mm}}l}   
\hline
\multicolumn{5}{c}{T-ReX Source} & 
\multicolumn{7}{c}{Spectral Fit} & 
 \multicolumn{8}{c}{X-ray Luminosities}\\
\multicolumn{5}{c}{\hrulefill} &
\multicolumn{7}{c}{\hrulefill} &    
\multicolumn{8}{c}{\hrulefill} \\
Label & CXOU J & Photon Flux & NetCts & $\chi_{\nu}^2$ & $N(H)^{\rm LMC}$ & $N(H)^{1}$ & $kT^1$ & $EM^1$ & $N(H)^{2}$ &  $kT^2$ & $EM^2$ &  
$L^{1,t}_{\rm X}$ & $L^{1,tc}_{\rm X}$ & $L^{2,t}_{\rm X}$ & $L^{2,tc}_{\rm X}$ & $L^{sc}_{\rm X}$ & $L^{hc}_{\rm X}$ & $L^t_{\rm X}$    & $L^{tc}_{\rm X}$  \\ [1pt]
%
 & & cm$^{-2}$\,s$^{-1}$ & count & & $10^{22}$cm$^{-2}$ & $10^{22}$cm$^{-2}$ & keV & log cm$^{-3}$ &  
                        $10^{22}$cm$^{-2}$ & keV & log cm$^{-3}$ &  
\multicolumn{8}{c}{log erg s$^{-1}$} \\
\hline
c5984 & 053838.40$-$690418.2 & $8.67 \times 10^{-8}$ &    38  &  1.10 &$0.54*$               &$0.012_{\cdots}^{+2.8}$    & $3.2_{\cdots}^{\cdots}$    & $ 54.9_{-0.2}^{+0.7}$    & $\cdots$ & $\cdots$   & $\cdots$  &  $\cdots$ & $\cdots$ & $\cdots$ & $\cdots$ &   31.70 &   31.70 &   31.88 &   32.00  \\ [1pt]
c5572 & 053835.56$-$690606.5 & $8.67 \times 10^{-8}$ &  18 &   0.67 &$0.90*$                &$0.0*$     & $0.88_{-0.3}^{+0.3}$ & $ 55.0_{-0.2}^{+0.2}$   & $\cdots$ & $\cdots$     & $\cdots$  &  $\cdots$ & $\cdots$ & $\cdots$ & $\cdots$ &   32.10 &   30.96 &   31.67 &   32.13  \\ [1pt]
c5987 & 053838.51$-$690621.8 & $8.47 \times 10^{-8}$ &  28 &  1.02 &$0.79*$                &$0.0*$    & $1.1_{-0.2}^{+0.3}$     & $ 55.0_{-0.2}^{+0.1}$    & $\cdots$ & $\cdots$    & $\cdots$   &  $\cdots$ & $\cdots$ & $\cdots$ & $\cdots$ &   31.99 &   31.11 &   31.70 &   32.04  \\ [1pt]
c6213 & 053839.84$-$690609.4 & $8.31 \times 10^{-8}$ &    24  &  0.93 &$0.79*$                &$0.66_{\cdots}^{+0.9}$     & $0.70_{-0.3}^{+0.3}$     & $ 55.3_{-0.4}^{+0.5}$   & $\cdots$ & $\cdots$   & $\cdots$  &  $\cdots$ & $\cdots$ & $\cdots$ & $\cdots$ &   31.99 &   30.96 &   31.67 &   32.02  \\ [1pt]
c3084 & 053752.08$-$690439.5 & $7.96 \times 10^{-8}$ &  43 & 1.15 &$0.62*$               &$0.0*$   & $0.63_{-0.2}^{+0.2}$     & $ 55.2_{-0.2}^{+0.2}$     & $\cdots$ & $\cdots$     & $\cdots$ &  $\cdots$ & $\cdots$ & $\cdots$ & $\cdots$ &   32.25 &   30.70 &   31.82 &   32.26  \\ [1pt]
c6472 & 053840.92$-$690554.9 & $7.93 \times 10^{-8}$ &   22 &  0.82 &$0.90*$                &$0.0*$    & $1.2_{-0.4}^{+0.5}$      & $ 55.0_{-0.2}^{+0.2}$    & $\cdots$ & $\cdots$    & $\cdots$  &  $\cdots$ & $\cdots$ & $\cdots$ & $\cdots$ &   31.93 &   31.18 &   31.66 &   32.00  \\ [1pt]
p1\_1021 & 053844.57$-$690512.4 & $7.89 \times 10^{-8}$ &  25 &  1.32 &$1.4*$                &$0.84_{-0.6}^{+1.0}$    & $1.0_{-0.3}^{+0.4}$      & $ 55.3_{-0.2}^{+0.2}$     & $\cdots$ & $\cdots$     & $\cdots$  &  $\cdots$ & $\cdots$ & $\cdots$ & $\cdots$ &   31.91 &   31.34 &   31.67 &   32.01 \\ [1pt]
c10271 & 053910.87$-$690613.7 & $7.55 \times 10^{-8}$ &    31  &  1.12 &$0.79*$                &$0.0*$     & $0.90_{-0.3}^{+0.3}$     & $ 54.9_{-0.2}^{+0.2}$     & $\cdots$ & $\cdots$  & $\cdots$  &  $\cdots$ & $\cdots$ & $\cdots$ & $\cdots$ &   31.97 &   30.87 &   31.60 &   32.00  \\ [1pt]
cc4651 & 053839.59$-$690559.3 & $7.50 \times 10^{-8}$ &    42  &  1.09 &$0.90*$                &$0.0*$     & $2.7_{-1.1}^{+5.0}$      & $ 55.1_{-0.2}^{+0.2}$    & $\cdots$ & $\cdots$   & $\cdots$  &  $\cdots$ & $\cdots$ & $\cdots$ & $\cdots$ &   31.83 &   31.76 &   31.92 &   32.10  \\ [1pt]
c8320 & 053848.07$-$690442.4 & $7.32 \times 10^{-8}$ &    18  &  1.10 &$1.6*$                &$1.6_{\cdots}^{+4.2}$        & $1.1_{-0.6}^{+1.2}$      & $ 55.4_{-0.4}^{+1.0}$    & $\cdots$ & $\cdots$  & $\cdots$ &  $\cdots$ & $\cdots$ & $\cdots$ & $\cdots$ &   31.68 &   31.46 &   31.64 &   31.88  \\ [1pt]
c7031 & 053842.21$-$690832.5 & $7.19 \times 10^{-8}$ &   28 &  1.02 &$1.5*$                &$2.2{\cdots}^{+3.9}$   & $1.7_{-1.0}^{\cdots}$    & $ 55.3_{-0.6}^{+0.7}$     & $\cdots$ & $\cdots$   & $\cdots$ &  $\cdots$ & $\cdots$ & $\cdots$ & $\cdots$ &   31.45 &   31.75 &   31.82 &   31.93  \\ [1pt]
c6170 & 053839.70$-$690608.6 & $7.10 \times 10^{-8}$ &  22 & 0.95 & $0.90*$ & $0.0*$ & $1.2_{-0.3}^{+0.5}$ & $55.0_{-0.2}^{+0.2}$ & $\cdots$ & $\cdots$    & $\cdots$   &  $\cdots$ & $\cdots$ & $\cdots$ & $\cdots$ & 31.90 & 31.21 & 31.66 & 31.98 \\ [1pt]
c7552 & 053843.42$-$690542.1 & $6.63 \times 10^{-8}$ & 27 & 1.04 & $0.67*$ & $0.39_{\cdots}^{+0.8}$ & $0.92_{-0.3}^{+1.0}$ & $54.9_{-0.3}^{+0.2}$ & $\cdots$ & $\cdots$    & $\cdots$   &  $\cdots$ & $\cdots$ & $\cdots$ & $\cdots$ & 31.78 & 30.91 & 31.55 & 31.83  \\ [1pt]
p1\_1530 & 053939.79$-$690430.5 & $6.57 \times 10^{-8}$ &   27 &  0.99 &$0.46*$               &$0.13_{\cdots}^{+1.5}$   & $0.72_{\cdots}^{+0.3}$     & $ 55.0_{-0.3}^{+1.7}$      & $\cdots$ & $\cdots$    & $\cdots$   &  $\cdots$ & $\cdots$ & $\cdots$ & $\cdots$ &   31.97 &   30.67 &   31.70 &   32.00  \\ [1pt]
c8591 & 053850.32$-$690538.3 &$6.16 \times 10^{-8}$ &   24 &  0.99 &$0.63*$               &$0.085_{\cdots}^{+0.4}$     & $0.86_{-0.3}^{+0.3}$     & $ 54.8_{-0.2}^{+0.2}$     & $\cdots$ & $\cdots$    & $\cdots$  &  $\cdots$ & $\cdots$ & $\cdots$ & $\cdots$ &   31.85 &   30.74 &   31.54 &   31.88  \\ [1pt]
c8052 & 053845.24$-$690546.3 & $6.15 \times 10^{-8}$ &   22  &  1.07 &$0.67*$               &$0.93_{\cdots}^{+1.6}$   & $0.73_{-0.4}^{+1.0}$     & $ 55.2_{-0.5}^{+0.9}$   & $\cdots$ & $\cdots$    & $\cdots$   &  $\cdots$ & $\cdots$ & $\cdots$ & $\cdots$ &   31.72 &   30.86 &   31.51 &   31.77  \\ [1pt]
c7254 & 053842.67$-$690635.9 & $6.02 \times 10^{-8}$ &    22  &  1.01 &$0.87*$   &$1.3_{\cdots}^{+1.7}$     & $0.34_{\cdots}^{+0.3}$    & $ 56.0_{-0.9}^{+0.6}$  & $\cdots$ & $\cdots$    & $\cdots$ &  $\cdots$ & $\cdots$ & $\cdots$ & $\cdots$ &   31.90 &   30.36 &   31.50 &   31.92  \\  [1pt]
c8204 & 053846.82$-$690603.0 & $5.69 \times 10^{-8}$ &    25 &  1.21 &$0.85*$                &$0.10_{\cdots}^{+1.3}$     & $0.58_{\cdots}^{+0.3}$     & $ 55.1_{-0.4}^{+1.7}$    & $\cdots$ & $\cdots$      & $\cdots$  &  $\cdots$ & $\cdots$ & $\cdots$ & $\cdots$ &   32.09 &   30.50 &   31.56 &   32.11  \\ [1pt]
 c3981 & 053812.09$-$690634.0 &$5.57 \times 10^{-8}$ &    23  &  1.15 &$1.1*$                &$1.8_{-0.9}^{+3.6}$        & $0.39_{\cdots}^{+0.3}$   & $ 56.0_{-0.9}^{+1.3}$  & $\cdots$ & $\cdots$    & $\cdots$  &  $\cdots$ & $\cdots$ & $\cdots$ & $\cdots$ &   31.77 &   30.63 &   31.43 &   31.80  \\ [1pt]
c8180 & 053846.49$-$690428.0 & $5.32 \times 10^{-8}$ &  23  &  0.67 &$0.88*$                &$1.5_{-0.7}^{+0.9}$   & $0.27\ddagger$     & $ 56.4_{-0.2}^{+0.3}$    & $\cdots$ & $\cdots$    & $\cdots$   &  $\cdots$ & $\cdots$ & $\cdots$ & $\cdots$ &   31.95 &   30.14 &   31.51 &   31.95 \\ [1pt]
c5374 & 053834.08$-$690421.4 & $4.68 \times 10^{-8}$ &    21 &  0.97 &$0.57*$               &$0.19_{\cdots}^{+0.6}$    & $0.79_{-0.4}^{+0.4}$     & $ 54.7_{-0.3}^{+0.2}$    & $\cdots$ & $\cdots$      & $\cdots$   &  $\cdots$ & $\cdots$ & $\cdots$ & $\cdots$ &   31.71 &   30.56 &   31.42 &   31.74 \\ [1pt]
c7235$^{b}$ & 053842.64$-$690536.7 & $4.64 \times 10^{-8}$ &     7 &    0.86 &$0.79*$   &$0.0*$    & $0.8*$   & $ 54.8_{\cdots}^{\cdots}$   & $\cdots$ & $\cdots$      & $\cdots$   &  $\cdots$ & $\cdots$ & $\cdots$ & $\cdots$ &   31.86 &   30.60 &   31.43 &   31.88  \\ [1pt]
c7528 & 053843.37$-$690547.9 & $4.62 \times 10^{-8}$ &    22  &  0.90 &$0.56*$               &$1.2_{-1.0}^{+2.1}$   & $0.44_{\cdots}^{+0.6}$   & $ 55.5_{-0.9}^{\cdots}$   & $\cdots$ & $\cdots$      & $\cdots$   &  $\cdots$ & $\cdots$ & $\cdots$ & $\cdots$  &   31.69 &   30.44 &   31.45 &   31.72  \\ [1pt]
cc4601 & 053839.04$-$690659.0 & $4.61 \times 10^{-8}$ &   16  &  1.06 &$1.4*$                &$0.50_{\cdots}^{\cdots}$     & $1.0_{\cdots}^{\cdots}$     & $ 54.9_{\cdots}^{\cdots}$   & $\cdots$ & $\cdots$      & $\cdots$   &  $\cdots$ & $\cdots$ & $\cdots$ & $\cdots$  &   31.64 &   30.93 &   31.31 &   31.72  \\ [1pt]
c6094 & 053839.32$-$690639.4 & $4.54 \times 10^{-7}$ &  13 &  1.07 &$1.7*$                &$3.1_{-1.6}^{+2.9}$        &  $1.0_{-0.3}^{+0.6}$      & $ 55.4_{-0.3}^{+0.3}$   &  $\cdots$ & $\cdots$      & $\cdots$   &  $\cdots$ & $\cdots$ & $\cdots$ & $\cdots$&   31.39 &   31.42 &   31.54 &   31.71  \\ [1pt]
c6164 & 053839.72$-$690623.7 & $4.40 \times 10^{-8}$ &    11 &  1.30 &$0.90*$                &$0.0_{\cdots}^{+5.1}$     & $1.3_{\cdots}^{+5.7}$      & $ 54.7_{-0.4}^{+0.3}$   & $\cdots$ & $\cdots$      & $\cdots$   &  $\cdots$ & $\cdots$ & $\cdots$ & $\cdots$  &   31.58 &   30.96 &   31.36 &   31.67  \\ [1pt]
c5695 & 053836.31$-$690608.1 & $4.17 \times 10^{-8}$ &   13 &  0.68 &$0.67*$               &$0.61_{\cdots}^{+1.2}$    & $0.66_{-0.3}^{+0.4}$     & $ 55.0_{-0.5}^{+0.4}$   & $\cdots$ & $\cdots$      & $\cdots$   &  $\cdots$ & $\cdots$ & $\cdots$ & $\cdots$  &   31.74 &   30.62 &   31.44 &   31.77  \\ [1pt]
c10098 & 053907.21$-$690745.9 & $4.08 \times 10^{-8}$ &   18 &  0.93 &$0.82*$                &$2.6_{-1.6}^{+3.5}$    & $0.27\ddagger$     & $ 56.5_{-7.0}^{+0.6}$   & $\cdots$ & $\cdots$      & $\cdots$   &  $\cdots$ & $\cdots$ & $\cdots$ & $\cdots$ &   31.57 &   30.23 &   31.31 &   31.59 \\ [1pt]
c5200 & 053832.33$-$690523.6 &$3.55 \times 10^{-8}$ &   13 &  0.96 &$0.63*$               &$0.0*$      & $0.32_{\cdots}^{+0.8}$   & $ 55.1_{-0.8}^{+0.8}$   & $\cdots$ & $\cdots$      & $\cdots$   &  $\cdots$ & $\cdots$ & $\cdots$ & $\cdots$ &   32.00 &   29.46 &   31.41 &   32.00  \\ [1pt]
c5633$^{b}$ & 053835.94$-$690609.2 & $3.25 \times 10^{-8}$ &     12 &  0.80 &$1.4*$  &$0.097_{\cdots}^{\cdots}$     & $0.80*$   & $ 54.9*$   & $\cdots$ & $\cdots$      & $\cdots$   &  $\cdots$ & $\cdots$ & $\cdots$ & $\cdots$  &   31.91 &   30.72 &   31.31 &   31.94  \\ [1pt]
cc7769$^{b}$ & 053916.35$-$690120.4 & $2.51 \times 10^{-8}$ &    6 &  0.97 & $1.2*$  &$0.0*$      & $0.80*$   & $ 54.5_{\cdots}^{\cdots}$ & $\cdots$ & $\cdots$      & $\cdots$   &  $\cdots$ & $\cdots$ & $\cdots$ & $\cdots$  &   31.57 &   30.31 &   31.00 &   31.59 \\ [1pt]
cc4200$^{b}$ & 053834.76$-$690500.7 & $2.27 \times 10^{-8}$ &     8 &  1.18 &$0.15*$               &$0.0*$   & $0.80*$    & $ 54.2*$   & $\cdots$ & $\cdots$      & $\cdots$   &  $\cdots$ & $\cdots$ & $\cdots$ & $\cdots$   &   31.27 &   30.02 &   31.16 &   31.29  \\ [1pt]
cc4181$^{b}$ & 053834.58$-$690605.7 & $2.16 \times 10^{-8}$ &  7 &  1.00 & $1.2*$  &$0.0*$     & $0.80*$           & $ 54.5_{\cdots}^{\cdots}$   & $\cdots$ & $\cdots$      & $\cdots$   &  $\cdots$ & $\cdots$ & $\cdots$ & $\cdots$&   31.58 &   30.33 &   30.99 &   31.61  \\ [1pt]
p1\_1274$^{b}$
                & 053858.66$-$690752.3 & $1.23 \times 10^{-8}$ &     4 &  1.38 &$1.1*$                &$0.0*$      & $0.80*$                & $ 54.2*$              & $\cdots$ & $\cdots$      & $\cdots$   &  $\cdots$ & $\cdots$ & $\cdots$ & $\cdots$ &   31.33 &   30.08 &   30.79 &   31.36  \\ [1pt]
 cc4294$^{b}$
                & 053835.68$-$690617.6 & $8.27 \times 10^{-9}$ &     3 &  1.19 &$0.87*$                &$0.0*$      & $0.80*$                & $ 53.9_{\cdots}^{\cdots}$ & $\cdots$ & $\cdots$      & $\cdots$   &  $\cdots$ & $\cdots$ & $\cdots$ & $\cdots$  &   30.95 &   29.69 &   30.49 &   30.97 \\ [1pt]
cc5800$^{b}$ & 053849.65$-$690855.2 & $6.50 \times 10^{-9}$  &        7 &  0.97 &$0.67*$               &$0.0*$  & $0.80*$   & $ 54.6_{\cdots}^{\cdots}$   &$\cdots$ & $\cdots$      & $\cdots$   &  $\cdots$ & $\cdots$ & $\cdots$ & $\cdots$   &   31.66 &   30.41 &   31.29 &   31.69  \\ [1pt]
cc7873$^{b}$
                & 053918.60$-$690748.3 & $3.91 \times 10^{-9}$ &    1 &  0.90 &$1.5*$                &$0.0*$      & $0.80*$                & $ 53.9_{\cdots}^{\cdots}$ & $\cdots$ & $\cdots$      & $\cdots$   &  $\cdots$ & $\cdots$ & $\cdots$ & $\cdots$  &   31.01 &   29.76 &   30.35 &   31.04 \\ [1pt]

\hline
\end{tabular}
\end{center}
\footnotesize{
b: For these sources, no reasonable fit could be obtained without assuming a plasma temperature.We chose a median value of 0.8~keV.\\
 More significant digits are used for uncertainties $<$0.1 in order to avoid large rounding errors; for consistency, the same number of significant digits is used for both lower and upper uncertainties. Uncertainties are missing when XSPEC was unable to compute them or when their values were so large that the parameter is effectively unconstrained.  Fits lacking uncertainties, fits with large uncertainties, and fits with frozen parameters should be viewed merely as splines to the data to obtain rough estimates of luminosities; the listed parameter values are not robust.}
\end{table}
\end{landscape}

\begin{landscape}
\begin{table}
\caption{Properties of early-type stars in the T-ReX point source catalogue for which  spectroscopic observations are  available. 
Bright X-ray variables are flagged with "var". Visual extinctions, $A_{\rm V}$ represent total extinctions, with an adopted Galactic foreground of $A_{\rm V}^{\rm MW}$ = 0.22 mag. Catalogues include R \citep{1960MNRAS.121..337F}, Sk \citep{1970CoTol..89.....S}, Mk  \citep{1985A&A...153..235M}, VFTS \citep{2011A&A...530A.108E}, MH \citep{1994AJ....107.1054M} HSH \citep{1995ApJ...448..179H}, SMB \citep{1999A&A...341...98S}, P \citep{1993AJ....106..560P},   ST \citep{1992A&AS...92..729S}, CCE \citep{2018A&A...614A.147C}, BAT \citep{1999A&AS..137..117B}.
Spectral type calibrations have been applied in some instances following \citet{2013A&A...558A.134D}, noted with ":".}
\label{A2}
\begin{center}
\begin{tabular}{l@{\hspace{1mm}}c@{\hspace{1mm}}l@{\hspace{1mm}}c@{\hspace{1mm}}l@{\hspace{1mm}}r@{\hspace{2mm}}r@{\hspace{1mm}}r@{\hspace{1mm}}r@{\hspace{1mm}}r@{\hspace{1mm}}r@{\hspace{1mm}}r@{\hspace{1mm}}r@{\hspace{1mm}}r@{\hspace{2mm}}l@{\hspace{1mm}}r@{\hspace{2mm}}l@{\hspace{2mm}}
l@{\hspace{2mm}}l@{\hspace{1mm}}l@{\hspace{2mm}}c@{\hspace{1mm}}c@{\hspace{2mm}}c@{\hspace{1mm}}c@{\hspace{1mm}}l@{\hspace{1mm}}l@{\hspace{1mm}}l}
\hline
T-ReX & HD      & R    & Sk  & Mk  & VFTS & MH & HSH & SMB & P & ST & CCE & BAT & Spectral & Ref & $A_{\rm V}$ & Ref & $\log L_{\rm Bol}$  & Ref & $\log L_{\rm X}$ & Var? & $\log L_{\rm X}/L_{\rm Bol}$ & $kT_{\rm m}$ & Hardness & Nature & Ref\\
Label &            &       &       &        &         &         &      &             &         &             &               &              & Type     &        &  mag           &        & $L_{\odot}$            &        & erg\,s$^{-1}$      & &                                              &      keV &     $\eta_{2}$       &             & \\
\hline
p1\_995 & $\cdots$             & $\cdots$      & $\cdots$     & 34    & $\cdots$       & 880 &  8    &  17     & 1134  &  $\cdots$ & 1766 & 116 & WN5h+WN5h & 1 & 2.0 & 1 & 6.70$\pm$0.1\phantom{0} & 1 & 35.31 & var & --4.98$\pm$0.1\phantom{0} & 3.3\phantom{0} & --0.05 & SB2 & 1 \\ 
p1\_752 & 269919a             &140a&  $\cdots$   & $\cdots$       & 507  & $\cdots$ &   $\cdots$     & $\cdots$        6   &  877  & $\cdots$ &  3191  & 101--2 & WC4+WN6+?   & 2 &  1.6 & 2 & 6.41$\pm$0.3: & 2 &  35.38 & var & --4.62$\pm$0.3: &  1.4\phantom{0} & --0.55  & $\cdots$   & $\cdots$   \\
p1\_610 & $\cdots$             &  $\cdots$      &$\cdots$     &  $\cdots$       & 399 & $\cdots$ & $\cdots$        & $\cdots$          & $\cdots$      &  $\cdots$  & $\cdots$      &    $\cdots$     & O9\,IIIn            & 3 &  1.7 & 4 & 4.81$\pm$0.04 & 5 & 34.97 & var & --3.42$\pm$0.04 & $\cdots$ & +0.65 & SB1 & 4 \\
p1\_893 & $\cdots$             &136c&  $\cdots$    & $\cdots$       &1025 & 681 & 10  &  $\cdots$         &  998  & $\cdots$   &     $\cdots$    & 112 & WN5h+? & 6 & 2.7 & 7 & 6.75$\pm$0.15 & 6 & 34.39 & var & --5.95$\pm$0.15 & 2.7\phantom{0} & --0.09 & SB? & 8\\ 
p1\_698 & $\cdots$            &  $\cdots$      & $\cdots$     &39    & 482 & 57 & 7     &  14       & 767      &  $\cdots$ & 2003        & 99  & O2.5\,If*/WN6+O3 & 9, 10 & 1.7 & 7 & 6.4\phantom{0}$\pm$0.1\phantom{0}  & 7 & 34.28  & var  & --5.71$\pm$0.1\phantom{0} & 1.6\phantom{0} & --0.28 & SB2 & 10-12\\ 
p1\_832 & \multicolumn{14}{l}{ {---}{---}{---} R136a-X1: see Table~\ref{multiple} for details {---}{---}{---}{---}{---}{---}{---}{---}{---}{---}{---}{---}{---}{---}{---}{---}{---}{---}{---}{---}{---}{---}{---}{---}{---}}  &  1.8 & 13 & 7.35$\pm$0.1\phantom{0} & 14 & 34.17  & var & --6.76$\pm$0.1\phantom{0} & 0.93 &  --0.48 & $\cdots$   & $\cdots$    \\
p1\_979 & $\cdots$            &   $\cdots$     &  $\cdots$      &33Sa &  870 & $\cdots$      & 18    & 44        & 1120  &  $\cdots$ &2177     &       & O3\,III(f*)          & 11 & 1.3 & 11 & 5.95$\pm$0.2\phantom{0}  & 15& 33.62 & var  & --5.92$\pm$0.2\phantom{0} & 0.42 & --0.41 & Single  & 11    \\ 
p1\_830 & $\cdots$            &139 &   $\cdots$      &  $\cdots$       & 527  &  $\cdots$ &   $\cdots$    & 2     &   952     &   $\cdots$ &    $\cdots$      & 107 & O6.5\,Iafc+O7\,Iaf & 3 & 1.6 & 16 & 6.49$\pm$0.06 & 17 & 33.62 & var  & --6.46$\pm$0.06 & 0.69 & --0.66 & SB2 & 17-18\\
p1\_1000 & $\cdots$           &   $\cdots$    & $\cdots$        &  33Na &  $\cdots$   & 887 & 16    & 33    &   1140   &  $\cdots$  & 1943 &   $\cdots$     & OC2.5\,If+O4\,V &19 & 1.8 & 19 & 6.28$\pm$0.2\phantom{0}  & 19 & 33.68  & var  & --6.18$\pm$0.2\phantom{0} & 1.4\phantom{0} & --0.35 & SB2 & 19 \\ 
p1\_1194 & \phantom{2}38282 & 144 &--69$^{\circ}$ 246 &$\cdots$  & $\cdots$ & $\cdots$ & $\cdots$ &  $\cdots$     &    9037  &  $\cdots$  & $\cdots$         & 118 & WN5-6h+WN6-7h & 20 & 0.6 & 20 & 6.72$\pm$0.1\phantom{0}  & 20 & 33.41 & var  & --6.89$\pm$0.1\phantom{0} & 1.4\phantom{0} & +0.18 & SB2 & 20 \\ 
p1\_296 & 269891& 130    & --69$^{\circ}$ 235 &    $\cdots$      &  $\cdots$      & $\cdots$ &   $\cdots$     &  $\cdots$     &  $\cdots$  &   $\cdots$    &$\cdots$    & 92  & WN/C+B1\,I      & 2 & 1.1 & 16 & 6.34$\pm$0.3\phantom{0}  & 16 & 33.66  &$\cdots$ & --6.27$\pm$0.3\phantom{0} & 6.3\phantom{0} & +0.41 & SB2 & 16 \\
cc4970 &   \multicolumn{14}{l}{ {---}{---}{---} R136a-X2: see Table~\ref{multiple} for details {---}{---}{---}{---}{---}{---}{---}{---}{---}{---}{---}{---}{---}{---}{---}{---}{---}{---}{---}{---}{---}{---}{---}{---}{---}} & 1.8 & 13 & 6.79$\pm$0.1\phantom{0} & 14 & 33.38 & var  & --7.00$\pm$0.1\phantom{0} & 1.2\phantom{0} & --0.15 & $\cdots$   & $\cdots$   \\
p1\_812 & $\cdots$           &   $\cdots$    &   $\cdots$     & 42 &   $\cdots$     & 374  &   2    &   $\cdots$        & 922   &  $\cdots$   & 2102        & 105 & O2\,If & 9 & 1.6 & 7 & 6.56$\pm$0.1\phantom{0}  & 7 & 33.34 & $\cdots$ & --6.81$\pm$0.1\phantom{0} & 0.86 & --0.30 &$\cdots$   & $\cdots$    \\ 
 p1\_1256 &  269928 & 145  & --69$^{\circ}$ 248 &  & 695     &  $\cdots$ & $\cdots$      &       $\cdots$   &  1788        &  $\cdots$ &  $\cdots$        & 119 & WN6h+O3.5\,If/WN7 & 21 & 1.4 & 21 & 6.64$\pm$0.2\phantom{0}  & 21 & 33.24  & var  & --6.81$\pm$0.2\phantom{0} & 0.86 & --0.30 & SB2 & 12,21 \\
p1\_1088 & 269926 & 146    & --69$^{\circ}$ 245 &  & 617 &   $\cdots$   & $\cdots$      &   $\cdots$     &   9033   &  $\cdots$&  $\cdots$         & 117 & WN5ha & 22 & 1.0 & 7 & 6.29$\pm$0.1\phantom{0}  & 7 &  33.13 & $\cdots$ & --6.75$\pm$0.1\phantom{0} & 0.92 & --0.45 & Single   & 23    \\
p1\_867 &    \multicolumn{14}{l}{ {---}{---}{---} R136a-X3: see Table~\ref{multiple} for details {---}{---}{---}{---}{---}{---}{---}{---}{---}{---}{---}{---}{---}{---}{---}{---}{---}{---}{---}{---}{---}{---}{---}{---}{---}} &  2.0 & 13 & 6.67$\pm$0.1\phantom{0} & 14 &33.30 &  $\cdots$ & --6.96$\pm$0.1\phantom{0} & 1.07 & --0.59 & $\cdots$   & $\cdots$   \\
p1\_754 &  269919b              & 140b &  $\cdots$    &  $\cdots$  & 509  &  $\cdots$ & $\cdots$       &  8     &   880     &  $\cdots$&  3174   & 103 & WN5h+O4\,V & 16 & 0.9 & 16 & 6.25$\pm$0.2\phantom{0}  & 16 & 33.22  &  var  & --6.62$\pm$0.2\phantom{0} & 4.2\phantom{0} & +0.14 & SB2 & 12,16 \\
c6981 &  \multicolumn{14}{l}{ {---}{---}{---} R136a-X4: see Table~\ref{multiple} for details {---}{---}{---}{---}{---}{---}{---}{---}{---}{---}{---}{---}{---}{---}{---}{---}{---}{---}{---}{---}{---}{---}{---}{---}{---}{---}}  & 1.5 & 13 & 5.82$\pm$0.3:\phantom{:}  & 2,24 & 33.04 & $\cdots$ & --6.37$\pm$0.3:\phantom{:} & 0.96 & --0.52 &  SB2 & 24 \\
p1\_745 & $\cdots$                 &   $\cdots$      & $\cdots$    &    $\cdots$     &    $\cdots$  & 201   & 28   & 53     &   860  &  $\cdots$ & 1912 &    $\cdots$    & O3\,V         & 25 & 1.6 & 26 & 5.74$\pm$0.12 & 26 & 33.04  & $\cdots$  & --6.29$\pm$0.12 & 0.78 & --0.59 & $\cdots$   & $\cdots$     \\ 
c7157 &   \multicolumn{14}{l}{ {---}{---}{---} R136a-X5: see Table~\ref{multiple} for details {---}{---}{---}{---}{---}{---}{---}{---}{---}{---}{---}{---}{---}{---}{---}{---}{---}{---}{---}{---}{---}{---}{---}{---}{---}} &  1.7 & 13 & 6.32$\pm$0.1\phantom{0} & 7 & 33.15 &$\cdots$   & --6.76$\pm$0.1\phantom{0} & 2.6\phantom{0} & --0.59 & $\cdots$   & $\cdots$   \\
 p1\_1234 & $\cdots$    &  $\cdots$   &  $\cdots$  &    $\cdots$       & 682     &  $\cdots$ & $\cdots$   &    $\cdots$       & 1732 &  $\cdots$ &  $\cdots$        &    $\cdots$    & WN5h     & 27 & 4.5 & 7 & 6.51$\pm$0.1\phantom{0}  & 7 & 33.04  & $\cdots$  & --7.06$\pm$0.1\phantom{0} & 2.0\phantom{0} & +0.07 & Single   & 27    \\
p1\_663 &  $\cdots$      &  $\cdots$    &  $\cdots$ &      $\cdots$      & 445   & $\cdots$ & $\cdots$   & 97        & 621  &  $\cdots$ & 2981 &   $\cdots$     & O3--4\,V+O4--7\,V & 3 & 1.9 & 7 & 5.80$\pm$0.15 & 26 & 32.95 & $\cdots$  & --6.44$\pm$0.15 & 0.89 & --0.40 & SB2 & 3,18 \\
p1\_924 & $\cdots$     &   $\cdots$   &  $\cdots$  &    35 & 545 & 742 & 12     & 23  &  1029   &   $\cdots$ &  1474    & 114 & O2\,If/WN5 & 9 &1.6 & 7 &  6.30$\pm$0.1\phantom{0}  & 7 & 32.71  & $\cdots$  & --7.18$\pm$0.1\phantom{0} & 0.80 & --0.58 & $\cdots$   & $\cdots$   \\ 
c7182 &     \multicolumn{14}{l}{ {---}{---}{---} R136a-X6: see Table~\ref{multiple} for details {---}{---}{---}{---}{---}{---}{---}{---}{---}{---}{---}{---}{---}{---}{---}{---}{---}{---}{---}{---}{---}{---}{---}{---}{---}} & 1.9 & 13 &  6.45$\pm$0.1\phantom{0} & 14 & 33.21  & $\cdots$  & --6.82$\pm$0.1\phantom{0} & 0.76 & --0.73 & $\cdots$   & $\cdots$   \\
p1\_786 & $\cdots$       &  $\cdots$    &  $\cdots$  & 37Wb & 1017 & 283 & 44& 88  &    897   &  $\cdots$  &  1374  &  104  & O2\,If/WN5 & 9 & 2.3 & 7 & 6.21$\pm$0.1\phantom{0}  & 7 & 32.76  & $\cdots$ & --7.04$\pm$0.1\phantom{0} & 1.0\phantom{0} & --0.53 & SB? & 28 \\
p1\_911 & $\cdots$    &  $\cdots$  &  $\cdots$ & 30 & 542 &   728 & 15   & 24      &    1018    &  $\cdots$    &  2999    &  113 & O2\,If/WN5 & 9 & 1.5 & 7 & 6.16$\pm$0.1\phantom{0}  & 7 & 32.57  & $\cdots$  & --7.18$\pm$0.1\phantom{0} &0.86 & --0.38 & SB & 12 \\ 
p1\_441 &  $\cdots$   & $\cdots$ &  $\cdots$ &    $\cdots$     & 217     &  $\cdots$ & $\cdots$      &     $\cdots$     &     $\cdots$     & 1--98  &       $\cdots$   &  $\cdots$     & O4\,V+O5\,V  & 3 & 0.7 & 2 & 5.79$\pm$0.13 & 17 & 32.63  & $\cdots$  & --6.74$\pm$0.13 & 0.50 & --0.95 & SB2 & 3,17 \\
p1\_480 &  $\cdots$  & $\cdots$ & $\cdots$   & $\cdots$     & 267       &  $\cdots$ & $\cdots$        &    $\cdots$      &    $\cdots$      &  $\cdots$ &     $\cdots$      &  $\cdots$     & O3\,III-I(n)f* & 3 & 1.2 & 7 & 5.96$\pm$0.1\phantom{0}  & 7 & 32.68  & $\cdots$  & --6.87$\pm$0.1\phantom{0} & 1.5\phantom{0} & --0.58 & SB1 & 3,18 \\ 
c7257 &    \multicolumn{14}{l}{ {---}{---}{---} R136a-X7: see Table~\ref{multiple} for details {---}{---}{---}{---}{---}{---}{---}{---}{---}{---}{---}{---}{---}{---}{---}{---}{---}{---}{---}{---}{---}{---}{---}{---}{---}}& 2.1 & 13 & 6.09$\pm$0.2\phantom{0} & 14 & 32.88  & $\cdots$   & --6.80$\pm$0.2\phantom{0} & 0.52 & --0.94 &   $\cdots$   & $\cdots$     \\ 
p1\_1186 & $\cdots$   &  $\cdots$&   $\cdots$  & 4 &  664     &  $\cdots$ & $\cdots$      & 65     &  1607   &  $\cdots$    & 774 &     $\cdots$    & O7\,II(f) & 3 & 1.6 & 7 & 5.53$\pm$0.1\phantom{0} & 5 & 32.68 & $\cdots$  & --6.44$\pm$0.1\phantom{0} & 2.4\phantom{0} & --0.41 & SB? & 3 \\
c5349 &  $\cdots$     &  $\cdots$    &  $\cdots$  &$\cdots$       & 404  & $\cdots$ & $\cdots$         &     $\cdots$      &    $\cdots$ &    $\cdots$      &  $\cdots$        & $\cdots$      & O3.5:\,V:+O5V: & 3, 42 & 1.6 & 7 & 5.91$\pm$0.1\phantom{0}  & 7 & 32.84 &$\cdots$   & --6.65$\pm$0.1\phantom{0} & 0.91 & --0.62 & SB2 & 3,42 \\ 
 p1\_795 &  $\cdots$    & $\cdots$  & $\cdots$   &    $\cdots$&     1019 & 320 & 38    & 70   & 912  &  $\cdots$ & 1608    &   $\cdots$     & O3\,V+O6\,V & 24 & 1.9 & 2 & 6.09$\pm$0.3:\phantom{:}  & 2 & 32.76 & $\cdots$  & --6.92$\pm$0.3:\phantom{:} & 0.65 & --0.90 & SB2 & 24 \\ 
c6974 &   $\cdots$        &  $\cdots$ &  $\cdots$  &  $\cdots$  &           522    & $\cdots$ & 82   & 134  &   921  &  $\cdots$ & 3030 &   $\cdots$    & O6\,II--Iab+O5.5\,V & 3 & 1.6 & 26 & 5.15$\pm$0.15 & 26 & 32.53 & $\cdots$  & --6.21$\pm$0.15 &  1.0\phantom{0} & --0.44 & SB2 & 3,29 \\
p1\_611 &    $\cdots$    &  135 &  $\cdots$ &   $\cdots$ & 402      &  $\cdots$ & $\cdots$       & 12      & 355   &   $\cdots$ &   $\cdots$       & 95 & WN5:+WN7 & 16 & 1.0 & 16 & 6.20$\pm$0.1\phantom{0}  & 16 & 32.99 & $\cdots$ & --6.79$\pm$0.1\phantom{0} & 9.3\phantom{0} & +0.87 & SB2 & 12,16 \\
 p1\_749 & $\cdots$    &  $\cdots$ &  $\cdots$ &    25 & 506 &  $\cdots$ & $\cdots$        & 19       & 871  &  $\cdots$  & 2395  &    $\cdots$  & ON2\,V          & 3 & 1.4 & 7 & 6.43$\pm$0.2\phantom{0}  & 30    & 32.79  & $\cdots$  &--7.23$\pm$0.2\phantom{0} & 0.70 & --0.85 & SB1 & 3 \\ 
\hline
\end{tabular}
\end{center}
\end{table}
\end{landscape}

\addtocounter{table}{-1}

\begin{landscape}
\begin{table}
\caption{continued}
\begin{center}
\begin{tabular}{l@{\hspace{1mm}}c@{\hspace{1mm}}l@{\hspace{1mm}}c@{\hspace{1mm}}l@{\hspace{1mm}}r@{\hspace{2mm}}r@{\hspace{1mm}}r@{\hspace{1mm}}r@{\hspace{1mm}}r@{\hspace{1mm}}r@{\hspace{1mm}}r@{\hspace{1mm}}r@{\hspace{1mm}}r@{\hspace{2mm}}l@{\hspace{1mm}}r@{\hspace{2mm}}l@{\hspace{2mm}}
l@{\hspace{2mm}}l@{\hspace{1mm}}l@{\hspace{2mm}}c@{\hspace{1mm}}c@{\hspace{2mm}}c@{\hspace{1mm}}c@{\hspace{1mm}}l@{\hspace{1mm}}l@{\hspace{1mm}}l}
\hline
T-ReX & HD      & R    & Sk  & Mk  & VFTS & MH & HSH & SMB & P & ST & CCE & BAT & Spectral & Ref & $A_{\rm V}$ & Ref & $\log L_{\rm Bol}$  & Ref & $\log L_{\rm X}$ & Var? & $\log L_{\rm X}/L_{\rm Bol}$ & $kT_{\rm m}$ & Hardness & Nature & Ref\\
Label &            &       &       &        &         &         &      &             &         &             &               &              & Type     &        &  mag           &        & $L_{\odot}$            &        & erg\,s$^{-1}$      & &                                              &      keV &  $\eta_{2}$         &             & \\
\hline
p1\_722 &  $\cdots$            &   $\cdots$      & $\cdots$                          &     $\cdots$   & 500    &  $\cdots$ & $\cdots$       &      $\cdots$  &    $\cdots$  &    $\cdots$  &     $\cdots$     &     $\cdots$    & O6.5\,IV+O6.5\,V & 3 & 0.8 & 2 & 5.55$\pm$0.07 & 17 & 32.36  & $\cdots$ & --6.78$\pm$0.07 & 1.1\phantom{0} & --0.54 & SB2 & 3,17 \\
c10289 &   38344& 147  &--69$^{\circ}$ 251&    $\cdots$    & 758 &  $\cdots$ & $\cdots$      &     $\cdots$      &      $\cdots$    &    $\cdots$   &     $\cdots$     & 122 & WN5h         & 22 & 1.6 & 7 & 6.36$\pm$0.1\phantom{0}  & 7   & 32.58 & $\cdots$  & --7.37$\pm$0.1\phantom{0} & 3.3\phantom{0} & +0.67 & Single   & 23   \\
p1\_955 &  $\cdots$            &   $\cdots$      &          $\cdots$                  &    $\cdots$    & 1031 & 815 & 25    & 39   &  1080 &    $\cdots$  & 2186    &   $\cdots$   &  O3\,V & 25 & 1.1 & 2 & 5.98$\pm$0.3:\phantom{:}  & 2 & 32.46  &$\cdots$  & --7.11$\pm$0.3:\phantom{:} & 0.97 & --0.82 & SB2 & 18,28 \\ 
c893 &  $\cdots$                  &    $\cdots$     &       $\cdots$                     &     $\cdots$  &   16       &  $\cdots$ &   $\cdots$     &     $\cdots$      &     $\cdots$   &   $\cdots$     &      $\cdots$    &   $\cdots$    & O2\,III-If*       & 3 & 1.2 & 7 & 6.12$\pm$0.1\phantom{0}  & 5    & 32.59  & $\cdots$  & --7.11$\pm$0.1\phantom{0} & 0.97 & --0.82 & Single & 3 \\
c7131 & $\cdots$                &     $\cdots$    &       $\cdots$                      & 37 & 1022 & 493 & 14 & 28         & 949      &    $\cdots$    & 1442   &    $\cdots$   & O3.5\,If/WN7 & 9 & 2.5 & 7 & 6.48$\pm$0.1\phantom{0}  & 7 & 32.38 & $\cdots$  & --7.69$\pm$0.1\phantom{0} & 0.55 & --0.65 & Single & 28 \\ 
p1\_1033 & $\cdots$            &    $\cdots$    &            $\cdots$                  &       $\cdots$   &    $\cdots$   & 926 & 43 & 83       &    1195  &    $\cdots$  &   2112 &   $\cdots$     &  O3\,V & 25 & 1.8 & 26 & 5.66$\pm$0.07 & 26 & 32.29 & $\cdots$ & --6.96$\pm$0.25 & 1.0\phantom{0} & --0.47 & $\cdots$   & $\cdots$    \\ 
p1\_1035 & $\cdots$            &    $\cdots$    &              $\cdots$                &     $\cdots$     &    $\cdots$ & $\cdots$  &   $\cdots$   &116      & 1222  &    $\cdots$    & 3180   &    $\cdots$    & O3--6\,V & 31 &1.6 & 26 &  5.06$\pm$0.05 & 26 & 32.41 & $\cdots$ &  --6.24$\pm$0.25 & 3.8\phantom{0} & +0.07 &  $\cdots$   & $\cdots$     \\
c7452 &     \multicolumn{14}{l}{ {---}{---}{---} R136: see Table~\ref{multiple} for details {---}{---}{---}{---}{---}{---}{---}{---}{---}{---}{---}{---}{---}{---}{---}{---}{---}{---}{---}{---}{---}{---}{---}{---}} & 2.2 & 13 & 5.91$\pm$0.2\phantom{0}  & 14 & 32.67 & $\cdots$ &--6.82$\pm$0.2\phantom{0} & 4.4\phantom{0} & --0.15 & $\cdots$ & $\cdots$  \\
c7018 &     \multicolumn{14}{l}{ {---}{---}{---} R136a-X8: see Table~\ref{multiple} for details {---}{---}{---}{---}{---}{---}{---}{---}{---}{---}{---}{---}{---}{---}{---}{---}{---}{---}{---}{---}{---}{---}} & 1.7 & 13 & 5.48$\pm$0.2\phantom{0}  & 14 & 32.27 & $\cdots$ & --6.80$\pm$0.2\phantom{0} & 0.89 & --0.48 & Single? & 14 \\
c6701 &  $\cdots$                &    $\cdots$    &             $\cdots$                 &      $\cdots$  & 1014 & 203 & 29 & 56      & 863      &       $\cdots$  &1956:       &    $\cdots$   & O3\,V+        & 25 & 2.4 & 7 & 6.22$\pm$0.2\phantom{0}  & 7     & 32.80 & $\cdots$ &  --7.01$\pm$0.2\phantom{0} & 0.35 & --1.00 & Single & 18 \\ 
p1\_987 & $\cdots$              &      $\cdots$   &          $\cdots$                   & 32   & 1034 & 878 & 13   & 21    & 1130   &    $\cdots$ &   3043   &     $\cdots$  & O7.5\,II     & 31 & 1.4 & 26 & 5.78$\pm$0.10 & 26 & 32.35 & $\cdots$ & --7.02$\pm$0.25 & 0.79 & --0.82 & Single & 28 \\ 
c6062 &  $\cdots$                &      $\cdots$   &          $\cdots$                    & 49 &  $\cdots$   & $\cdots$    &    $\cdots$      & 30     &  691  &    $\cdots$  & 1261     & 98 & WN6(h) &  32  & 1.4 & 2 & 6.7\phantom{0} $\pm$0.2\phantom{0}  & 33 & 32.41 & $\cdots$ & --7.88$\pm$0.2\phantom{0} & 4.1\phantom{0} & +0.79 & Single & 12 \\
c4876 &  $\cdots$               &     $\cdots$    &                $\cdots$              &   $\cdots$   & 352   &  $\cdots$ &  $\cdots$      &       $\cdots$   &    $\cdots$ &    $\cdots$        &        $\cdots$     &   $\cdots$    & O4.5\,Vz:+O5.5\,Vz: & 3 & 0.8 & 2 & 5.38$\pm$0.04 & 17 & 32.17 & $\cdots$ & --6.79$\pm$0.04 & 0.96 & --0.82 & SB2 & 3,17-18 \\ 
p1\_682 & $\cdots$             &     $\cdots$    &          $\cdots$                   & 36 &  468 &   17 &   $\cdots$     &   86  &  706   &    $\cdots$ &1749       &   $\cdots$   & O2\,V((f*))+? & 3 & 2.2 & 7 & 6.17$\pm$0.2\phantom{0}  & 30 & 32.70 & $\cdots$ & --7.06$\pm$0.2\phantom{0} & 0.59 & --0.95 & Single & 3 \\  
cc4968 &  \multicolumn{14}{l}{ {---}{---}{---} R136a-X9: see Table~\ref{multiple} for details {---}{---}{---}{---}{---}{---}{---}{---}{---}{---}{---}{---}{---}{---}{---}{---}{---}{---}{---}{---}{---}{---}}  & 1.5 & 13 & 5.15$\pm$0.2\phantom{0}  & 14 & 32.25 & $\cdots$ & --6.49$\pm$0.2\phantom{0} & 1.8\phantom{0} & --0.35 & Single & 14 \\
p1\_760 & $\cdots$               &      $\cdots$  &               $\cdots$              &    $\cdots$   & 512  &  $\cdots$ &  $\cdots$      & 68      & 885    &    $\cdots$ & 1199     &  $\cdots$& O2\,V--III & 3 & 1.9 & 7 & 6.04$\pm$0.2\phantom{0}  & 7 & 32.32 & $\cdots$ & --7.31$\pm$0.2\phantom{0} & 1.1\phantom{0} & --0.75 & SB1 & 3 \\
p1\_866 &  $\cdots$              &      $\cdots$   &          $\cdots$                   &    $\cdots$   &     $\cdots$    & 620 & 54   & 109   &  978    &    $\cdots$ & 1963     &  $\cdots$  & O4:  & 26  & 2.6 & 26 & 5.96$\pm$0.11 & 26 & 32.22 & $\cdots$ & --7.33$\pm$0.11 & 0.83 & --0.56 & $\cdots$   & $\cdots$   \\ 
c8160 & $\cdots$                   &      $\cdots$  &         $\cdots$                    & 13  & 599  & $\cdots$ &  $\cdots$      & 40    & 1311   &    $\cdots$ & 1433     &  $\cdots$  & O3\,III(f*) & 3 & 1.4 & 7 & 6.01$\pm$0.10 & 5 & 32.37 & $\cdots$ & --7.23$\pm$0.10 & 0.44 & --0.95 & SB1 & 3 \\
p1\_637 &   $\cdots$              &    $\cdots$   &          $\cdots$                    & 53 & 427  & $\cdots$  & $\cdots$   & 51     &     $\cdots$       &    $\cdots$ & 389      & 96 & WN8(h)           & 22 & 2.2 & 7 & 6.13$\pm$0.1\phantom{0}  & 7 & 32.45 & $\cdots$ & --7.27$\pm$0.1\phantom{0} & 7.2\phantom{0} & +0.23 & Single & 12 \\
c6030 & $\cdots$                   &    $\cdots$    &           $\cdots$                  & 51 & 457   &   $\cdots$ & $\cdots$    & 50     & 666     &    $\cdots$ & 603     & 97  & O3.5\,If/WN7 & 9 & 2.3 & 7 & 6.2\phantom{0} $\pm$0.1\phantom{0}  & 7 & 32.16 & $\cdots$ & --7.63$\pm$0.1\phantom{0} & 1.1\phantom{0} & --0.24 & Single & 12 \\
c7410 & $\cdots$                   &    $\cdots$     &            $\cdots$               & 35N & 1026 & 716 & 41  & 76  &  1013  &    $\cdots$ &     1494       &     $\cdots$  & O3\,III+ & 25 & 1.9 & 7 & 5.83$\pm$0.2\phantom{0}  & 7 & 32.32 & $\cdots$ & --7.10$\pm$0.2\phantom{0} & 0.71 & --0.83 & Single & 28 \\ 
p1\_695 & $\cdots$                &     $\cdots$    &           $\cdots$                &    $\cdots$    &    $\cdots$      &  53 & 63 &105   & 761    &    $\cdots$ &2077 &  $\cdots$      & O3--6\,V &  31 & 2.2 & 26 & 5.87$\pm$0.15 & 26 & 32.27 &$\cdots$  & --7.19$\pm$0.15 & 1.6\phantom{0} & --0.45 &$\cdots$   & $\cdots$    \\ 
c6973 &  $\cdots$                   &      $\cdots$  &            $\cdots$               &37Wa& 1021 & 339 & 11 &  25    &      917     &    $\cdots$ & 1349 &      $\cdots$   & O4\,If$^{+}$ & 25 & 2.2 & 7 & 6.34$\pm$0.1\phantom{0}  & 7 & 32.26 &$\cdots$  & --7.67$\pm$0.1\phantom{0} & 0.49 & --0.86 & Single & 28 \\ 
c6443 & $\cdots$                    &     $\cdots$    &          $\cdots$                &   28a      &  $\cdots$ &  $\cdots$      &   $\cdots$   & 80 &   805    &    $\cdots$ & 2447   &  $\cdots$     & O5--6\,V & 34 & 1.8 & 26 & 5.53$\pm$0.05 & 26 & 32.34 & $\cdots$ & --6.78$\pm$0.05 & 1.0\phantom{0} & --0.79 & SB?   & 35  \\ 
c8112 & $\cdots$                    &     $\cdots$    &         $\cdots$                 & 12     &  591 &  $\cdots$ &  $\cdots$     & 11 & 1257   &    $\cdots$ &     1279     &  $\cdots$   & B0.2\,Ia & 36  & 1.8 & 26 & 5.91$\pm$0.15 & 37 & 32.28 & $\cdots$ & --7.22$\pm$0.15 & 0.76 & --0.86 & SB: & 36,38 \\
c7859 & $\cdots$                    &    $\cdots$     &         $\cdots$                 & 26    & 562  &  $\cdots$ &  $\cdots$    & 32  & 1150   &    $\cdots$ & 2819 &     $\cdots$     & O4\,V      & 39 & 1.6 & 7 & 6.05$\pm$0.2\phantom{0}  & 7 & 31.97 & $\cdots$ & --7.67$\pm$0.2\phantom{0} & 0.33 & --0.64 & $\cdots$   & $\cdots$   \\
c6370 &  $\cdots$                   & 134  &              $\cdots$           &    $\cdots$   & 1001 &   71 & $\cdots$   &   $\cdots$      &   786   &    $\cdots$ & 1978   & 100  & WN6(h) & 32 & 1.5 & 7 & 6.2\phantom{0} $\pm$0.1\phantom{0}  & 7 & 32.46 & $\cdots$ & --7.33$\pm$0.1\phantom{0} & 8.6\phantom{0} & +0.64 & Single & 12 \\ 
p1\_553 & $\cdots$                 & 133  &          $\cdots$               &   $\cdots$   & 333   & $\cdots$ & $\cdots$     &    $\cdots$     &    42       &    $\cdots$ &      $\cdots$    &       $\cdots$    & O9\,II+O9.2\,V & 3,42 &  0.8 & 7 & 5.88$\pm$0.1\phantom{0} & 5 & 32.20 & $\cdots$ & --7.27$\pm$0.1\phantom{0} & 0.63 & --0.95 & SB2 & 3,42 \\
c7907 & $\cdots$                    &    $\cdots$     &     $\cdots$                     & 23  & 566   &   $\cdots$ & $\cdots$    & 42   &1163    &    $\cdots$ &    $\cdots$      &      $\cdots$    & O3\,III(f*) & 3 & 1.3 & 7 & 5.83$\pm$0.1\phantom{0} & 5 & 32.31 & $\cdots$ & --7.11$\pm$0.1\phantom{0} & 0.38 & --0.96 & Single & 3 \\
 p1\_718 & $\cdots$                &   $\cdots$     &       $\cdots$                   &   $\cdots$    &    $\cdots$   & 108  & 51  & 89    & 812    &    $\cdots$ &     $\cdots$       &    $\cdots$  &  O3\,V & 25 & 1.5 & 2 & 5.71$\pm$0.3:\phantom{:}  & 2 & 32.16 & $\cdots$ & --7.14$\pm$0.3:\phantom{:} & 0.86 & --0.82 & $\cdots$   & $\cdots$   \\ 
p1\_766 &  $\cdots$.               &     $\cdots$     &       $\cdots$                  &    $\cdots$    & 513  &   $\cdots$ & $\cdots$  & 266  &      $\cdots$     &    $\cdots$ &      2799      &   $\cdots$   & O6--7\,II & 3 & 1.3 & 2 & 5.00$\pm$0.1\phantom{0} & 5 & 31.91 & $\cdots$ & --6.68$\pm$0.1\phantom{0} & 0.83 & --0.13 & Single & 3\\
p1\_672 & $\cdots$                 &     $\cdots$    &           $\cdots$              &     $\cdots$    & 455 &  $\cdots$ & $\cdots$     & 108   &   661  &    $\cdots$ & 1572 &    $\cdots$   & O5:V & 3 & 1.7 & 7 & 5.63$\pm$0.2\phantom{0} & 7 & 32.07 & $\cdots$ & --7.15$\pm$0.2\phantom{0} & 1.3\phantom{0} & --0.63 & SB1 & 3 \\
p1\_937 &  $\cdots$               &     $\cdots$    &          $\cdots$                & 35Sa & 1028 & 749 & 23  & 37     &   1036 &    $\cdots$ & 1423 &    $\cdots$  & O3\,III(f*) & 25 & 1.5 & 7 & 6.09$\pm$0.2\phantom{0}  & 7 & 32.43 & $\cdots$ & --7.25$\pm$0.2\phantom{0} & 0.31 & --1.00 & Single & 28 \\ 
cc5383 &  $\cdots$                &      $\cdots$      &        $\cdots$               &    $\cdots$     & 579  &   $\cdots$ & $\cdots$    & 171  &  1201 &    $\cdots$ & 3116  &   $\cdots$  & O9:((n)) & 3 & 1.1 & 2 & 5.46$\pm$0.02 & 5 & 31.89 & $\cdots$ & --7.16$\pm$0.02 & 1.3\phantom{0} & --0.17 & SB? & 3 \\
p1\_1062 & $\cdots$             &    $\cdots$        &        $\cdots$               & 14N &  601 &  986 &  $\cdots$    & 91  & 1317 &    $\cdots$ &  1890   &   $\cdots$    & O5--6\,V & 3 & 1.6 & 30 & 5.55$\pm$0.18 & 30 & 31.99 & $\cdots$ & --7.15$\pm$0.18 & 0.44 & --0.91 & Single & 3 \\
\hline
\end{tabular}
\end{center}
\end{table}
\end{landscape}

\addtocounter{table}{-1}

\begin{landscape}
\begin{table}
\begin{center}
\caption{(continued)}
\begin{tabular}{l@{\hspace{1mm}}c@{\hspace{1mm}}l@{\hspace{1mm}}c@{\hspace{1mm}}l@{\hspace{1mm}}r@{\hspace{2mm}}r@{\hspace{1mm}}r@{\hspace{1mm}}r@{\hspace{1mm}}r@{\hspace{1mm}}r@{\hspace{1mm}}r@{\hspace{1mm}}r@{\hspace{1mm}}r@{\hspace{2mm}}l@{\hspace{1mm}}r@{\hspace{2mm}}l@{\hspace{2mm}}
l@{\hspace{2mm}}l@{\hspace{1mm}}l@{\hspace{2mm}}c@{\hspace{1mm}}c@{\hspace{2mm}}c@{\hspace{1mm}}c@{\hspace{1mm}}l@{\hspace{1mm}}l@{\hspace{1mm}}l}
\hline
T-ReX & HD      & R    & Sk  & Mk  & VFTS & MH & HSH & SMB & P & ST & CCE & BAT & Spectral & Ref & $A_{\rm V}$ & Ref & $\log L_{\rm Bol}$  & Ref & $\log L_{\rm X}$ & Var? & $\log L_{\rm X}/L_{\rm Bol}$ & $kT_{\rm m}$ & Hardness & Nature & Ref\\
Label &            &       &       &        &         &         &      &             &         &             &               &              & Type     &        &  mag           &        & $L_{\odot}$            &        & erg\,s$^{-1}$      & &                                              &      keV &  $\eta_{2}$          &             & \\
\hline
c5984 &  $\cdots$                  &    $\cdots$        &         $\cdots$             &   $\cdots$      &  $\cdots$ & $\cdots$     &     $\cdots$   &   $\cdots$     & 649 &    $\cdots$ &       $\cdots$     &    $\cdots$     & O8--9\,V & 34 & 1.2 & 2 & 5.00$\pm$0.3:\phantom{:}  & 2 & 32.00 & $\cdots$ & --6.59$\pm$0.3:\phantom{:} & 3.2\phantom{0} & +0.00 & $\cdots$   & $\cdots$    \\
c5572 & $\cdots$                  &   $\cdots$    &          $\cdots$                  &     $\cdots$   & 416   & $\cdots$   &    $\cdots$   & 93    & 467    &    $\cdots$  & 1700  &   $\cdots$   & O8.5\,V             & 29 & 1.8 & 26 & 5.66$\pm$0.05 & 26   & 32.13 & $\cdots$ & --7.12$\pm$0.05 & 0.88 & --0.87 &$\cdots$   & $\cdots$    \\
c5987 & $\cdots$                  &   $\cdots$   &            $\cdots$                & 50  & 450      &  $\cdots$ & $\cdots$      & 34     & 643    &    $\cdots$ & 1293   &  $\cdots$  & O9.7\,III:+O7:: & 3 & 1.6 & 26 & 5.69$\pm$0.1\phantom{0}  & 17 & 32.04 & $\cdots$ & --7.23$\pm$0.1\phantom{0}& 1.1\phantom{0} & --0.77 & SB2 & 3,17 \\
c6213 & $\cdots$                  &   $\cdots$    &           $\cdots$                 &   $\cdots$     &     $\cdots$  & $\cdots$     &    $\cdots$    & 151   & 743  &    $\cdots$  & 1523  &   $\cdots$  & O6:             & 26 & 1.6 & 26 & 5.00$\pm$0.29                   & 26 & 32.02 &  $\cdots$ & --6.57$\pm$0.29 & 0.70 & --0.85 & EB & 40\\ 
c3084 & $\cdots$                  &    $\cdots$    &          $\cdots$                 &     $\cdots$   & 186     &   $\cdots$ &  $\cdots$   &    $\cdots$   &    $\cdots$    &        $\cdots$   &     $\cdots$      &   $\cdots$   & B1\,IV          & 36 & 1.3 & 26 & 4.45$\pm$0.3:\phantom{:}     & 2 & 32.26 &$\cdots$  & --5.78$\pm$0.3:\phantom{:} & 0.63 & --0.95 & Single  & 36 \\
c6472 & $\cdots$                   &   $\cdots$   &             $\cdots$               &     $\cdots$   &    $\cdots$  & 116      & 95   & 155   &    $\cdots$ &   $\cdots$     & 1920   &     $\cdots$  & O4:        & 26 & 1.8 & 26 & 6.00$\pm$0.15 & 26  & 32.00 & $\cdots$ & --7.59$\pm$0.15 & 1.2\phantom{0} & --0.70 & $\cdots$   & $\cdots$   \\
p1\_1021 & $\cdots$               &    $\cdots$  &             $\cdots$              &     $\cdots$   & 564    & $\cdots$ &   $\cdots$     & 253  &  $\cdots$     &    $\cdots$     & 3200  &     $\cdots$   & O6--8\,V & 3 & 2.6 & 30 & 5.33$\pm$0.13 & 30 & 32.01 & $\cdots$ & --6.91$\pm$0.13 & 1.0\phantom{0} & --0.58 & Single & 3 \\
c10271 & $\cdots$                 &    $\cdots$   &             $\cdots$              &    $\cdots$     & 755   &  $\cdots$ &   $\cdots$     &     $\cdots$   & 2041 &    $\cdots$ &     $\cdots$       &   $\cdots$    & O3\,Vn((f*)) & 3 & 1.6 & 7 & 5.65$\pm$0.2\phantom{0}  & 30 &  32.00 & $\cdots$ & --7.24$\pm$0.2\phantom{0} & 0.90 & --0.86 & Single & 3 \\
cc4651 & $\cdots$                 &    $\cdots$    &            $\cdots$              &    $\cdots$      &     $\cdots$ & 25   & 157 & 255 &       $\cdots$    &    $\cdots$ & 1858   &     $\cdots$  & O7: & 26 & 1.8 & 26 & 4.79$\pm$0.06 & 26 & 32.10 & $\cdots$ & --6.28$\pm$0.06 & 2.7\phantom{0} & --0.08 &  $\cdots$   & $\cdots$     \\
c8320 &  $\cdots$                 &     $\cdots$   &             $\cdots$             &      $\cdots$    & 621 & $\cdots$ & $\cdots$       &      $\cdots$  & 1429  &    $\cdots$ &    $\cdots$       &   $\cdots$    & O2\,V((f)*)z & 3 & 3.0 & 7 & 6.22$\pm$0.2\phantom{0}  & 30 & 31.88 & $\cdots$ & --7.93$\pm$0.2\phantom{0} & 1.1\phantom{0} & --0.25 & V. Comp & 3\\
c7031 &  $\cdots$                   &     $\cdots$   &         $\cdots$               &     $\cdots$    & 526  &  $\cdots$ &  $\cdots$      &     $\cdots$   & 925    &    $\cdots$ &   $\cdots$       &    $\cdots$  & O8.5\,I(n)fp+? & 3 & 2.9 & 2 & 5.71$\pm$0.3:\phantom{:}  & 2 & 31.93 & $\cdots$ & --7.37$\pm$0.3:\phantom{:} & 1.7\phantom{0}& +0.33 & SB1 & 3\\
c6170 & $\cdots$                         & $\cdots$         & $\cdots$     & $\cdots$ &  $\cdots$ & 28          &  138          &  144                & 729                    & $\cdots$          & 1535         & $\cdots$       & O6--7 & 24 &1.8 & 26 & 5.34$\pm$0.07 & 26 & 31.98 & $\cdots$ & --6.95$\pm$0.07 & 1.2\phantom{0} & --0.66 & $\cdots$ & $\cdots$ \\ 
c7552 & $\cdots$                         & $\cdots$         & $\cdots$     & $\cdots$ &  1030      & $\cdots$ &  $\cdots$   &  335                & 1043                 & $\cdots$          & $\cdots$     & $\cdots$.     & O9\,V & 41 &1.4 & 2   & 4.56$\pm$0.3:\phantom{:} & 2 & 31.83 & $\cdots$ & --6.32$\pm$0.3: & 0.92 & --0.77 & $\cdots$ & $\cdots$  \\     
p1\_1530 & $\cdots$                &    $\cdots$    &        $\cdots$                 &    $\cdots$     & 830  &  $\cdots$ &  $\cdots$     &     $\cdots$   & 2270   &    $\cdots$ &   $\cdots$       &     $\cdots$   & O5--6\,V     & 3 & 1.0 & 2 & 5.05$\pm$0.3:\phantom{:}  & 2 & 32.00 & $\cdots$ & --6.64$\pm$0.3:\phantom{:} & 0.72 & --0.89 & SB & 3 \\   
c8591     & $\cdots$                &    $\cdots$    &        $\cdots$                 &  8   & 648    &  $\cdots$ &  $\cdots$     & 58   & 1531   &    $\cdots$ & 2780   &   $\cdots$  & O5.5\,IV(f) & 3 & 1.3 & 30 & 5.66$\pm$0.13 & 30 & 31.88 & $\cdots$ & --7.37$\pm$0.13 & 0.86 & --0.86 & SB & 3 \\
 c8052    &  $\cdots$                &    $\cdots$    &       $\cdots$                  &   $\cdots$    & 585    &  943 &  $\cdots$     & 31   & 1231    &    $\cdots$ & 2193  &  $\cdots$   & O7\,V(n) & 3 & 1.4 & 26 & 5.61$\pm$0.25 & 26 & 31.77 & $\cdots$ & --7.43$\pm$0.25  & 0.73 & --0.77      & SB1 & 3 \\ 
c7254 &  $\cdots$                     &     $\cdots$   &       $\cdots$                 &    $\cdots$   & 532    &  $\cdots$ & $\cdots$      & 104    & 974     &    $\cdots$ & 963   &  $\cdots$    & O3.5:\,V(n)+B\,III-I & 3,42 & 1.8 & 7 & 5.74$\pm$0.2\phantom{0} & 7 & 31.92 & $\cdots$ & --7.41$\pm$0.2\phantom{0} & 0.34 & --0.93 & SB2 & 3,42\\
c8204 &  $\cdots$                     &    $\cdots$   &          $\cdots$              & 14  & 608    & 999 &  $\cdots$       & 64        & 1350   &    $\cdots$ & 1827   &    $\cdots$   & O4\,III(f)     & 3 & 1.7 & 7 & 5.86$\pm$0.1\phantom{0}  & 7    & 32.11 & $\cdots$ & --7.34$\pm$0.1\phantom{0} & 0.58 & --0.93 & SB1 & 3 \\ 
c3981 &  $\cdots$                     &    $\cdots$     &         $\cdots$             &    $\cdots$   & 259   & $\cdots$  &   $\cdots$     &     $\cdots$     &    $\cdots$ &      $\cdots$          &     $\cdots$       &     $\cdots$      & O6\,Iaf & 3 & 2.1 & 7 & 6.1\phantom{0}$\pm$0.1\phantom{0}  & 7 & 31.80 &$\cdots$  & --7.89$\pm$0.1\phantom{0} & 0.39 & --0.87 & SB1 & 3 \\
 c8180 &  $\cdots$                   &  $\cdots$     &     $\cdots$                   & 10  & 603  & $\cdots$ &  $\cdots$    &    $\cdots$         & 1341       &    $\cdots$ &   $\cdots$       &       $\cdots$    & O4\,III:+OB: & 3,42 & 1.8 & 7 & 5.98$\pm$0.1\phantom{0}  & 7 & 31.95 & $\cdots$ & --7.62$\pm$0.1\phantom{0} & 0.27 & --0.98 & SB2 & 3,42 \\
c5374 & $\cdots$             &   $\cdots$    &       $\cdots$               & 55  &  406 &  $\cdots$ &  $\cdots$       &     $\cdots$      & 370         &    $\cdots$ &    $\cdots$     &      $\cdots$    & O6\,Vnn & 3& 1.2 & 7 & 5.48$\pm$0.2\phantom{0}  & 7 & 31.74 & $\cdots$ & --7.33$\pm$0.2\phantom{0} & 0.79 & --0.87 & Single & 3 \\ 
c7235 &  $\cdots$                    &    $\cdots$    &        $\cdots$               &   $\cdots$    & $\cdots$ &     $\cdots$   &    $\cdots$      & 71      & 975         &    $\cdots$ & 2748   &   $\cdots$     & O6--7\,V & 34 &1.6 & 26 & 5.58$\pm$0.25 & 26 & 31.88 & $\cdots$ & --7.29$\pm$0.25 & $\cdots$ & $\cdots$ & Single  & 35  \\
c7528 &  $\cdots$                    &     $\cdots$    &       $\cdots$                &   $\cdots$   & 1029 & 764 & 56   & 87      &  1042      &    $\cdots$ & 2128  &   $\cdots$   & O3.5\,I+OB & 41 & 1.2 & 26 & 5.59$\pm$0.3: & 2 & 31.72 & $\cdots$ &--7.46$\pm$0.3: & 0.44 &  --0.88 & SB? & 28 \\ 
cc4601 &   $\cdots$                  &   $\cdots$     &        $\cdots$               &     $\cdots$ & 460   & $\cdots$ &  $\cdots$     & 278    &  674   &    $\cdots$ & 290     &  $\cdots$   & O7.5\,V+O7.5\,V & 3 & 2.6 & 26 & 5.37$\pm$0.05  & 26  & 31.72 & $\cdots$ & --7.24$\pm$0.05 & 1.0\phantom{0} & --0.67 & SB2 & 3,35 \\
c6094 &  $\cdots$                     &     $\cdots$     &      $\cdots$                &   $\cdots$   &  465  &   $\cdots$ &  $\cdots$   & 365    &  700   &    $\cdots$ & 830     &   $\cdots$  & On             & 3 & 3.2 & 26 & 5.57$\pm$0.10 & 5      & 31.71 & $\cdots$ & --7.45$\pm$0.10 & 1.0\phantom{0} & +0.03 & Single & 3 \\
c6164 &  $\cdots$                     &     $\cdots$     &          $\cdots$            &    $\cdots$  &     $\cdots$  & $\cdots$   &  $\cdots$      & 75      &  724   &    $\cdots$ & 1274   &  $\cdots$    & O7\,III & 34 & 1.8 & 26 & 5.74$\pm$0.25 & 26 & 31.67 & $\cdots$ & --7.66$\pm$0.25 & 1.3\phantom{0} & --0.62 & $\cdots$   & $\cdots$    \\
c5695 &  $\cdots$                     &      $\cdots$     &        $\cdots$             &   $\cdots$   &     $\cdots$     &    $\cdots$  & $\cdots$ & 82      &  $\cdots$         &    $\cdots$ & 1632    &   $\cdots$    & O8: & 26 & 1.4 & 26 & 5.17$\pm$0.05 & 26 & 31.77 & $\cdots$ & --6.99$\pm$0.05 & 0.66 & --0.86& $\cdots$   & $\cdots$    \\
c10098 &  $\cdots$                    &      $\cdots$    &          $\cdots$            &    $\cdots$  &  746  &  $\cdots$ & $\cdots$    &    $\cdots$       &   $\cdots$        &    $\cdots$ &      $\cdots$      &   $\cdots$    & O6\,Vnn & 3 & 1.7 & 30 & 5.29$\pm$0.24 & 30 & 31.59 & $\cdots$ & --7.29$\pm$0.24 & 0.27 & --0.91 & Single & 3 \\
c5200 & $\cdots$                        &     $\cdots$     &     $\cdots$                &    $\cdots$  & 385    &   $\cdots$ & $\cdots$   & 84     & 288    &    $\cdots$ & 2451    &   $\cdots$   & O4--5\,V     & 3 & 1.3 & 30 & 5.55$\pm$0.29 & 30 & 32.00 & $\cdots$ & --7.14$\pm$0.29 & 0.32 &  --1.00 & SB & 3 \\
c5633 &  $\cdots$                        &      $\cdots$   &      $\cdots$               & 54 & 420    & $\cdots$ &  $\cdots$    & 16     & 488    &    $\cdots$ & 1689    &  $\cdots$    & B0.5\,Ia & 36 & 2.6 & 26 & 5.84$\pm$0.15 & 37 & 31.94 & $\cdots$ & --7.49$\pm$0.15 & $\cdots$ & $\cdots$ & Single & 36 \\
cc7769 & $\cdots$                        &     $\cdots$   &       $\cdots$              &   $\cdots$  & 768    & $\cdots$ &    $\cdots$  &       $\cdots$    &    $\cdots$ &  $\cdots$         &     $\cdots$       &    $\cdots$   & O8\,Vn         & 3 & 2.3 & 30 & 5.09$\pm$0.22 & 30 & 31.59 & $\cdots$ & --7.09$\pm$0.22 & $\cdots$ & $\cdots$ & SB2? & 3 \\
cc4200 & $\cdots$                        &     $\cdots$    &    $\cdots$                &     $\cdots$ & 411   & $\cdots$ &   $\cdots$   & 409    &    $\cdots$       &    $\cdots$ &      $\cdots$      &     $\cdots$  & B1--3\,V--III     & 36 & 0.5 & 2 & 3.57$\pm$0.4:\phantom{:}  & 2 & 31.29 & $\cdots$ & --5.87$\pm$0.4:\phantom{:} & $\cdots$ & $\cdots$ & Single &  36 \\
 cc4181 &  $\cdots$                     &      $\cdots$   &   $\cdots$                &  $\cdots$  & 410    &   $\cdots$ &  $\cdots$   & 288    &     409  &    $\cdots$ & 1864 &   $\cdots$    & O7--8\,V         & 3 & 2.4 & 30 & 5.14$\pm$0.13 & 30 &31.61 & $\cdots$ & --7.12$\pm$0.13 & $\cdots$ & $\cdots$ & V. Comp & 3 \\
p1\_1274 & $\cdots$                   &      $\cdots$     &      $\cdots$              &  $\cdots$ &   703 & $\cdots$ & $\cdots$       &         $\cdots$ & 1828    &    $\cdots$ &    $\cdots$       &    $\cdots$    & O7:\,V+O8:\,V & 3 & 2.1 & 2 & 4.76$\pm$0.3:\phantom{:}  & 2 & 31.36 & $\cdots$ & --6.99$\pm$0.3:\phantom{:} & $\cdots$ & $\cdots$  & SB2 & 3 \\
cc4294 &  $\cdots$                      &      $\cdots$    &     $\cdots$                &  $\cdots$ & 417    & $\cdots$ & $\cdots$      & 204   &     $\cdots$        &    $\cdots$ & 1399   &   $\cdots$  & B2\,Ib    &  36  & 1.7 & 2 & 4.51$\pm$0.15 & 37 & 30.97 & $\cdots$ & --7.13$\pm$0.15 & $\cdots$ & $\cdots$  & Single & 36 \\
cc5800  & $\cdots$                       &        $\cdots$  &      $\cdots$              & $\cdots$   & 640    & $\cdots$ & $\cdots$     &     $\cdots$     &1489    &    $\cdots$ &       $\cdots$     &    $\cdots$    & B2\,V      & 36 & 1.4 & 2 & 4.03$\pm$0.3:\phantom{:}  & 2 &31.69 & $\cdots$ & --5.93$\pm$0.3:\phantom{:} & $\cdots$ & $\cdots$ & Single & 36 \\
cc7873 & $\cdots$                        &      $\cdots$    &     $\cdots$&  $\cdots$  & 774   & $\cdots$   & $\cdots$    &  $\cdots$        &   $\cdots$          &  $\cdots$        &    $\cdots$ &     $\cdots$    & O7.5\,IVp+O9.5:V & 3 & 2.8 & 2 & 4.97$\pm$0.3:\phantom{:} & 2 & 31.04 &$\cdots$  & --7.52$\pm$0.3:\phantom{:} & $\cdots$ & $\cdots$   & SB2 & 3 \\
 \hline
\end{tabular}
\end{center}
{\footnotesize
1: \citet{2019MNRAS.484.2692T} 
2: \citet{2013A&A...558A.134D} 
3:  \citet{2014A&A...564A..40W} 
4: \citet{2015A&A...579A.131C} 
5: \citet{2017A&A...600A..81R} 
6: \citet{2010MNRAS.408..731C} 
7: \citet{2014A&A...570A..38B} 
8: \citet{2009MNRAS.397.2049S} 
9: \citet{2011MNRAS.416.1311C} 
10: A.M.T.~Pollock et al. (in prep) 
11: \citet{2005ApJ...627..477M} 
12: \citet{2008MNRAS.389..806S} 
13: \citet{2016MNRAS.458..624C} 
14: \citet{2020MNRAS.499.1918B} 
15: \citet{2012A&A...543A..95R} 
16: \citet{2019A&A...627A.151S} 
17: \citet{2020A&A...634A.118M} 
18: \citet{2017A&A...598A..84A} 
19: \citet{2022MNRAS.510.6133B} 
20: \citet{2021A&A...650A.147S} 
21: \citet{2017A&A...598A..85S} 
22: \citet{2011A&A...530A.108E} 
23: \citet{2003MNRAS.338.1025F} 
24: \citet{2002ApJ...565..982M} 
25: \citet{1998ApJ...493..180M} 
26: \citet{2021A&A...648A..65C} 
27: \citet{2011A&A...530L..14B} 
28: \citet{2012A&A...546A..73H} 
29: \citet{2012ApJ...748...96M} 
30: \citet{2017A&A...601A..79S} 
31: \citet{1997ApJS..112..457W} 
32: \citet{1997A&A...320..500C} 
33: \citet{2014A&A...565A..27H} 
34: \citet{1999A&AS..137...21B} 
35: \citet{2009AJ....137.3437B} 
36: \citet{2015A&A...574A..13E} 
37: \citet{2015A&A...575A..70M} 
38 :\citet{2021MNRAS.507.5348V} 
39: \citet{1985A&A...153..235M} 
40: \citet{2011AcA....61..103G} 
41: D.J. Lennon (priv. comm.), 
42: T.~Shenar et al. submitted


}
\end{table}
\end{landscape}

\section{LMC baseline abundances}

\begin{landscape}
\begin{table}
\caption{LMC abundances (X/H by number) adopted for XSPEC fitting with respect to $Z_{\odot}$ from \citet[AGS09]{2009ARA&A..47..481A}. 
Values shown in parentheses are excluded from the average, due to issues with enhancements owing to mixing (e.g. N), depletion on dust grains (Fe) or other concerns. 0.5 $Z_{\odot}$ is adopted for Co}
\label{abundances}
\begin{center}
\begin{footnotesize}
\begin{tabular}{
l@{\hspace{1mm}} 
c@{\hspace{2mm}} 
c@{\hspace{1mm}}c@{\hspace{1mm}}c@{\hspace{1mm}}c@{\hspace{2mm}} 
c@{\hspace{1mm}}c@{\hspace{1mm}}c@{\hspace{2mm}}  
c@{\hspace{1mm}}c@{\hspace{2mm}} 
c@{\hspace{2mm}} 
c@{\hspace{2mm}} 
c@{\hspace{2mm}}c} 
\hline
Element & SNR& \multicolumn{4}{c}{{---}{---}{---} H\,{\sc ii} regions {---}{---}{---}} & \multicolumn{3}{c}{{---}{---} B stars {---}{---}} & \multicolumn{2}{c}{--- F supergiants --- } & \multicolumn{2}{c}{<LMC>}   & Solar   & <LMC>\\
              & D19 & G99 & P03/SCT17 & T03 & L08   & K00/K05 & T07/H07    & D18  & HAS95 & A01 & X/H &  logX/H+12 & AGS09 &$Z_{\odot}$ \\
\hline
He & $\cdots$ & $\cdots$ & 8.5$\times 10^{-2}$ & $\cdots$ & $\cdots$ & 8.1$\times 10^{-2}$ & $\cdots$ & $\cdots$ &  $\cdots$ & $\cdots$ & 8.3$\times 10^{-2}$ & 10.92 & 8.5$\times 10^{-2}$ & 1.0\phantom{0} \\
C & 1.2$\times 10^{-4}$ & 7.9$\times 10^{-5}$ & 1.0$\times 10^{-4}$ & $\cdots$ & $\cdots$ & 9.5$\times 10^{-5}$ & $\cdots$ & $\cdots$ & 1.2$\times 10^{-4}$ & 8.9$\times 10^{-5}$ & 1.0$\times 10^{-4}$ & 8.01 & 2.7$\times 10^{-4}$  & 0.38 \\
N & (1.6$\times 10^{-5}$) & 7.9$\times 10^{-6}$ & 1.6$\times 10^{-5}$ & $\cdots$ & $\cdots$ & 1.0$\times 10^{-5}$   & 7.9$\times 10^{-6}$ & $\cdots$ & $\cdots$ & $\cdots$ & 1.1$\times 10^{-5}$ & 7.03 & 6.8$\times 10^{-5}$ & 0.16 \\
O & 2.2$\times 10^{-4}$ & 2.5$\times 10^{-5}$ & 3.2$\times 10^{-4}$ & 2.2$\times 10^{-4}$ & $\cdots$ & 2.5$\times 10^{-4}$ & 2.3$\times 10^{-4}$ & $\cdots$ & (2.8$\times 10^{-4}$) & (3.2$\times 10^{-4}$) & 2.5$\times 10^{-4}$ & 8.40 & 4.9$\times 10^{-4}$ & 0.51 \\
Ne & 4.0$\times 10^{-5}$ &  4.0$\times 10^{-5}$ & 6.8$\times 10^{-5}$ & 4.6$\times 10^{-5}$ & 5.8$\times 10^{-5}$ & $\cdots$ & $\cdots$ & $\cdots$ & $\cdots$ & $\cdots$ & 5.0$\times 10^{-5}$ & 7.70 & 8.5$\times 10^{-5}$ & 0.59 \\
Na & $\cdots$ & $\cdots$ & $\cdots$ & $\cdots$ & $\cdots$ & $\cdots$ & $\cdots$ & $\cdots$ & 1.5$\times 10^{-6}$ & 7.9$\times 10^{-7}$ & 1.2$\times 10^{-6}$ & 6.1\phantom{0} & 1.7$\times 10^{-6}$  & 0.7\phantom{0} \\
Mg & 1.5$\times 10^{-5}$ & $\cdots$ & $\cdots$ &  $\cdots$ & $\cdots$ & 1.4$\times 10^{-5}$ & 1.2$\times 10^{-5}$ & 1.0$\times 10^{-5}$ & 1.6$\times 10^{-5}$ & $\cdots$ & 1.3$\times 10^{-5}$ & 7.13 & 4.0$\times 10^{-5}$ & 0.34 \\
Al & $\cdots$ & $\cdots$ & $\cdots$ &  $\cdots$ & $\cdots$ & 7.2$\times 10^{-7}$ & $\cdots$ & $\cdots$ & 2.2$\times 10^{-6}$ & $\cdots$ & 1.5$\times 10^{-6}$ & 6.2\phantom{0} & 2.8$\times 10^{-6}$  & 0.5\phantom{0} \\
Si & 1.3$\times 10^{-5}$ &   5.0$\times 10^{-6}$ & $\cdots$ & $\cdots$ & $\cdots$ & 1.2$\times 10^{-5}$ & 1.6$\times 10^{-5}$ & $\cdots$ & (3.7$\times 10^{-5}$)  & $\cdots$ & 1.1$\times 10^{-5}$ & 7.06 & 3.2$\times 10^{-5}$ & 0.35 \\
S & 9.5$\times 10^{-6}$ &  5.0$\times 10^{-6}$ & 9.8$\times 10^{-6}$ & 5.9$\times 10^{-6}$ & 5.9$\times 10^{-6}$ & $\cdots$ & $\cdots$ & $\cdots$ & $9.9\times 10^{-6}$ & $\cdots$ & 7.7$\times 10^{-6}$ & 6.88 & 1.3$\times 10^{-5}$ & 0.58 \\
Cl & 9.1$\times 10^{-8}$ &  $\cdots$  & 6.6$\times 10^{-8}$ & $\cdots$ & $\cdots$ & $\cdots$ & $\cdots$ & $\cdots$ & $\cdots$ & $\cdots$ & 7.9$\times 10^{-8}$ & 4.9\phantom{0} & 3.2$\times 10^{-7}$ &  0.2\phantom{0} \\ 
Ar & 6.2$\times 10^{-7}$ &   1.6$\times 10^{-6}$ & 1.8$\times 10^{-6}$ & 1.4$\times 10^{-6}$ & 2.1$\times 10^{-6}$ & $\cdots$ & $\cdots$ & $\cdots$ & $\cdots$ & $\cdots$ & 1.5$\times 10^{-6}$ & 6.2\phantom{0} & 2.5$\times 10^{-6}$ & 0.6\phantom{0} \\
Ca & 1.0$\times 10^{-6}$ & $\cdots$ & $\cdots$ & $\cdots$ & $\cdots$ & $\cdots$ & $\cdots$ & $\cdots$ & 1.6$\times 10^{-6}$ & $\cdots$ & 1.3$\times 10^{-6}$ & 6.1\phantom{0} & 2.2$\times 10^{-6}$  & 0.6\phantom{0} \\
Cr & 2.2$\times 10^{-7}$ & $\cdots$ & $\cdots$ & $\cdots$ & $\cdots$ & $\cdots$ & $\cdots$ & $\cdots$ & 2.6$\times 10^{-7}$ & $\cdots$ & 2.4$\times 10^{-7}$ & 5.4\phantom{0} & 4.4$\times 10^{-7}$  & 0.6\phantom{0} \\
Fe & 2.1$\times 10^{-5}$ &  $\cdots$ & (2.5$\times 10^{-6}$)  & $\cdots$ & (1.7$\times 10^{-6}$) & 1.2$\times 10^{-5}$ & 1.7$\times 10^{-5}$ & $\cdots$ & 1.7$\times 10^{-5}$ & $\cdots$ & 1.7$\times 10^{-5}$ & 7.23 & 3.2$\times 10^{-5}$ & 0.54 \\
Ni & 7.9$\times 10^{-7}$  & $\cdots$ & $\cdots$ & $\cdots$ & $\cdots$ & $\cdots$ & $\cdots$ & $\cdots$ & 9.2$\times 10^{-7}$ & $\cdots$ & 8.6$\times 10^{-7}$ & 5.9\phantom{0} & 1.7$\times 10^{-6}$  & 0.5\phantom{0} \\
\hline
\end{tabular}
\end{footnotesize}
\end{center} 
HAS95: \citet{1995A&A...293..347H};
G99: \citet{1999IAUS..190..266G}; 
K00: \citet{2000A&A...353..655K} (He/Al/Fe only); 
K05: \citet{}; (excl He/Al/Fe) 
A01: \citet{2001A&A...367..605A}; 
P03: \citet{2003ApJ...584..735P} (excl. C/O);
T03: \citet{2003MNRAS.338..687T}; 
T07: \citet{2007A&A...471..625T}; 
H07: \citet{2007A&A...466..277H}; 
L08: \citet{2008ApJ...680..398L}; 
SCT17: \citet{2017MNRAS.467.3759T} (C/O recomb lines for 30~Dor); 
D18: \citet{2018A&A...615A.101D}; 
D19: \citet{2019AJ....157...50D} 
\end{table}
\end{landscape}


\setcounter{section}{19}
\addtocounter{table}{-1}

\begin{landscape}
\begin{table}
\begin{center}
\caption{Upper limits to observed X-ray luminosities $L^{t}_{\rm X}$ for luminous ($\log L/L_{\odot} \geq 5$) early-type stars excluded from the T-ReX point source catalogue, in RA order. Conversions from photon flux to attenuated luminosity are obtained from early-type stars in the T-ReX PSC, namely $\log L_{\rm X}^{t}$ = $\log$ (PhotonFlux/cm$^{-2}$\,s$^{-1}$) + (38.81$\pm$0.16). Spectral type calibrations have been applied in some instances following \citet{2013A&A...558A.134D}, noted with ":". Limits on intrinsic X-ray luminosities are estimated from average attenuation corrections of 0.31$\pm$0.17 dex. Catalogues include 
R \citep{1960MNRAS.121..337F}, 
Sk \citep{1970CoTol..89.....S}, 
BI \citep{1975A&AS...21..109B}, 
Mk  \citep{1985A&A...153..235M}, 
VFTS \citep{2011A&A...530A.108E}, 
MH \citep{1994AJ....107.1054M},
HSH \citep{1995ApJ...448..179H}, 
SMB \citep{1999A&A...341...98S}, 
P \citep{1993AJ....106..560P},  
ST \citep{1992A&AS...92..729S},
CCE \citep{2018A&A...614A.147C}, 
BAT \citep{1999A&AS..137..117B}. T-ReX PSC sources in close proximity ($\sim$1-2$''$) to the star $\ddag$.}
\label{A3}
\begin{tabular}{
c@{\hspace{1mm}}c@{\hspace{3mm}} 
r@{\hspace{1mm}} 
c@{\hspace{1mm}}r@{\hspace{1mm}} 
r@{\hspace{1mm}}r@{\hspace{2mm}} 
r@{\hspace{1mm}}r@{\hspace{1mm}}r@{\hspace{1mm}} 
r@{\hspace{1mm}}r@{\hspace{1mm}} 
r@{\hspace{1mm}}r@{\hspace{1mm}}  
r@{\hspace{1mm}}l@{\hspace{2mm}} 
c@{\hspace{1mm}}l@{\hspace{2mm}} 
c@{\hspace{2mm}}c@{\hspace{2mm}} 
c@{\hspace{2mm}}  
l@{\hspace{1mm}}l@{\hspace{1mm}}r} 
\hline
RA           & Dec            &  R     & Sk  & BI  & Mk & VFTS & MH & HSH & SMB & P & ST & CCE & BAT & Spect. & Ref & $\log L_{\rm Bol}$& Ref & Photon Flux   & $\log L^{t}_{\rm X}$  &$\log L^{tc}_{\rm X}/L_{\rm Bol}$ & Nature & Ref & PSC\\
h m s       & deg m s         &       &       &       &          &    &          &              &         &             &            &           &     & Type  &       &  $L_{\odot}$              &       & cm$^{-2}$\,s$^{-1}$ & erg\,s$^{-1}$ &             &    &    & source$\ddag$ \\
\hline
05 36 55.24 & --69 11 37.6 & $\cdots$ & --69$^{\circ}$ 224 & $\cdots$ & $\cdots$ & 3 & $\cdots$ &$\cdots$ & $\cdots$ &  $\cdots$ & $\cdots$ & $\cdots$ & $\cdots$ & B1\,Ia+ & 1 & 6.03$\pm$0.10 & 2-3 &
\multicolumn{3}{c}{Variable background (survey field edge)} & $\cdots$ & $\cdots$ & $\cdots$\\ 
05 37 07.51 & --69 05 06.9 & $\cdots$ & $\cdots$ & $\cdots$ & $\cdots$ & 14 & $\cdots$ &$\cdots$ & $\cdots$ & $\cdots$ & $\cdots$ & $\cdots$ & $\cdots$ & O8.5\,Vz & 4 & 5.16$\pm$0.16 & 3,5 & $<$2.41$\times 10^{-7}$ & $<$32.19 & $<$--6.25$\pm$0.28 & SB & 4 & $\cdots$\\
\hline
\end{tabular}
\end{center}
\end{table}
\end{landscape}

\addtocounter{table}{-1}

\begin{landscape}
\begin{table}
\begin{center}
\caption{(continued)}
\begin{tabular}{
c@{\hspace{1mm}}c@{\hspace{3mm}} 
r@{\hspace{1mm}} 
c@{\hspace{1mm}}r@{\hspace{1mm}} 
r@{\hspace{1mm}}r@{\hspace{2mm}} 
r@{\hspace{1mm}}r@{\hspace{1mm}}r@{\hspace{1mm}} 
r@{\hspace{1mm}}r@{\hspace{1mm}} 
r@{\hspace{1mm}}r@{\hspace{1mm}}  
r@{\hspace{1mm}}l@{\hspace{2mm}} 
c@{\hspace{1mm}}l@{\hspace{2mm}} 
c@{\hspace{2mm}}c@{\hspace{2mm}} 
c@{\hspace{2mm}}  
l@{\hspace{1mm}}l@{\hspace{1mm}}r} 
\hline
RA           & Dec            &  R     & Sk  & BI  & Mk & VFTS & MH & HSH & SMB & P & ST & CCE & BAT & Spect. & Ref & $\log L_{\rm Bol}$& Ref & Photon Flux   & $\log L^{t}_{\rm X}$  &$\log L^{tc}_{\rm X}/L_{\rm Bol}$ & Nature & Ref & PSC\\
h m s       & deg m s         &       &       &       &          &    &          &              &         &             &            &           &     & Type  &       &  $L_{\odot}$              &       & cm$^{-2}$\,s$^{-1}$ & erg\,s$^{-1}$ &             &    &    & source$\ddag$ \\
\hline

05 37 46.08& --69 09 08.8 & $\cdots$ & $\cdots$ & $\cdots$ & $\cdots$ & 145 & $\cdots$ &$\cdots$ & $\cdots$ & $\cdots$ & $\cdots$ & $\cdots$ & $\cdots$ & O8fp & 4 & 5.63$\pm$0.03 & 10&  \multicolumn{3}{c}{Variable background (N157B SNR)} & SB1 & 4 & $\cdots$\\

05 37 46.19  & --69 09 10.4 & $\cdots$ & $\cdots$ & $\cdots$ & $\cdots$ & 147 & $\cdots$ &$\cdots$ & $\cdots$ & $\cdots$ & $\cdots$ & $\cdots$ & 91 & WN6(h) & 20 & 5.83$\pm$0.2\phantom{0} & 14 &  \multicolumn{3}{c}{Variable background (N157B SNR)} & Single & 15& $\cdots$ \\
\hline
\end{tabular}
\end{center}
\end{table}
\end{landscape}

\addtocounter{table}{-1}

\begin{landscape}
\begin{table}
\begin{center}
\caption{(continued)}
\begin{tabular}{
c@{\hspace{1mm}}c@{\hspace{3mm}} 
r@{\hspace{1mm}} 
c@{\hspace{1mm}}r@{\hspace{1mm}} 
r@{\hspace{1mm}}r@{\hspace{2mm}} 
r@{\hspace{1mm}}r@{\hspace{1mm}}r@{\hspace{1mm}} 
r@{\hspace{1mm}}r@{\hspace{1mm}} 
r@{\hspace{1mm}}r@{\hspace{1mm}}  
r@{\hspace{1mm}}l@{\hspace{2mm}} 
c@{\hspace{1mm}}l@{\hspace{2mm}} 
c@{\hspace{2mm}}c@{\hspace{2mm}} 
c@{\hspace{2mm}}  
l@{\hspace{1mm}}l@{\hspace{1mm}}r} 
\hline
RA           & Dec            &  R     & Sk  & BI  & Mk & VFTS & MH & HSH & SMB & P & ST & CCE & BAT & Spect. & Ref & $\log L_{\rm Bol}$& Ref & Photon Flux   & $\log L^{t}_{\rm X}$  &$\log L^{tc}_{\rm X}/L_{\rm Bol}$ & Nature & Ref & PSC\\
h m s       & deg m s         &       &       &       &          &    &          &              &         &             &            &           &     & Type  &       &  $L_{\odot}$              &       & cm$^{-2}$\,s$^{-1}$ & erg\,s$^{-1}$ &             &    &    & source$\ddag$ \\
\hline

 05 38 32.28 & --69 05 44.6 & $\cdots$ & $\cdots$ & $\cdots$ & $\cdots$ & 382 & $\cdots$ &$\cdots$ & 226 & $\cdots$ & $\cdots$ & 2901 & $\cdots$ & O4--5\,V((fc))z & 4 & 5.31$\pm$0.13 & 3,5 &  $<$7.48$\times 10^{-8}$ & $<$31.68 & $<$--6.90$\pm$0.26 & $\cdots$ & $\cdots$ & $\cdots$\\
 05 38 32.54 & --69 04 32.0& $\cdots$ & $\cdots$ & $\cdots$ & 58b & 386 & $\cdots$ &$\cdots$ & $\cdots$ & 294 & $\cdots$ & $\cdots$ & $\cdots$ & O9\,V(n)+B1\,V: & 4,37 & 5.41$\pm$0.3: & 6 &  $<$7.96$\times 10^{-8}$ & $<$31.71 & $<$--6.98$\pm$0.38 & SB2 & 4,37 & $\cdots$\\
\hline
\end{tabular}
\end{center}
\end{table}
\end{landscape}

\addtocounter{table}{-1}

\begin{landscape}
\begin{table}
\begin{center}
\caption{(continued)}
\begin{tabular}{
c@{\hspace{1mm}}c@{\hspace{3mm}} 
r@{\hspace{1mm}} 
c@{\hspace{1mm}}r@{\hspace{1mm}} 
r@{\hspace{1mm}}r@{\hspace{2mm}} 
r@{\hspace{1mm}}r@{\hspace{1mm}}r@{\hspace{1mm}} 
r@{\hspace{1mm}}r@{\hspace{1mm}} 
r@{\hspace{1mm}}r@{\hspace{1mm}}  
r@{\hspace{1mm}}l@{\hspace{2mm}} 
c@{\hspace{1mm}}l@{\hspace{2mm}} 
c@{\hspace{2mm}}c@{\hspace{2mm}} 
c@{\hspace{2mm}}  
l@{\hspace{1mm}}l@{\hspace{1mm}}r} 
\hline
RA           & Dec            &  R     & Sk  & BI  & Mk & VFTS & MH & HSH & SMB & P & ST & CCE & BAT & Spect. & Ref & $\log L_{\rm Bol}$& Ref & Photon Flux   & $\log L^{t}_{\rm X}$  &$\log L^{tc}_{\rm X}/L_{\rm Bol}$ & Nature & Ref & PSC\\
h m s       & deg m s         &       &       &       &          &    &          &              &         &             &            &           &     & Type  &       &  $L_{\odot}$              &       & cm$^{-2}$\,s$^{-1}$ & erg\,s$^{-1}$ &             &    &    & source$\ddag$ \\
\hline

05 38 40.85 & --69 06 57.5  & $\cdots$ & $\cdots$ & $\cdots$ & $\cdots$ & 491 & $\cdots$ &$\cdots$ & 213 & 803 & $\cdots$ & 276 & $\cdots$ & O6\,V((fc)) & 4 & 5.43$\pm$0.16 & 3,5 & $<$8.66$\times 10^{-8}$ & $<$31.75 & $<$--6.96$\pm$0.27 & SB? & 4& $\cdots$ \\
 05 38 41.04 & --69 04 24.8 & $\cdots$ & $\cdots$ & $\cdots$ & 21 & 492 & $\cdots$ &$\cdots$ & $\cdots$ & 830 & $\cdots$ & $\cdots$ & $\cdots$ & O8\,V+O9.5:\,V & 4 & 5.28$\pm$0.3: & 6 &$<$7.68$\times 10^{-8}$ & $<$31.70 & $<$--6.86$\pm$0.38 & SB2 & 4 & $\cdots$\\
\hline
\end{tabular}
\end{center}
\end{table}
\end{landscape}

\addtocounter{table}{-1}

\begin{landscape}
\begin{table}
\begin{center}
\caption{(continued)}
\begin{tabular}{
c@{\hspace{1mm}}c@{\hspace{3mm}} 
r@{\hspace{1mm}} 
c@{\hspace{1mm}}r@{\hspace{1mm}} 
r@{\hspace{1mm}}r@{\hspace{2mm}} 
r@{\hspace{1mm}}r@{\hspace{1mm}}r@{\hspace{1mm}} 
r@{\hspace{1mm}}r@{\hspace{1mm}} 
r@{\hspace{1mm}}r@{\hspace{1mm}}  
r@{\hspace{1mm}}l@{\hspace{2mm}} 
c@{\hspace{1mm}}l@{\hspace{2mm}} 
c@{\hspace{2mm}}c@{\hspace{2mm}} 
c@{\hspace{2mm}}  
ll@{\hspace{1mm}}l@{\hspace{1mm}}r} 
\hline
RA           & Dec            &  R     & Sk  & BI  & Mk & VFTS & MH & HSH & SMB & P & ST & CCE & BAT & Spect. & Ref & $\log L_{\rm Bol}$& Ref & Photon Flux   & $\log L^{t}_{\rm X}$  &$\log L^{tc}_{\rm X}/L_{\rm Bol}$ & Nature & Ref & PSC\\
h m s       & deg m s         &       &       &       &          &    &          &              &         &             &            &           &     & Type  &       &  $L_{\odot}$              &       & cm$^{-2}$\,s$^{-1}$ & erg\,s$^{-1}$ &             &    &    & source$\ddag$ \\
\hline

05 38 42.96 & --69 06 05.2  & $\cdots$ & $\cdots$ & $\cdots$ & $\cdots$ & $\cdots$&  695 & 57    & $\cdots$ &  $\cdots$ & $\cdots$ & $\cdots$ & $\cdots$ & O3\,III(f*)    & 29 & 5.71$\pm$0.3: & 6 &$<$3.97$\times 10^{-7}$&$<$32.41&$<$--6.58$\pm$0.38&$\cdots$  & $\cdots$ & $\cdots$\\ 

05 38 43.04 & --69 03 44.9 & $\cdots$ & $\cdots$ & $\cdots$ & $\cdots$ & 537 & $\cdots$ &$\cdots$ & $\cdots$ & 1022 & $\cdots$ & $\cdots$ & $\cdots$ & O5\,V((fc))z & 4 & 5.19$\pm$0.13 & 3,5 & $<$5.91$\times 10^{-8}$ & $<$31.58 & $<$--6.89$\pm$0.26 & $\cdots$ & $\cdots$ & $\cdots$\\
\hline
\end{tabular}
\end{center}
\end{table}
\end{landscape}

\addtocounter{table}{-1}

\begin{landscape}
\begin{table}
\begin{center}
\caption{(continued)}
\begin{tabular}{
c@{\hspace{1mm}}c@{\hspace{3mm}} 
r@{\hspace{1mm}} 
c@{\hspace{1mm}}r@{\hspace{1mm}} 
r@{\hspace{1mm}}r@{\hspace{2mm}} 
r@{\hspace{1mm}}r@{\hspace{1mm}}r@{\hspace{1mm}} 
r@{\hspace{1mm}}r@{\hspace{1mm}} 
r@{\hspace{1mm}}r@{\hspace{1mm}}  
r@{\hspace{1mm}}l@{\hspace{2mm}} 
c@{\hspace{1mm}}l@{\hspace{2mm}} 
c@{\hspace{2mm}}c@{\hspace{2mm}} 
c@{\hspace{2mm}}  
ll@{\hspace{1mm}}l@{\hspace{1mm}}r} 
\hline
RA           & Dec            &  R     & Sk  & BI  & Mk & VFTS & MH & HSH & SMB & P & ST & CCE & BAT & Spect. & Ref & $\log L_{\rm Bol}$& Ref & Photon Flux   & $\log L^{t}_{\rm X}$  &$\log L^{tc}_{\rm X}/L_{\rm Bol}$ & Nature & Ref & PSC\\
h m s       & deg m s         &       &       &       &          &    &          &              &         &             &            &           &     & Type  &       &  $L_{\odot}$              &       & cm$^{-2}$\,s$^{-1}$ & erg\,s$^{-1}$ &             &    &    & source$\ddag$ \\
\hline

 05 38 51.20 & --69 05 59.4 & $\cdots$ & $\cdots$ & $\cdots$ & 6 & 656 & $\cdots$ &$\cdots$ & 61 & 1563 & $\cdots$ & 1979 & $\cdots$ & O7.5\,III(n)((f))p & 4 & 5.34$\pm$0.05 & 23 & $<$7.62$\times 10^{-8}$ & $<$31.69 & $<$--6.93$\pm$0.24 & SB1 & 4 & $\cdots$\\

05 38 51.31 & --69 04 08.5 & $\cdots$ & $\cdots$ & $\cdots$ & $\cdots$ & 657 & $\cdots$ &$\cdots$ & $\cdots$ & 1573 & $\cdots$ & $\cdots$ & $\cdots$ & O7\,II:(f)+OB: & 4,37 & 5.48$\pm$0.3: & 6 & $<$5.14$\times 10^{-8}$ & $<$31.52 & $<$--7.24$\pm$0.38 & SB2: & 4,37 & $\cdots$\\
\hline
\end{tabular}
\end{center}
{\footnotesize
1: \citet{2015A&A...574A..13E} 
2: \citet{2015A&A...575A..70M} 
3 :\cite{2018Sci...359...69S} 
4:  \citet{2014A&A...564A..40W} 
5: \citet{2017A&A...601A..79S} 
6: \citet{2013A&A...558A.134D} 
7: \citet{2014A&A...565A..27H} 
8: \citet{2003MNRAS.338.1025F} 
9: \citet{2017A&A...600A..81R} 
10: \citet{2020A&A...634A.118M} 
11: \citet{1992A&AS...92..729S} 
12: \citet{2011AcA....61..103G} 
13: \citet{1997A&A...320..500C} 
14: \citet{2014A&A...570A..38B} 
15: \citet{2008MNRAS.389..806S} 
16: \citet{2016AcA....66..421P} 
17: \citet{1990ApJ...348..471S} 
18: \citet{2002A&A...392..653C} 
19: \citet{2001MNRAS.324...18B} 
20: \citet{2011A&A...530A.108E} 
21 :\citet{2021MNRAS.507.5348V} 
22: \citet{2021A&A...648A..65C} 
23: \citet{1999A&AS..137...21B} 
24: \citet{2009AJ....137.3437B} 
25: \citet{2012ApJ...748...96M} 
26: \citet{2012A&A...546A..73H} 
27: D.J. Lennon (priv. comm.);  
28: \citet{2018A&A...615A.101D} 
29: \citet{1998ApJ...493..180M} 
30: \citet{2020MNRAS.499.1918B} 
31: \citet{1993AJ....106..560P} 
32: \citet{2016MNRAS.458..624C} 
33: \citet{1997ApJS..112..457W} 
34: \citet{1998ApJ...503..278S} 
35: \citet{2012A&A...542A..50D} 
36: \citet{Shenar2022} 
37. T.~Shenar et al. submitted

}
 \end{table}
\end{landscape}

\bsp	
\label{lastpage}
\end{document}